\documentclass[prd,aps,nofootinbib,superscriptaddress,preprintnumbers,epsf,psf]{revtex4}

\usepackage[a4paper]{geometry}
\usepackage{graphicx}% Include figure files
\usepackage{dcolumn}% Align table columns on decimal point
\usepackage{bm}% bold math
\usepackage{booktabs}
\usepackage{siunitx}
\usepackage{amsmath}
\usepackage{amssymb}
\usepackage{nicefrac}
\usepackage{float}
\usepackage{url}
\usepackage[normalem]{ulem}
\usepackage{natbib}
\usepackage{xcolor}
\usepackage{hypernat}
\usepackage[colorlinks,
            bookmarksnumbered=true,
            bookmarks=true,
            backref=page,
            hyperfootnotes=true,
            ]
           {hyperref}
  \hypersetup{
    linktocpage = true,
    linkcolor = blue,
    urlcolor = blue,
    colorlinks = true,
    linkcolor = teal,
    anchorcolor = blue,
    citecolor = teal,
    urlcolor = teal,
    pdftitle = {Hunting BFKL in semi-hard reactions at the LHC},                
    pdfauthor = {F.G. Celiberto},                 
    pdfsubject = {High-energy phenomenology},               
    pdfcreator = {F.G. Celiberto},               
    pdfproducer = {F.G. Celiberto},
    pdfstartview = {XYZ null null 1.25},
    pdfkeywords = {hep-ph, QCD, BFKL, DGLAP, LHC, semi-hard processes, inclusive processes, jets, jet phsysics, hadrons, hadron physics, hep-ph, phenomenology, physics, hep-th, theoretical physics, theory}, 
             }
\usepackage{backrefx}
  \renewcommand{\backrefpagesname}{Cited on page~}
  \renewcommand{\backrefpagesnames}{Cited on pages~}
\usepackage{doi}

\newcommand{\beq}{\begin{equation}}
\newcommand{\eeq}{\end{equation}}
\newcommand{\bea}{\begin{eqnarray}}
\newcommand{\eea}{\end{eqnarray}}
\newcommand{\real}{{\sf I}\kern-.12em{\sf R}}
\newcommand{\comp}{{\sf I}\kern-.50em{\sf C}}
\newcommand{\unity}{{\sf I}\kern-.54em{\sf 1}}
\newcommand{\as}{\alpha_s}
\newcommand{\LQCD}{\Lambda_{\rm QCD}}
\newcommand{\MSb}{\overline{\rm MS}}
\newcommand{\drv}{{\rm d}}
\newcommand{\CnLLA}{{\cal C}_n^{\rm LLA}}
\newcommand{\CnNLA}{{\cal C}_n^{\rm NLA}}
\newcommand{\CnDGLAP}{{\cal C}_n^{\rm DGLAP}}
\newcommand{{\Jethad}}{\tt JETHAD}
\newcommand{\Lexa}{{\tt LExA}}
\newcommand{\Hell}{{\tt HELL}}
\newcommand{\Apfel}{{\tt APFEL}}

\def\spose#1{\hbox to 0pt{#1\hss}}
\def\ltapprox{\mathrel{\spose{\lower 3pt\hbox{$\mathchar"218$}}
 \raise 2.0pt\hbox{$\mathchar"13C$}}}

\newcounter{appcnt}

\makeatletter

\renewcommand*{\p@subsection}{}

\renewcommand*{\p@subsubsection}{}
\makeatother

\begin{document}

\setcounter{secnumdepth}{4}
\setcounter{tocdepth}{4}

\title{\Large Hunting BFKL in semi-hard reactions at the LHC}
% \author{Andr\`ee Dafne Bolognino}
% \email{ad.bolognino@unical.it}
% \affiliation{Dipartimento di Fisica, Universit\`a della Calabria, I-87036 Arcavacata di Rende, Cosenza, Italy}
% \affiliation{INFN, Gruppo collegato di Cosenza, I-87036 Arcavacata di Rende,
% Cosenza, Italy}
\author{\large \bf Francesco Giovanni Celiberto}
\email{francescogiovanni.celiberto@unipv.it}
\affiliation{Dipartimento di Fisica, Universit\`a degli Studi di Pavia, I-27100 Pavia, Italy}
\affiliation{INFN, Sezione di Pavia, I-27100 Pavia, Italy}
\affiliation{European Centre for Theoretical Studies in Nuclear Physics and Related Areas (ECT*), I-38123 Villazzano, Trento, Italy}
\affiliation{Fondazione Bruno Kessler (FBK), I-38123 Povo, Trento, Italy}
%\affiliation{INFN-TIFPA Trento Institute of Fundamental Physics and Applications, I-38123 Povo, Trento, Italy}
% \author{Dmitry Yu. Ivanov}
% %\email{d-ivanov@math.nsc.ru}
% \affiliation{Sobolev Institute of Mathematics, 630090 Novosibirsk, Russia}
% \affiliation{Novosibirsk State University, 630090 Novosibirsk, Russia}
% \author{Alessandro Papa}
% %\email{alessandro.papa@fis.unical.it}
% \affiliation{Dipartimento di Fisica dell'Universit\`a della Calabria \\
% I-87036 Arcavacata di Rende, Cosenza, Italy}
% \affiliation{INFN - Gruppo collegato di Cosenza, I-87036 Arcavacata di Rende,
% Cosenza, Italy}
%\author{...}

% \date{\today}% It is always \today, today,
%              %  but any date may be explicitly specified

\begin{abstract}
\vspace{0.50cm}
\hrule \vspace{0.75cm}
The agreement between calculations inspired by the resummation of energy logarithms, known as BFKL approach, and experimental data in the semi-hard sector of QCD has become manifest after a wealthy series of phenomenological analyses. However, the contingency that the same data could be concurrently portrayed at the hand of fixed-order, DGLAP-based calculations, has been pointed out recently, but not yet punctually addressed. Taking advantage of the richness of configurations gained by combining the acceptances of CMS and CASTOR detectors, we give results in the full next-to-leading logarithmic approximation of cross sections, azimuthal correlations and azimuthal distributions for three distinct semi-hard processes, each of them featuring a peculiar final-state exclusiveness. Then, making use of disjoint intervals for the transverse momenta of the emitted objects, {\it i.e.} $\kappa$\textit{-windows}, we clearly highlight how high-energy resummed and fixed-order driven predictions for semi-hard sensitive observables can be decisively discriminated in the kinematic ranges typical of current and forthcoming analyses at the LHC. The scale-optimization issue is also introduced and explored, while the uncertainty coming from the use of different PDF and FF set is deservedly handled. Finally, a brief overview of {\Jethad}, a numerical tool recently developed, suited for the computation of inclusive semi-hard reactions is provided.
\vspace{0.75cm} \hrule
\vspace{0.75cm}
{
 \setlength{\parindent}{0pt}
}
\end{abstract}

\maketitle

\newpage

\hrule

\begingroup
 \hypersetup{linktoc = all, 
             linkcolor = black
             }
 \tableofcontents
\endgroup

\vspace{0.75cm} \hrule \vspace{0.75cm}

\parskip 0.10cm

\section{Introductory remarks}
\label{introd}

\begin{figure}[t]
 \centering
 \includegraphics[scale=0.65]{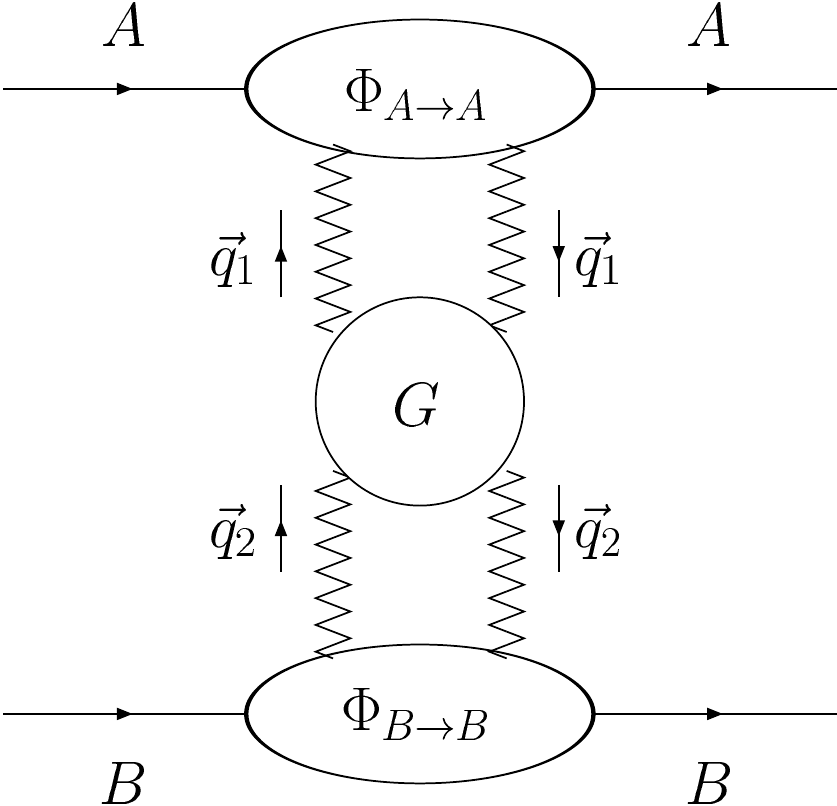}
 \caption[]
 {Diagrammatic representation of the imaginary part of the BFKL amplitude.}
 \label{fig:BFKL-amplitude}
\end{figure}

The search for evidence of New Physics is in the viewfinder of current and forthcoming analyses at the Large Hadron Collider (LHC). This is the best time to shore up our knowledge of strong interactions though, the high luminosity and the record energies reachable widening the horizons of kinematic sectors uninvestigated so far. A broad class of processes, called \emph{diffractive semi-hard} reactions~\cite{Gribov:1984tu}, \emph{i.e} where the scale hierarchy, $s \gg \{ Q^2 \} \gg \Lambda_{\rm QCD}^2$ ($s$~is the squared center-of-mass energy, $\{ Q \}$ a (set of) hard scale(s) characteristic of the process and $\Lambda_{\rm QCD}$ the QCD scale), is stringently preserved, gives us a faultless chance to test perturbative QCD (pQCD) in new and quite original ways\footnote{Diffractive reactions are characterized by colorless exchanges (\emph{i.e.} vacuum quantum numbers) in the $t$-channel, this leading to large, non exponentially suppressed rapidity intervals. Depending on the number and on the rapidity of final-state objects, we classify the given process as single or double diffraction (\emph{dissociation}), central diffraction (\emph{double-Pomeron} exchange), and so on. It is worth to note that not all semi-hard processes fall in the diffractive class. As an example, di-jet production in \emph{ultra-peripheral} collisions of heavy ions can easily feature a semi-hard ordering and also large rapidity intervals among the recoiling hadrons and the di-jet system, but the condition of vacuum quantum numbers exchanged in the $t$-channel is violated. 
In that case, in a kinematic configuration where photons carry just a fraction of the energy of the corresponding parent nuclei, observable jet transverse momenta are much larger than $\LQCD$ but, at the same time, much smaller than $\sqrt{s}$. Therefore, although this process is not diffractive, a strict semi-hard scale ordering is respected.
From here on out, we will implicitly refer to diffractive scatterings when considering semi-hard production channels.}.

In the kinematical regime, known as \emph{Regge limit}, where $s$ is much larger than the Mandelstam variable~$t$, fixed-order calculations in pQCD based on collinear factorization fail to catch the effect of large-energy logarithmic contributions, entering the perturbative series with a power increasing with the order and thus balancing the slightness of the strong coupling, $\alpha_s$. The Balitsky--Fadin--Kuraev--Lipatov (BFKL) approach~\cite{BFKL_1,BFKL_2,BFKL_3,BFKL_4} serves as the most adequate tool to perform the all-order resummation of these large-energy logarithms 
both in the leading approximation (LLA), namely of terms proportional to $(\alpha_s \ln s)^n$
and the next-to-leading approximation (NLA), namely of terms proportional to $\alpha_s (\alpha_s \ln s)^n$.
In the BFKL formalism, the imaginary part the amplitude of a hadronic process (thence its cross section, in the case of inclusive final states, via the \emph{optical theorem}~\cite{optical_theorem_Newton}) is elegantly expressed via a suitable convolution, pictorically depicted in Fig.~\ref{fig:BFKL-amplitude}, between two impact factors, portraying the transition from each parent particle to the final-state object(s) belonging to its \emph{fragmentation region}, and a process-independent Green's function. The BFKL Green's function depends on energy and its evolution is regulated by an integral equation, whose kernel is known at the next-to-leading order (NLO) both for forward
scattering~\cite{Fadin:1998py,Ciafaloni:1998gs}, \emph{i.e.} for $t = 0$ and color singlet in the $t$-channel, and for any fixed (not growing with energy) $t$ and any possible two-gluon color configuration in the $t$-channel~\cite{Fadin:1998jv,Fadin:2000kx,Fadin:2004zq,Fadin:2005zj}.
Conversely, impact factors strictly depend on the final state and on the hard scale, but not on the energy. Thus, the number of processes which can be investigated via the BFKL resummation is shortened by the narrow list of available impact factors, just few of them being calculated at the NLO: 1) colliding-parton (quarks and gluons) impact
factors~\cite{Fadin:1999de,Fadin:1999df,Ciafaloni:1998kx,Ciafaloni:1998hu,Ciafaloni:2000sq}, which represent the landmark for the calculation of the 2) forward-jet impact
factor~\cite{Bartels:2001ge,Bartels:2002yj,Caporale:2011cc,Ivanov:2012ms,Colferai:2015zfa} and of
the 3) forward charged-light hadron one~\cite{Ivanov:2012iv}, 4) the impact
factor specific of the ($\gamma^* \to$ LVM)
transition~\cite{Ivanov:2004pp} at leading twist, where LVM stands for light vector meson, and 5) and the one detailing the ($\gamma^* \to \gamma^*$)
subprocess~\cite{Bartels:2000gt,Bartels:2001mv,Bartels:2002uz,Bartels:2003zi,Bartels:2004bi,Fadin:2001ap,Balitsky:2012bs}.

With the aim of deepening our understanding of the high-energy regime, a notable variety of semi-hard reactions (see Ref.~\cite{Celiberto:2017ius} for applications) has been proposed so far.

In particular, jet impact factors have been widely employed to describe the inclusive hadroproduction of two jets tagged with high transverse momenta
and large separation in rapidity (Mueller--Navelet configuration~\cite{Mueller:1986ey}),
for which several phenomenological studies have been carried out so far~\cite{Marquet:2007xx,Colferai:2010wu,Caporale:2012ih,Ducloue:2013wmi,Ducloue:2013bva,Caporale:2013uva,Ducloue:2014koa,Caporale:2014gpa,Ducloue:2015jba,Caporale:2015uva,Celiberto:2015yba,Celiberto:2015mpa,Celiberto:2016ygs,Celiberto:2016vva,Caporale:2018qnm,Chachamis:2015crx}.
The comparison of theoretical distributions on the azimuthal-angle distance between the two jets, $\varphi = \varphi_{J_1} - \varphi_{J_2} - \pi$, with experimental analyses~\cite{Khachatryan:2016udy} recently conducted by the Compact-Muon-Solenoid (CMS) collaboration at \textsc{Cern}, has addressed \emph{projections} on azimuthal-angle observables and ratios between any two of them,
\begin{equation}
 \label{Rnm_introd}
 R_{n0} \equiv \langle \cos (n \varphi) \rangle \; , \quad
 R_{nm} \equiv \frac{\langle \cos (n \varphi) \rangle}{\langle \cos (m \varphi) \rangle} \; ,
\end{equation}
as very favorable observables in the search for distinct signals of the onset of BFKL dynamics.

The same impact factors, taken at the leading-order (LO) of the perturbative expansion, enter the calculation of
the so-called \emph{generalized azimuthal correlations}, 
\begin{equation}
 \label{generalized_azimuthal_correlations_introd}
 C_{n_1 \cdots n_{N-1}} =
 \left\langle 
 \prod_{k=1}^{N-1} \cos\left(n_k \, \varphi_{[k,k+1]}\right)
 \right\rangle = \hspace{-0.1cm} 
 \int_0^{2\pi} \hspace{-0.4cm} \drv \varphi_1 
 \cdots \hspace{-0.1cm} 
 \int_0^{2\pi} \hspace{-0.4cm} \drv \varphi_N
 \prod_{k=1}^{N-1} \cos\left(n_k \,  \varphi_{[k,k+1]}\right)
 \drv \sigma^{N\text{-jet}}
\end{equation}
and of their ratios, defined in the framework of the inclusive multi-jet ($N$-jet) hadroproduction~\cite{Caporale:2015vya,Caporale:2015int,Caporale:2016soq,Caporale:2016vxt,Caporale:2016xku,Celiberto:2016vhn,Caporale:2016djm,Caporale:2016lnh,Caporale:2016zkc} at LHC collision energies. In Eq.~(\ref{generalized_azimuthal_correlations_introd}), $\varphi_{[k,k+1]} = \varphi_k - \varphi_{k+1} - \pi$, with $\varphi_k$ the azimuthal angle of the $k^{\rm th}$ emitted jet, and $\drv \sigma^{N\text{-jet}}$ stands for the differential partonic cross section.

Then, forward-hadron impact factors allowed us for a complete NLA study of the inclusive detection of two identified hadrons (composed of light quarks and well separated in rapidity)~\cite{Celiberto:2016hae,Celiberto:2016zgb,Celiberto:2017ptm}, and of the inclusive hadron-jet hadroproduction~\cite{Bolognino:2018oth,Bolognino:2019yqj,Bolognino:2019cac,Celiberto:2020rxb}.

The theoretical description of the ($\gamma^*$-$\gamma^*$) total cross section relies on the ($\gamma^* \to \gamma^*$) impact factor. Although being recognized as a privileged channel for the manifestation of the BFKL dynamics, comparisons between the large number of BFKL predictions~\cite{Brodsky:2002ka,Brodsky:1998kn,Caporale:2008is,Zheng:2013uja,Chirilli:2014dcb,Ivanov:2014hpa,Deak:2020zay} with the only available data from the LEP2 \textsc{Cern} experiment were ineffective due to the relatively low center-of-mass energy and the limited accuracy of the detector.
Quite recently~\cite{Ermolaev:2017ttc,Ermolaev:2017uhy}, the amplitude for the light-by-light elastic scattering via a single-quark loop was calculated in the so-called \emph{double-logarithmic} approximation (DLA). This approach\footnote{As a meaningful remark, we note that \emph{sub-eikonal} corrections, inevitably neglected by BFKL, are instead genuinely included in DLA-based calculations. Thus, DLA resummation comes out as a powerful tool to investigate polarization effects in the high-energy/small-$x$ regime.  In this context, the Bartels--Ermolaev--Ryskin (BER) approach permits us to calculate flavor non-singlet~\cite{Bartels:1995iu} and singlet contributions~\cite{Bartels:1996wc} to the $g_1$ helicity structure function in the Deep Inelastic Scattering (DIS) at $\mbox{small-}x$. Another way to resum double-logarithmic powers in the definition of small-$x$ polarized parton densities is represented by the Kovchegov--Pitonyak--Sievert (KPS) formalism~\cite{Kovchegov:2015pbl,Kovchegov:2016weo,Kovchegov:2016zex,Kovchegov:2017jxc,Kovchegov:2017lsr}, whose evolution equations are built in terms of (polarized) \emph{Wilson lines}~\cite{Wilson:1974sk} and account for saturation effects. Originally aimed at the study of quark and gluon helicity distributions, the KPS scheme was then extended to valence-quark transversity~\cite{Kovchegov:2018zeq} and orbital angular momentum~\cite{Kovchegov:2019rrz} in the small-$x$ limit.} accounts for the resummation of those contributions proportional to $(\alpha_s \ln s \ln s)^n$ that, at variance with gluon interactions in the BFKL ladder, appear in the quark-antiquark exchange~\cite{Gorshkov:1966ht,Gorshkov:1966hu,Kirschner:1982xw,Kirschner:1983di}, giving rise to a different asymptotic behavior in the power of $s$ with respect to the single-logarithmic trend of the BFKL Pomeron (for more details, see Refs.~\cite{Bartels:2003dj,Bartels:2004mu}).

The exclusive leptoproduction of two LVMs was investigated by encoding the ($\gamma^* \to$~LVM) impact factors in the definition of the BFKL amplitude.
In particular, first studies at Born level were done in Ref.~\cite{Pire:2005ic} with longitudinally polarized virtual photons, and in Ref.~\cite{Segond:2007fj} by accounting also for the photon transverse polarization. Then, a LLA treatment supplemented by improvements of the LO BFKL kernel was proposed in Ref.~\cite{Enberg:2005eq}, while a full NLA analysis was performed in Refs.~\cite{Ivanov:2005gn,Ivanov:2006gt}.

LO impact factors for the production of forward heavy-quark pairs~\cite{Ginzburg:1996vq,Bolognino:2019yls} represent the key ingredients for phenomenological analyses with partial NLA resummation effects, by including next-to-leading logarithmic corrections to the BFKL Green's function, of the heavy-quark di-jet photo-~\cite{Celiberto:2017nyx,Bolognino:2019ouc} and hadroproduction~\cite{Bolognino:2019yls,Bolognino:2019ouc}, tracing the path towards a prospective NLA BFKL treatment of heavy-flavored-meson production channels. With the same accuracy, $J/\Psi$-jet~\cite{Boussarie:2017oae}, Higgs-jet~\cite{Celiberto:2020tmb} and heavy-light di-jet~\cite{Bolognino:2021mrc} correlations have been proposed, while a full LLA description, together with consistency-inspired collinear improvements~\cite{Kwiecinski:1996td} to the BFKL kernel, of the forward Drell--Yan dilepton production plus a backward jet has been recently conducted~\cite{Golec-Biernat:2018kem}.

All the possibilities presented above fall into a distinctive family of reactions, where at least two final-state particles are always emitted with large mutual separation in rapidity, together (in the inclusive channel) with a secondary, undetected gluon system. Another interesting subclass of semi-hard processes is represented by those final states featuring the tag of a single forward object in lepton-proton or proton-proton scatterings. This engaging configuration offers the chance to define an unintegrated, \emph{transverse-momentum-dependent} gluon distribution (UGD) in the proton.
This scheme is known as {\em high-energy factorization}\footnote{Some ambiguities on the definition of the high-energy factorization, also known as $\kappa$-factorization, arose in the literature over the course of time. Originally developed in Refs.~\cite{Catani:1990eg,Levin:1991ry}, it consists in a scheme, valid in the high-energy limit, where amplitudes (or cross sections) factorize into a convolution among off-shell matrix elements and transverse-momentum-dependent parton distribution functions. Then, in the BFKL approach, the matrix element corresponds to the forward impact factor describing the emission of the final-state particle, while the parton content, almost entirely driven by gluon evolution, is embodied by the UGD.}.
In its standard definition, the analytic structure of the UGD is given in terms of a convolution between the BFKL gluon Green's function and a soft proton impact factor. Being a non-perturbative quantity, the UGD is not well known and several parametrizations have been developed so far (see, for instance, Refs.~\cite{Blumlein:1995eu,Ryskin:1995wz,Kwiecinski:1997ee,GolecBiernat:1998js,GolecBiernat:1999qd,Bartels:2000hv,Jung:1998mi,Jung:2000hk,Ivanov:2000cm,Ivanov:2003iy,Andersen:2003xj,Kimber:2001sc,Watt:2003mx,Kutak:2012rf,Hentschinski:2012kr,Hautmann:2013tba,Hautmann:2017fcj,Hautmann:2017xtx,Martinez:2018jxt,Andersen:2006pg,Andersson:2002cf,Angeles-Martinez:2015sea}).

A phenomenological UGD, incorporating NLA BFKL contributions together with a relatively simple model for the proton impact factor, was originary employed in the analysis of DIS structure functions~\cite{Hentschinski:2012kr} and then in the study of single bottom-quark hadroproduction~\cite{Chachamis:2015ona} and of quarkonium-state ($J/\Psi$ or $\Upsilon$) photoproduction~\cite{Bautista:2016xnp,Garcia:2019tne}. The same model allowed us to improve the description of the inclusive forward Drell--Yan dilepton production at the LHC~\cite{Celiberto:2018muu}, previously investigated with LLA accuracy and in the dipole formalism with saturation corrections~\cite{GolecBiernat:2010de,Motyka:2014lya,Basso:2015pba,Schafer:2016qmk,Brzeminski:2016lwh,Motyka:2016lta}. On the other hand, in Refs.~\cite{Bolognino:2018rhb,Bolognino:2018mlw,Bolognino:2019bko,Celiberto:2019slj} it was highlighted how comparisons with HERA data on helicity-dependent observables for the forward polarized $\rho$-meson electroproduction permit to constrain the $\kappa$-shape of the UGD (pioneering studies on the diffractive production of $\rho$ mesons in the high-energy factorization can be found in Refs.~\cite{Anikin:2009bf,Anikin:2011sa,Besse:2012ia,Besse:2013muy}). Quite recently~\cite{Bolognino:2019pba}, the study of helicity-conserving amplitudes and cross sections was extended to the case of the single forward $\phi$-meson emission, gauging the effect of the inclusion of effective strange-quark masses and calculating the ($\gamma^* \to \phi$) massive impact factor~\cite{Cisek:2010jk} in the light-cone wave-function approach~\cite{Terentev:1976jk,Berestetsky:1977zk,Kondratyuk:1979gj,Bjorken:1970ah,Lepage:1980fj,Brodsky:1997de,Heinzl:1998kz}.

Not enough high center-of-mass energies, leading to insufficiently large rapidity distances among the final-state detected particles, had been so far the weakness point in the search for unambiguous signals of the high-energy resummation. In addition, too inclusive observables were considered\footnote{An astonishing example is the growth with energy of DIS structure functions at small-$x$, where a fair agreement with HERA data was obtained both by BFKL predictions~\cite{Hentschinski:2013id} and by fixed-order based calculations. On the other side, strong statements have been made in Ref.~\cite{Ball:2017otu}, highlighting how a BFKL-like resummation, encoded in the evolution of collinear parton distributions, clearly improves the description of HERA data, in comparison with a pure fixed-order approach.}. 
The current and forthcoming record energies, together with the superior azimuthal-angle resolution of detectors provided by the LHC, offer us a peerless opportunity to disentangle the BFKL dynamics from other resummation mechanisms.

The first analysis on azimuthal-angle decorrelations in the Mueller--Navelet channel, conducted by the CMS collaboration at $\sqrt{s} = 7$ TeV~\cite{Khachatryan:2016udy}, gave a valid indication that the considered kinematic domain lies \emph{in between} the sectors described by the BFKL and the Dokshitzer--Gribov--Lipatov--Altarelli--Parisi (DGLAP)~\cite{DGLAP,DGLAP_2,DGLAP_3,DGLAP_4,DGLAP_5} evolution, whereas clearer manifestations of high-energy signatures are expected to be more definite at increasing collision energies. Then, in a recent pioneering study~\cite{Celiberto:2015yba,Celiberto:2015mpa}, it was pointed out how the pure adoption of partially \emph{asymmetric} configurations for the transverse momenta of the two tagged jets allows for a clear separation between BFKL-resummed and fixed-order predictions. More in general, the use of $\kappa$-asymmetric windows has various benefits. On the one side, it allows to dampen the Born contribution,
thus heightening effects of the additional undetected hard-gluon radiation. This enhances the imprints of BFKL with respect to the DGLAP ones in the cross section, suppressing at the same time possible instabilities observed in fixed-order calculations at the NLO~\cite{Andersen:2001kta,Fontannaz:2001nq}. On the other side, violation of the energy-momentum conservation in the NLA is definitely quenched with respect to what happens in the LLA (see Ref.~\cite{Ducloue:2014koa}).

\begin{figure}[t]
 \centering
 \includegraphics[scale=0.4]{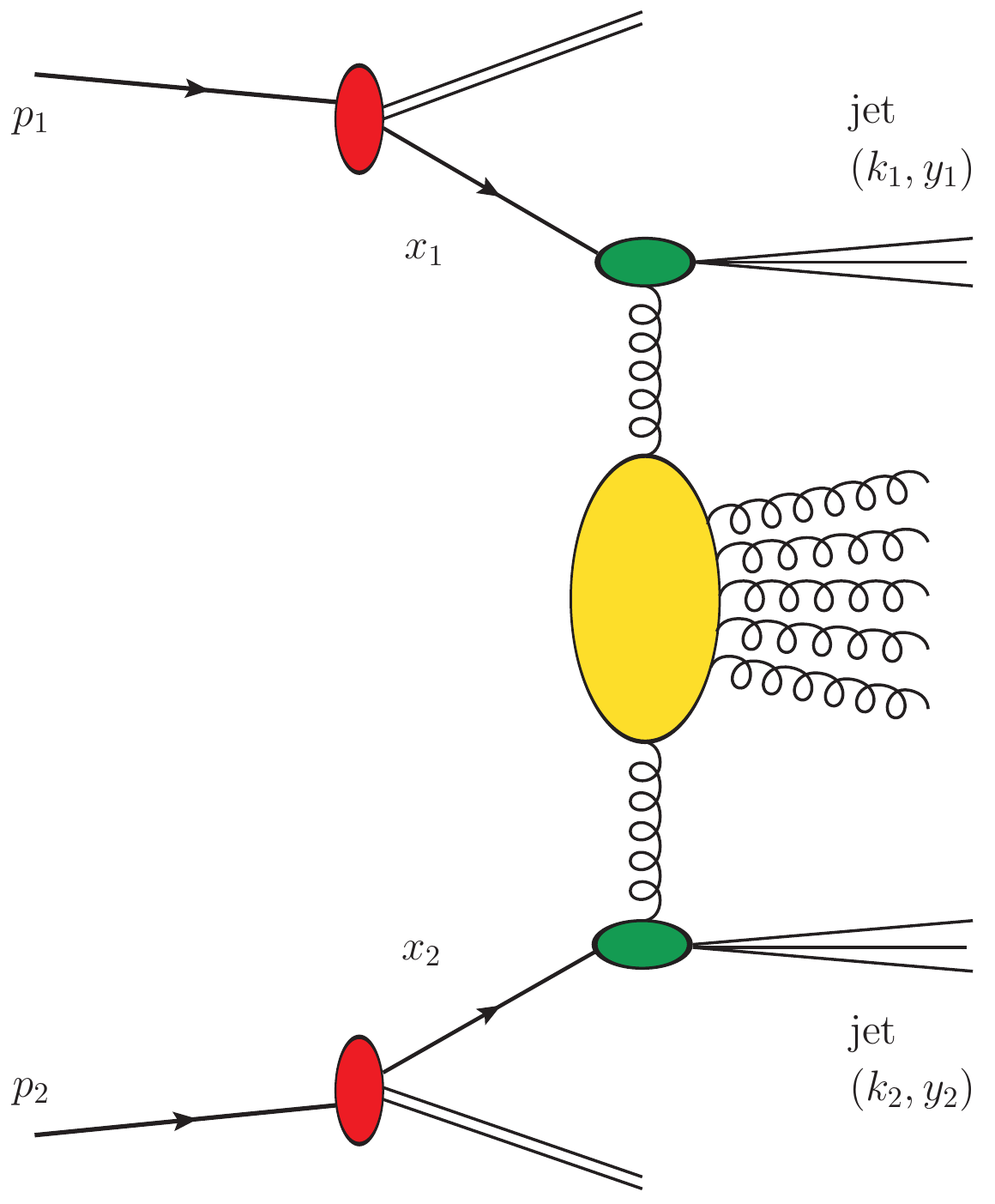}
 \hspace{0.75cm}
 \includegraphics[scale=0.4]{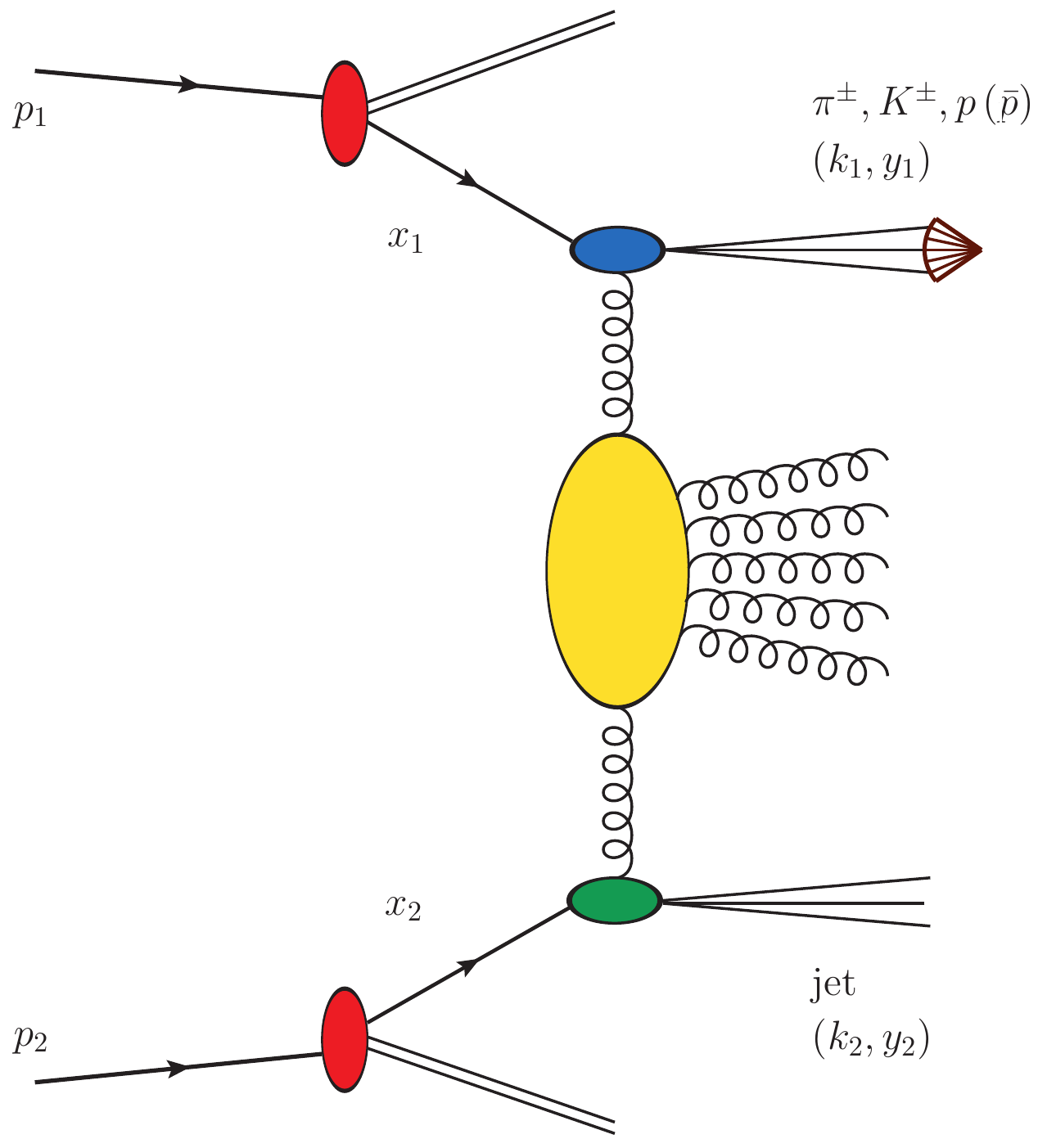}
 \hspace{0.75cm}
 \includegraphics[scale=0.4]{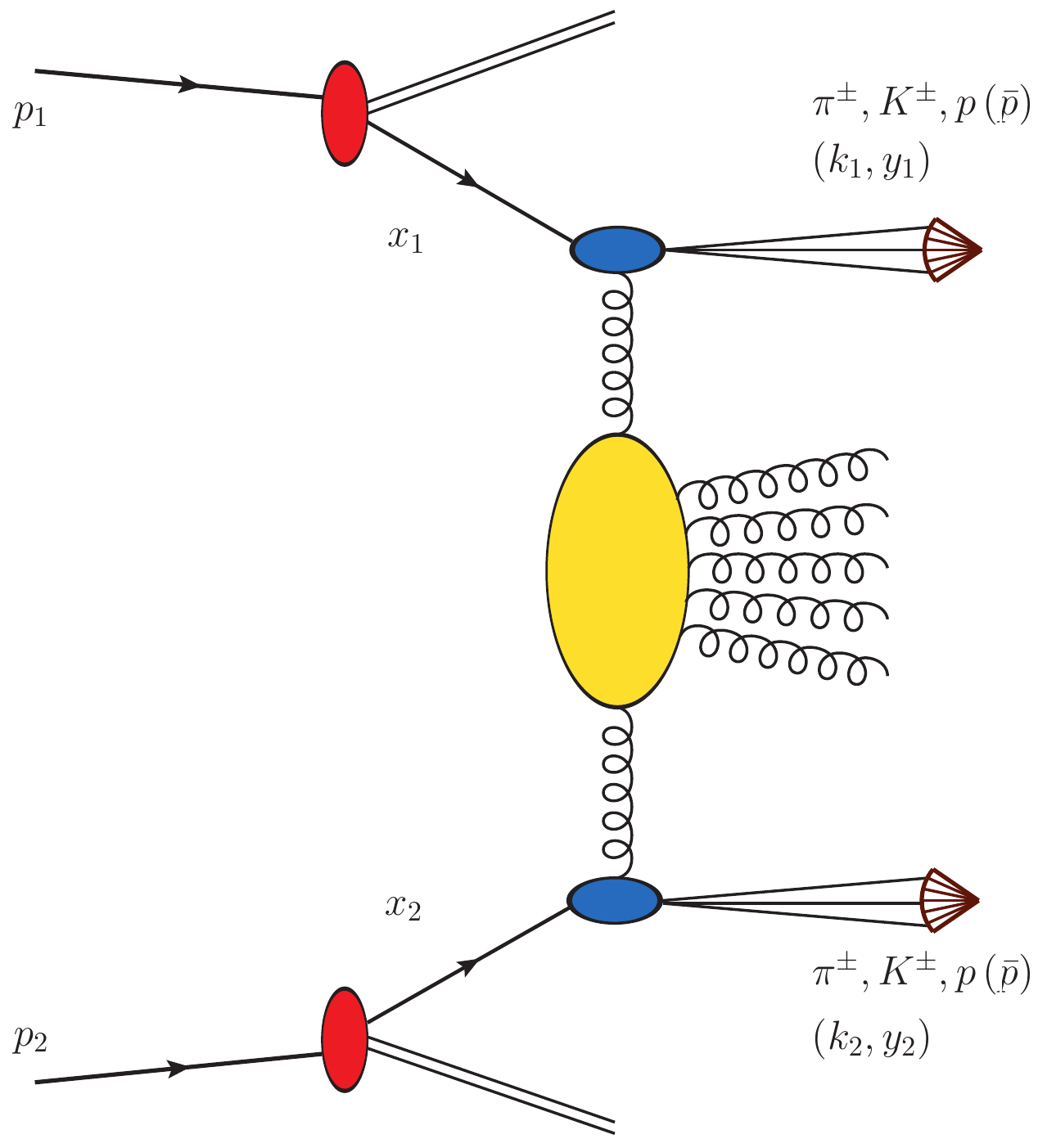}
 \\ \vspace{0.25cm}
 a) Mueller--Navelet jet channel \hspace{1.00cm}
 b) Inclusive hadron-jet production \hspace{1.00cm}
 c) Inclusive di-hadron production
 \caption[]
 {Diagrammatic representation of the three semi-hard processes investigated. The BFKL gluon Green's function is schematically represented in yellow, while the hard factor of the jet (hadron) impact factor is portrayed by the green (blue) blob. Collinear PDFs describing the incoming partons are given in red, whereas the final-state hadron FFs are pictorically expressed through the bordeaux arrowheads.}
 \label{fig:processes}
\end{figure}

The aim of this paper is to improve and extend the analysis previously discussed, setting a common framework for the description of inclusive semi-hard reactions and studying the rapidity-behavior of azimuthal correlations and distributions for a selection of three of them, namely: a) Mueller--Navelet jet, b) hadron-jet and c) di-hadron production at the LHC~(Fig.~\ref{fig:processes}). Our choice falls into a subclass of those, hadron-initiated processes, for which the theoretical description can be afforded with full NLA accuracy. Each kind of final-state object brings with itself some unique and distinctive features. On the one hand, jet emissions can be detected either in the barrel and in the endcap CMS calorimeters or in the CASTOR very-backward detector (the acronym stands for ``Centauro And Strange Object Research''), allowing for a wider rapidity range. They also permit to compare different parametrizations for parton distribution functions (PDFs) and for jet algorithms. On the other hand, hadron detection is feasible only inside the barrel calorimeter of CMS, this leading to a significant reduction in rapidity with respect to jets. It gives the possibility, however, to probe and constrain not only PDFs, but also fragmentation functions (FFs) entering the expression of the hadronic cross section. The limiting acceptances imposed by the hadron identification can be compensated by considering the concurrent emission of a forward (backward) hadron and a backward (forward) jet, as recently proposed in a first work on hadron-jet correlations~\cite{Bolognino:2018oth}.
Thus, in our study we take advantage of the window of opportunites offered by all the possible final-state combinations given in Fig.~\ref{fig:processes}, shaped on CMS and CASTOR acceptances.

Imposing disjoint intervals for final-state emissions in the transverse-momentum plane ($\kappa$\textit{-windows}) to shed light on the transition region between the two resummations, we provide indisputable evidence that the onset of BFKL dynamics can be disentangled from DGLAP-sensitive contaminations at current energies and kinematic configurations of prospective experimental studies at the LHC.

At the same time, we address the issue related to the ``optimal'' value for the renormalization scale (together with a supplementary analysis on single hadron-species detection) and gauge the effect of the uncertainty coming from the use of different PDF and FF parametrizations.

Finally, we present the main features of {\Jethad}, an object-based interface we have been developing, pursuing the goal to provide the scientific community with a common reference software for the phenomenological study of inclusive semi-hard processes.

The paper reads as follows: in
Section~\ref{theory} a general framework for the three considered reactions is set; in
Section~\ref{results} results for observables of interest are presented and discussed;
Section~\ref{conclusions} brings our conclusions.

\section{Theoretical setup}
\label{theory}

A comprehensive expression of the processes under exam reads (see panels of Fig.~\ref{fig:processes}):
\begin{equation}
\label{processes}
 {\rm proton}(p_1) + {\rm proton}(p_2) \to O_1(k_1, y_1) + X + O_2(k_2 , y_2) \;,
\end{equation}
where $O_{i=1,2} = \{{\rm hadron}, {\rm jet}\}$ are the final-state objects\footnote{In the case of hadron emission, the \emph{sum} over charged-light hadrons, $\pi^\pm$, $K^\pm$, $p(\bar{p}),$ is intended. A complementary analysis, focused on the effect of the scale choice for single hadron-species detection, is given in Section~\ref{had_species}.}, emitted with high transverse momenta, $|\vec k_{1,2}|$~$\equiv$~$\kappa_{1,2}$~$\gg$~$\LQCD$, and large rapidity separation, $ Y \equiv y_1 - y_2$, while $X$ is a secondary, undetected hadronic system.
The protons' momenta, $p_{1,2}$ are taken as Sudakov vectors satisfying $p^2_{1,2} = 0$ and $(p_1 p_2) = s/2$,  allowing for a suitable decomposition of the momenta of the produced objects:
\begin{equation}
\label{sudakov}
 k_{1,2} = x_{1,2} p_{1,2} + \frac{\vec k_{1,2}^2}{x_{1,2} s} p_{2,1} + k_{{1,2 \perp}} \;, \qquad k_{1,2 \perp}^2 = - \vec k_{1,2}^2 \equiv - \kappa_{1,2}^2 \; .
\end{equation}

In the center-of-mass system, the final-state longitudinal momentum fractions, $x_{1,2}$, are related to the respective rapidities through the expressions
$y_{1,2} = \pm \frac{1}{2}\ln\frac{x_{1,2}^2 s}
{\kappa_{1,2}^2}$,
so that $\drv y_{1,2} = \pm \frac{\drv x_{1,2}}{x_{1,2}}$, 
and $Y = y_1 - y_2 = \ln\frac{x_1 x_2 s}{\kappa_1\kappa_2}$, here the
space part of the four-vector $p_1$ parallel to the beam axis being taken positive.

We can present a general formula for the processes under consideration~(Eq.~(\ref{processes})), making use of collinear factorization:
\begin{equation}
\label{sigma_collinear}
 \frac{\drv \sigma}{\drv x_1 \drv x_2 \drv^2 \vec k_1 \drv^2 \vec k_2}
 =\sum_{\alpha,\beta=q,{\bar q},g}\int_0^1 \drv x_1 \int_0^1 \drv x_2\ f_\alpha\left(x_1,\mu_{F1}\right)
 \ f_\beta\left(x_2,\mu_{F2}\right)
 \frac{\drv {\hat\sigma}_{\alpha,\beta}\left(\hat s,\mu_{F1,2}\right)}
 {\drv x_1 \drv x_2 \drv^2 \vec k_1 \drv^2 \vec k_2}\;,
\end{equation}
where the ($\alpha, \beta$) indices run over the parton types 
(quarks $q = u, d, s, c, b$;
antiquarks $\bar q = \bar u, \bar d, \bar s, \bar c, \bar b$; 
or gluon $g$), $f_{\alpha,\beta}\left(x, \mu_{F1,2} \right)$ are the initial proton PDFs; 
$x_{1,2}$ stand for the longitudinal fractions of the partons involved in the hard
subprocess, whereas $\mu_{F1}$~($\mu_{F2}$) is the factorization scale characteristic of the fragmentation region of the upper (lower) parent proton in panels of Fig.~\ref{fig:processes};
$\drv \hat\sigma_{\alpha,\beta}\left(\hat s,\mu_{F1,2} \right)$ denotes
the partonic cross section and $\hat s \equiv x_1x_2s$ is the squared center-of-mass energy of the partonic collision.

\subsection{Cross section in the NLA BFKL}
\label{cross_section}

We can decompose the cross section (see Ref.~\cite{Caporale:2012ih} for further details) into a suitable Fourier expansion of the azimuthal coefficients, ${\cal C}_n$, 
getting so:
\begin{equation}
\frac{\drv \sigma}
{\drv y_1 \drv y_2\, \drv \kappa_1 \, \drv \kappa_2 \drv \varphi_1 \drv \varphi_2}
=\frac{1}{(2\pi)^2}\left[{\cal C}_0 + 2 \sum_{n=1}^\infty  \cos (n\varphi )\,
{\cal C}_n\right]\, ,
\end{equation}
where $\varphi \equiv \varphi_1-\varphi_2-\pi$, with $\varphi_{1,2}$ the azimuthal angles of the tagged objects, $O_{1,2}$, and the Jacobian for the change of variables has been taken into account.
The zeroth coefficient, ${\cal C}_{n = 0}$, gives the $\varphi$-averaged contribution to the cross section, while the other ones, ${\cal C}_{n \neq 0}$, are simply called \textit{azimuthal-correlation} coefficients. 
It is known that various expressions for ${\cal C}_n$, equivalent in the NLA BFKL approximation, exist (for an extensive study of them and of their peculiarities, see Ref.~\cite{Caporale:2014gpa}). For the purposes of the analysis proposed in this work, it is sufficient to consider just one representation, the so-called \textit{exponentiated} one, where the following global formula for the ${\cal C}_n \equiv \CnNLA$ coefficients holds with NLA BFKL accuracy:
\[
 \CnNLA \equiv \int_0^{2\pi} \drv \varphi_1\int_0^{2\pi} \drv \varphi_2\,
 \cos (n \varphi) \,
 \frac{\drv \sigma}{\drv y_1 \drv y_2\, \drv \kappa_1 \, \drv \kappa_2 \drv \varphi_1 \drv \varphi_2}\;
\]
\[
 = \frac{x_1 x_2}{\kappa_1 \kappa_2}
 \int_{-\infty}^{+\infty} \drv \nu \, e^{\bar \alpha_s(\mu_R)\left\{\chi(n,\nu)+\bar\alpha_s(\mu_R)
 \left[\bar\chi(n,\nu)+\nicefrac{\beta_0}{(8 N_c)}\chi(n,\nu)\left[-\chi(n,\nu)+\nicefrac{10}{3}+4\ln\left(\nicefrac{\mu_R}{\sqrt{\kappa_1 \kappa_2}}\right)\right]\right]\right\}}
\]
\[
 \times \, \alpha_s^2(\mu_R) c_1(n,\nu,\kappa_1, x_1)[c_2(n,\nu,\kappa_2,x_2)]^*\,
\]
\begin{equation}
\label{BFKL_Cn}%\nonumber
 \times \, \left\{1
 +\alpha_s(\mu_R)\left[\frac{\hat c_1(n,\nu,\kappa_1,x_1)}{c_1(n,\nu,\kappa_1, x_1)}
 +\left[\frac{\hat c_2(n,\nu,\kappa_2, x_2)}{c_2(n,\nu,\kappa_2,x_2)}\right]^*
% \right]
% \right.
%  \end{equation}
%  \begin{equation}\nonumber
% \left. 
 +  \bar \alpha_s(\mu_R) %\alpha_s^2(\mu_R) 
 \, Y %\ln\left(\frac{x_1 x_2 s}{s_0}\right)
 \frac{\beta_0}{4 \pi}\chi(n,\nu)f(\nu)
 \right]
 \right\} \;,
\end{equation}
with $n$ is an integer larger than one and $\bar \alpha_s(\mu_R) \equiv \alpha_s(\mu_R) N_c/\pi$, where $N_c$ is the color number and $\beta_0$ the first coefficient of the QCD $\beta$-function, responsible for running-coupling effects (Eq.~(\ref{as_parameters})).
Details on the BFKL kernel and on the impact factors can be found in Sections~\ref{BFKL_kernel} and~\ref{LO_IFs}, respectively. It is worth to remark that the formula given in Eq.~(\ref{BFKL_Cn}) is obtained in the so-called $(n,\nu)$\textit{-representation}, \textit{i.e.} taking the projection onto the eigenfunctions of the \textit{LO BFKL kernel} (for more details, see Refs.~\cite{Ivanov:2005gn,Ivanov:2006gt}).
For the sake of completeness, we give the expression for the azimuthal coefficients in the LLA BFKL approximation:
\begin{equation}
\label{LLA_Cn}%\nonumber
 \CnLLA = \frac{x_1 x_2}{\kappa_1 \kappa_2}
 \int_{-\infty}^{+\infty} \drv \nu \, e^{\bar \alpha_s(\mu_R)\chi(n,\nu)}
 \, \alpha_s^2(\mu_R) c_1(n,\nu,\kappa_1, x_1)[c_2(n,\nu,\kappa_2,x_2)]^* \;.
\end{equation}
All these formul{\ae} for the azimuthal coefficients are obtained in the $\MSb$ renormalization scheme. Corresponding expressions in the MOM scheme and with scale-optimization setting will be given in Section~\ref{blm}.

\subsubsection{LO and NLO BFKL kernel}
\label{BFKL_kernel}

The expression in Eq.~(\ref{BFKL_Cn}) for the LO BFKL characteristic function or \textit{LO BFKL kernel} is
\begin{equation}
 \label{kernel_LO}
 \chi\left(n,\nu\right) = 2\psi\left(1\right) - \psi\left(\frac{n}{2} + \frac{1}{2} + i \nu \right) - \psi\left(\frac{n}{2} + \frac{1}{2} - i\nu \right) \, ,
\end{equation}
with $\psi(z) \equiv \Gamma^\prime
(z)/\Gamma(z)$ the logarithmic derivative of the Gamma function, while
$\bar\chi(n,\nu)$, calculated in Ref.~\cite{Kotikov:2000pm} (see also Ref.~\cite{Kotikov:2002ab}), stands for the \textit{NLO correction} to the BFKL kernel, handily represented in the form:
\[
 \bar \chi(n,\nu)\,=\, - \frac{1}{4}\left\{\frac{\pi^2 - 4}{3}\chi(n,\nu) - 6\zeta(3) - \chi^{\prime\prime}(n,\nu)  + \,2\,\phi(n,\nu) + \,2\,\phi(n,-\nu)
\right.
\]
\begin{equation}
 \label{kernel_NLO}
 \left.
 +\frac{\pi^2\sinh(\pi\nu)}{2\,\nu\, \cosh^2(\pi\nu)}
 \left[
 \left(3+\left(1+\frac{n_f}{N_c^3}\right)\frac{11+12\nu^2}{16(1+\nu^2)}\right)
 \delta_{n0}
 -\left(1+\frac{n_f}{N_c^3}\right)\frac{1+4\nu^2}{32(1+\nu^2)}\delta_{n2}
\right]\right\} \, ,
\end{equation}
where $n_f$ is the flavor number, and
\[
 \phi(n,\nu)\,=\,-\int\limits_0^1 \drv x\,\frac{x^{-1/2+i\nu+n/2}}{1+x}\left\{\frac{1}{2}\left(\psi^\prime\left(\frac{n+1}{2}\right)-\zeta(2)\right)+\mbox{Li}_2(x)+\mbox{Li}_2(-x)\right.
\]
\[
\left.
 +\ln x\left[\psi(n+1)-\psi(1)+\ln(1+x)+\sum_{k=1}^\infty\frac{(-x)^k}{k+n}\right]+\sum_{k=1}^\infty\frac{x^k}{(k+n)^2}\left[1-(-1)^k\right]\right\}
\]
\[
 =\sum_{k=0}^\infty\frac{(-1)^{k+1}}{k+(n+1)/2+i\nu}\left\{\psi^\prime(k+n+1)-\psi^\prime(k+1)\right.
\]
\begin{equation}
\label{kernel_NLO_phi}
 \left.
 +(-1)^{k+1}\left[\beta_{\psi}(k+n+1)+\beta_{\psi}(k+1)\right]-\frac{\psi(k+n+1)-\psi(k+1)}{k+(n+1)/2+i\nu}\right\} \; ,
\end{equation}
\begin{equation}
\label{kernel_NLO_phi_beta_psi}
 \beta_{\psi}(z)=\frac{1}{4}\left[\psi^\prime\left(\frac{z+1}{2}\right)
 -\psi^\prime\left(\frac{z}{2}\right)\right] \; ,
\end{equation}
\begin{equation}
\label{dilog}
\mbox{Li}_2(z) \equiv \int\limits_0^z \drv u \,\frac{\ln(1-u)}{-u} \; .
\end{equation}

\subsubsection{LO impact factors in the $(n,\nu)$-representation}
\label{LO_IFs}

Two kinds of LO impact factors, generically indicated in Eq.~(\ref{BFKL_Cn}) as $c_{i}(n,\nu, \kappa_i,x_i) \equiv \{ c^{(H)}(n,\nu), c^{(J)}(n,\nu) \}$, where the labels $(H)$ and $(J)$ refer to the jet and the hadron case, respectively, will be considered.

The LO forward-hadron impact factor, $c^{(H)}(n,\nu)$, consists in the convolution over the parton longitudinal fraction $x$ between the quark/antiquark/gluon PDFs and the FFs portraying the detected hadron:
\[
 c^{(H)}(n,\nu,\kappa_H,x_H) = 2 \sqrt{\frac{C_F}{C_A}}(\kappa_H^2)^{i\nu-1/2}\,\int_{x_H}^1\frac{\drv x}{x}\left( \frac{x}{x_H}\right)
% ^{2 i\nu-1+\bar\alpha_s(\mu_R)\chi(n,\nu)}
 ^{2 i\nu-1} 
\]
\begin{equation}
\label{hadron_IF_LO}
 \times \, \left[\frac{C_A}{C_F}f_g(x)D_g^h\left(\frac{x_H}{x}\right) + \sum_{\alpha=q,\bar q}f_\alpha(x)D_\alpha^h\left(\frac{x_H}{x}\right)\right] \; ,
\end{equation}
with $C_F = (N_c^2-1)/(2N_c)$ and $C_A \equiv N_c$ the Casimir factors connected to gluon emission from a quark and from a gluon, respectively.
Correspondingly, $c^{(J)}(n,\nu)$ is the LO forward-jet impact factor in the $(n,\nu)$-repre\-sen\-ta\-tion:
\begin{equation}
\label{jet_IF_LO}
 c^{(J)}(n,\nu,\kappa_J,x_J)=2\sqrt{\frac{C_F}{C_A}}(\kappa_J^2)^{i\nu-1/2}\,\left(\frac{C_A}{C_F}f_g(x_J) + \sum_{\alpha=q,\bar q}f_\alpha(x_J)\right) \; ,
\end{equation}
and the $f(\nu)$ function in the last term of Eq.~(\ref{BFKL_Cn}) reads
\begin{equation}
\label{fnu}
 f(\nu) = \frac{i}{2} \left[ \frac{\drv}{\drv \nu} \ln\left(\frac{c_1(n,\nu)}{[c_2(n,\nu)]^*}\right) + \ln\left(\kappa_1^2 \kappa_2^2\right) \right] \; ,
\end{equation}
after fixing the final state, \textit{i.e.} having selected one of the processes in Fig.~\ref{fig:processes}.
The remaining objects are NLO corrections to impact factors, $\hat c_{i}(n, \nu, \kappa_{i}, x_{i}) \equiv \{ \hat c^{(H)}(n,\nu), \hat c^{(J)}(n,\nu) \}$, their expressions being given in Appendix~\hyperlink{app:hadron_IF_NLO_link}{A} (hadron) and in Appendix~\hyperlink{app:jet_IF_NLO_link}{B} (jet).

\subsection{High-energy DGLAP}
\label{DGLAP}

With the aim of providing a systematic comparison between BFKL-inspired predictions and fixed-order calculations, we derive a DGLAP-like general formula, valid for all the considered processes (Fig.~\ref{fig:processes}), where the $\CnDGLAP$ azimuthal coefficients are introduced as truncation to the ${\cal O}(\alpha_s^3)$ order of the corresponding NLA BFKL ones, $\CnNLA$, up to the inclusion of terms beyond the LO accuracy. This allows us to catch
the leading-power asymptotic features of a genuine NLO DGLAP description, neglecting at the same time those factors which are quenched by inverse powers of the energy of the partonic subprocess. Although being an alternative (and approximated) way to the standard, fixed-order DGLAP analysis, such procedure results to be adequate in the region of high-rapidity distance, $Y$, investigated in this work, as well as easy and flexible to implement.

Starting from Eq.~(\ref{BFKL_Cn}), one gets the DGLAP limit by expanding the exponentiated BFKL kernel up to the first order in $\as$, and keeping impact factors at NLO.
Our DGLAP master formula reads
\[
 \CnDGLAP \equiv \frac{x_1 x_2}{\kappa_1 \kappa_2}
 \int_{-\infty}^{+\infty} \drv \nu \, \alpha_s^2(\mu_R) c_1(n,\nu,\kappa_1, x_1)[c_2(n,\nu,\kappa_2,x_2)]^*\,
\]
\begin{equation}
\label{DGLAP_Cn}%\nonumber
 \times \, \left\{1+\alpha_s(\mu_R)\left[Y \frac{C_A}{\pi} \chi(n,\nu)+\frac{\hat c_1(n,\nu,\kappa_1,x_1)}{c_1(n,\nu,\kappa_1, x_1)}
 +\left[\frac{\hat c_2(n,\nu,\kappa_2, x_2)}{c_2(n,\nu,\kappa_2,x_2)}\right]^* 
 %+ \bar \alpha_s(\mu_R) %\alpha_s^2(\mu_R) 
 %\, Y %\ln\left(\frac{x_1 x_2 s}{s_0}\right)
 %\frac{\beta_0}{4 \pi}\chi(n,\nu)f(\nu)
 \right]
 \right\} \;,
\end{equation}
where the BFKL exponentiated kernel has been replaced by its expansion up to terms proportional to $\as(\mu_R)$.

We stress that, although the $\nu$-integral in Eq.~(\ref{DGLAP_Cn}) leads to distributions in the transverse-momentum space rather than ordinary functions, final expressions for the DGLAP azimuthal coefficients become regular when integrated over the final-state phase space, as in Eq.~(\ref{int_Cn}). This allows us to safely calculate them via a multidimensional strategy, where $\nu$ and phase-space integration are simultaneously performed (for more details, see Section~\ref{numerics}).

\subsection{Strong-coupling setting}
\label{strong_coupling}

We use a two-loop running-coupling setup with $\alpha_s\left(M_Z\right)=0.11707$ and five quark flavors, $n_f$, active. Its expression in the $\MSb$ scheme reads:
\begin{equation}
\label{as_MSb}
 \as(\mu_R) \equiv \as^{(\MSb)}(\mu_R) = \frac{\pi}{\beta_0 \lambda(\mu_R)} \left( 4 - \frac{\beta_1}{\beta_0^2} \frac{\ln \lambda(\mu_R)}{\lambda(\mu_R)} \right) \;,
\end{equation}
where
\begin{equation}
\label{as_parameters}
 \lambda(\mu_R) = 2 \ln \frac{\mu_R}{\LQCD} \;, \qquad
 \beta_0 = 11 - \frac{2}{3} n_f \;, \qquad
 \beta_1 = 102 - \frac{38}{3} n_f \;.
\end{equation}
It is possible to obtain the corresponding espression for the strong coupling in the MOM scheme, whose definition is related to the 3-gluon vertex (an essential ingredient in the BFKL framework), by performing the following finite renormalization:
\begin{equation}
\label{as_MOM}
 \as^{(\MSb)} = \as^{\rm (MOM)} \left( 1 +  \frac{T^{\beta} + T^{\rm conf}}{\pi} \as^{\rm (MOM)} \right) \;,
\end{equation}
with
\[
\label{T_bc}
T^{\beta}=-\frac{\beta_0}{2}\left( 1+\frac{2}{3}I \right) \; ,
\]
\begin{equation}
T^{\rm conf}= \frac{C_A}{8}\left[ \frac{17}{2}I +\frac{3}{2}\left(I-1\right)\xi
+\left( 1-\frac{1}{3}I\right)\xi^2-\frac{1}{6}\xi^3 \right] \; ,
\end{equation}
where 
%$C_A \equiv N_c$ is the color factor associated with gluon emission from a gluon, 
$I = -2 \int_0^1 \drv u \frac{\ln \left( u \right)}{u^2 - u + 1} \simeq 2.3439$ and $\xi$ 
is a gauge parameter, fixed at zero in the following.

The infrared improvement of the running coupling via well-known procedures, as the Webber parametrization~\cite{Webber:1998um}, here is not needed, since energy scales are always in the perturbative region. They are strictly related to transverse-momentum ranges (see Section~\ref{observables}), their large values protecting us from a region where the \emph{diffusion pattern}~\cite{Bartels:1993du} (see also Refs.~\cite{Caporale:2013bva,Ross:2016zwl}) becomes relevant.

\subsection{PDF and FF selection}
\label{PDF_FF}

Potential sources of uncertainty are expected to arise from the choice of particular PDF and FF sets, rather then other ones. While preliminary tests done using the three most popular NLO PDF parametrizations ({\tt MMHT 2014}~\cite{Harland-Lang:2014zoa}, {\tt CT 2014}~\cite{Dulat:2015mca} and {\tt NNPDF3.0}~\cite{Ball:2014uwa}) have shown no significant discrepancy in the kinematic regions of our interest, this does not hold anymore for FFs, where the selection of distinct sets leads to non-negligible effects. This feature was pointed out in a first phenomenological analysis~\cite{Celiberto:2016hae} on $\varphi$-averaged cross sections for the inclusive di-hadron production, with the hadrons' ranges tailored on the acceptances of the CMS detector, and then confirmed also for azimuthal correlations (\textit{i.e.} ratios of phase-space integrated azimuthal coefficients, as explained in Section~\ref{observables}) in the case of inclusive hadron-jet emission~\cite{Bolognino:2018oth}, when the jet is tagged inside the CASTOR backward detector.
The simplest strategy to address this issue is in line with the recommendations suggested by the PDF4LHC community~\cite{Butterworth:2015oua} for Standard Model phenomenology, and consists in selecting an individual PDF set and taking its convolution with a given FF. This method will be stringently applied in our analysis on the scale optimization (Section~\ref{scale_optimization}) and on our results for azimuthal distributions (Section~\ref{azimuthal_distribution}), where we will give predictions for the {\tt MMHT 2014}~\cite{Harland-Lang:2014zoa} NLO PDF together with the {\tt NNFF1.0}~\cite{Bertone:2017tyb} NLO FF.
Conversely, we will adopt a more extensive solution when gauging the effect discussed above for the comparison of our BFKL results with fixed-order DGLAP calculations, which is actually the main outcome of this paper (Section~\ref{core}). On the one hand, as for the PDFs, we will choose the NLO {\tt PDF4LHC15{\textunderscore}100} parametrization~\cite{Butterworth:2015oua}, which represents a statistical combination of the three sets mentioned before~\cite{Harland-Lang:2014zoa,Dulat:2015mca,Ball:2014uwa}. On the other hand, we will build our own combination for the FFs by merging predictions obtained with the four following NLO parametrizations: {\tt AKK 2008}~\cite{Albino:2008fy}, {\tt DSS 2007}~\cite{deFlorian:2007aj,deFlorian:2007ekg}, {\tt HKNS 2007}~\cite{Hirai:2007cx} and {\tt NNFF1.0}~\cite{Bertone:2017tyb}. It is worth to remark that 
our prescription refers to the average of final results for our observables, while no new FF set has been created.
A higher-level analysis, based on the so-called \textit{bootstrap replica method}~(see Ref.~\cite{Forte:2002fg} for a combined study including fit procedures
inspired by neural-network techniques), is employed just in one case, namely our \emph{theory-versus-esperiment} study for azimuthal correlations in the Mueller--Navelet channel at~$\sqrt{s}~=~7$~TeV (Section~\ref{th_vs_exp}).
We make use of a sample of 100 replicas of the central value of the {\tt PDF4LHC15{\textunderscore}100} set. This collection of replicas is obtained by randomly altering the central value with a Gaussian background featuring the original variance.
The extensive application of the replica method to all the considered observables goes beyond our scope and is left to further investigations which will include all the systematic effects.

\subsection{Jet algorithm and event selection}
\label{jet_algorithm}

Several types of reconstruction algorithms can be implemented in the definition of the (NLO) jet impact factor, although the choice of a particular one instead of the others seems not to produce a crucial effect on semi-hard typical observables, as pointed out in Section~2.3 of Ref~\cite{Ducloue:2013wmi}. The most popular jet-reconstruction functions essentially belong to two distinct classes (for further details, see \emph{e.g.}, Refs.~\cite{Chekanov:2002rq,Salam:2009jx} and Refs. therein): \emph{cone-type} and \emph{sequential-clustering} algorithms (the well known (\emph{anti-})$\kappa_\perp$ kind \cite{Catani:1993hr,Cacciari:2008gp} falls into this last family). We will adopt a simpler version, \emph{infrared-safe} up to NLO perturbative accuracy and suited to numerical calculations, given in Ref.~\cite{Ivanov:2012ms} in terms of the so-called ``small-cone'' approximation 
(SCA)~\cite{Furman:1981kf,Aversa:1988vb}, {\it i.e.} for small-jet cone aperture in the rapidity-azimuthal angle plane (see Ref.~\cite{Colferai:2015zfa} for a detailed study on different jet algorithms and their small-cone, approximated versions).

A potential issue, highlighted first~\cite{Caporale:2014gpa} in the case of Mueller--Navelet jet production (left panel of Fig.~\ref{fig:processes}) but present in any process featuring jet emission in the final state, is related with the experimental event selection in a situation when more than one jet is tagged in a single event. As an example, let us consider events with three 
objects in the final state, two of them being jets emitted in the forward region (with 
large positive rapidities, $y_{J,1} > y_{J,2} >0$), and the third one being a backward object (hadron or jet, in our analyses), with negative rapidity, $y_O < 0$. Traditionally, as in the current CMS analysis, such event would be classified as a unique one, with a single forward jet plus another, backward particle detected, the largest rapidity interval being, $Y = y_{J,1} - y_O$. This selection 
criterion is suited to experimental analyses, but it does not match the theoretical definition in NLA BFKL calculations. Examining the derivation of the NLO jet impact factor~\cite{Bartels:2001ge,Caporale:2011cc}, it follows that our calculation describes an \emph{inclusive}, forward-in-rapidity jet production in the fragmentation region of the parent hadron, whereas possible additional parton radiation is attributed to 
the inclusive hadron system, $X$ (Eq.~(\ref{processes})). The issue may be clarified either from the experimental side, by adapting the
jet-selection paradigm, or from the theoretical side, by improving the NLO calculation, as proposed in Ref.~\cite{Colferai:2016inu}.

\section{Phenomenology}
\label{results}

\subsection{Integration over the final-state phase space}
\label{observables}

Starting from Eqs.~(\ref{BFKL_Cn}) and~(\ref{DGLAP_Cn}), we consider the \emph{integrated} coefficients
over the phase space for the two emitted objects, $O_{1,2}(\kappa_{1,2},y_{1,2})$, while their rapidity distance, $Y = y_1 - y_2$, is kept fixed: 
\begin{equation}
\label{int_Cn}
 C_n^{\rm [resum]}(s,Y) = 
 \int_{\kappa_1^{\rm min}}^{\kappa_1^{\rm max}} \drv \kappa_1
 \int_{\kappa_2^{\rm min}}^{\kappa_2^{\rm max}} \drv \kappa_2
 \int_{y_1^{\rm min}}^{y_1^{\rm max}} \drv y_1
   \int_{y_2^{\rm min}}^{y_2^{\rm max}} \drv y_2
 \, \delta \left( Y - (y_1 - y_2) \right)
 \, {\cal C}_n^{\rm [resum]} \, ,
\end{equation}
where the superscript $\rm [resum]$ stands indistinctly for LLA, NLA or DGLAP. This allows us to impose and study different ranges in transverse momenta and rapidities, based on realistic kinematic configurations used in experimental analyses at the LHC. 
Let us introduce the following representation for final-state particles, which will facilitate us to distinguish, process by process, the considered ranges:
\begin{alignat}{2}
\label{final_state_processes}
 & {\rm a)} \; \mbox{Mueller--Navelet:} \qquad
 && O_1(\kappa_1,y_1) + X + O_2(\kappa_2,y_2) \; \equiv \; {\rm jet}(\kappa_{J,1},y_{J,1}) + X + {\rm jet}(\kappa_{J,2},y_{J,2}) \; ; \nonumber \\
 & {\rm b)} \; \mbox{hadron-jet:} \qquad
 && O_1(\kappa_1,y_1) + X + O_2(\kappa_2,y_2) \; \equiv \; {\rm hadron}(\kappa_H,y_H) + X + {\rm jet}(\kappa_J,y_J) \; ; \\
 & {\rm c)} \; \mbox{di-hadron:} \qquad
 && O_1(\kappa_1,y_1) + X + O_2(\kappa_2,y_2) \; \equiv \; {\rm hadron}(\kappa_{H,1},y_{H,1}) + X + {\rm hadron}(\kappa_{H,2},y_{H,2}) \; . \nonumber
\end{alignat}
Here labels a), b) and c) refer to the respective panels in Fig.~\ref{fig:processes}, while a slight, process-suitable change of notation in the final-state variables has been made.
For all the considered reactions, the maximum values for the transverse momenta are determined by general requirements, based on kinematics. 
On the one side, in the case of CMS (di-)hadron emission, $\kappa_H^{\rm max}$ is constrained by the lower cutoff of the used FF set and is always fixed at $\kappa_{H,\rm{CMS}}^{\rm max} \simeq 21.5$~GeV. 
On the other side, when (at least) a jet is tagged inside CMS or CASTOR, the value for the upper cutoff, $\kappa_J^{\rm max}$, follows from the given bounds in rapidity, at fixed center-of-mass energy, $\sqrt{s}$ and will be individually discussed for each kinematic configuration introduced below. The rapidity interval, $Y$, is everywhere taken positive: $0 < Y \leq y^{\rm max}_1
- y^{\rm min}_2$.

\subsubsection{Symmetric CMS configuration}
\label{sCMS}

The first kinematic configuration we take into account, namely the \textbf{\textit{symmetric CMS}} choice,  prescribes that both final-state particles, $O_{1,2}$, are tagged in the CMS detector, their transverse momenta and rapidities confined in symmetric ranges, characteristic of the experimental studies recently conducted~\cite{Khachatryan:2016udy}:
\begin{alignat*}{4}
 & {\rm a)} \; \mbox{Mueller--Navelet:} \;\,
 && 35 \mbox{ GeV} < \kappa_{J \, 1,2} < \kappa_J^{\rm max} \;\, &\mbox{and} \;\,
 & |y_{J \, 1,2}| < 4.7 \; , \; Y < 9.4 \; .
\end{alignat*}
We will give results in this configuration just for Mueller--Navelet jet production (left panel of Fig.~\ref{fig:processes}), comparing LLA and NLA BFKL predictions with experimental data at $\sqrt{s} = 7$ TeV and for $\kappa_J^{\rm max} \equiv \kappa_{J,{\rm (CMS)}}^{\rm max}= 60$ GeV.

\subsubsection{Asymmetric CMS configuration}
\label{aCMS}

The second range we include in our study is the \textbf{\textit{asymmetric CMS}} configuration and consists in requiring the emitted objects, $O_{1,2}$, to be detected in disjoint intervals for the transverse momenta, {\it i.e.} $\kappa$\textit{-windows}. In the hadron-jet production case (central panel in Fig.~\ref{fig:processes}), the final-state asymmetry if further enriched by the distinct rapidity ranges which characterize the hadron and the jet, respectively. So one has:
\begin{alignat*}{4}
 & {\rm a)} \; \mbox{Mueller--Navelet:} \;\,
 && 35 \mbox{ GeV} < \kappa_{J,1} < 45 \mbox{ GeV} < \kappa_{J,2} < \kappa_J^{\rm max} \;\, &\mbox{and} \;\,
 &|y_{J,1,2}| < 4.7 \; , \; Y < 9.4 \; ;
 \\
 & {\rm b)} \; \mbox{hadron-jet:} \;\,
 && 5 \mbox{ GeV} < \kappa_H < \kappa_H^{\rm max} < 35 \mbox{ GeV} < \kappa_J < \kappa_J^{\rm max} \;\, &\mbox{and} \;\, & |y_H| < 2.4 \; , \; |y_J| < 4.7 \; , \; Y < 7.1 \; ;
 \\
 & {\rm c)} \; \mbox{di-hadron:} \;\,
 && 5 \mbox{ GeV} < \kappa_{H,1} < 10  \mbox{ GeV} < \kappa_{H,2} < \kappa_H^{\rm max} \;\, &\mbox{and} \;\,
 & |y_{H,1,2}| < 2.4 \; , \; Y < 4.8 \; .
\end{alignat*}
We will perform our analysis for $\sqrt{s} = 13$ TeV, $\kappa_H^{\rm max} \equiv \kappa_{H,\rm{CMS}}^{\rm max} \simeq 21.5$ GeV and $\kappa_J^{\rm max} \equiv \kappa_{J,{\rm CMS}}^{\rm max} = 60$ GeV. This last value has been kept equal with respect to the symmetric case at $\sqrt{s} = 7$ TeV (Section~\ref{sCMS}).

\subsubsection{CASTOR-jet configuration}
\label{CASTORj}

This peculiar \textbf{\textit{CASTOR-jet}} range selection~\cite{CMS:2016ndp}, valid only for Muller--Navelet and hadron-jet production (first two panels of Fig.~\ref{fig:processes}), introduces the possibility to tag a backward jet inside the CASTOR (CST) calorimeter, together with another object (jet or hadron) detected by CMS. This peculiar configuration allows one to reach the largest values for the rapidity separation, $Y$, taking advantage of the very backward rapidity range provided by CASTOR: 
\begin{alignat*}{4}
 & {\rm a)} \; \mbox{Mueller--Navelet:} \;\,
 && 10 \mbox{ GeV} < \kappa_{J,2} < \kappa_{J,{\rm CST}}^{\rm max} < 20 \mbox{ GeV} < \kappa_{J,1} < \kappa_{J,{\rm CMS}}^{\rm max} \;\, &\mbox{and} \;\,
 & |y_{J,1}| < 4.7 \; , \; -6.6 < y_{J,2} < -5.2 \; , \; Y < 11.3 \; ;
 \\
 & {\rm b)} \; \mbox{hadron-jet:} \;\,
 && 5 \mbox{ GeV} < \kappa_H < 10 \mbox{ GeV} < \kappa_J < \kappa_{J,{\rm CST}}^{\rm max} \;\, &\mbox{and} \;\, & |y_H| < 2.4 \; , \; -6.6 < y_{J} < -5.2 \; , \; Y < 9 \; .
\end{alignat*}
Results will be shown for $\sqrt{s} = 13$ TeV, $\kappa_{J,{\rm CST}}^{\rm max} \simeq 17.68$ GeV and $\kappa_{J,{\rm CMS}}^{\rm max} = 60$ GeV.

\subsection{Final-state observables}
\label{observables}

The phenomenological observables of our interest fall into the following three classes:

\begin{enumerate}

 \item
  $\varphi$-averaged \emph{cross section}, $C_0$, and the ratio, $R_0^{\rm NLA/LLA}$, between $C_0$ calculated in the full NLA BFKL accuracy and its counterpart in the LLA BFKL approximation;

 \item
  Azimuthal-correlation moments (or simply \emph{azimuthal correlations}), $R_{nm} \equiv C_n/C_m $, with the ratios of the form $R_{n0}$ among them having the immediate physical interpretation of averaged values of the cosine of $n$-multiples of $\varphi$, $\langle \cos(n \varphi) \rangle$;

 \item
  \emph{Azimuthal distribution} of the two emitted objects, which actually represents one of the most directly accessible observables in experimental analyses:
  \begin{equation}
  \label{eq_azimuthal_distribution}
   \frac{1}{\sigma} \frac{\drv \sigma}{\drv \varphi} = \frac{1}{2 \pi} \left\{ 1 + 2 \sum_{n=1}^{\infty} \cos(n \varphi) \langle \cos(n \varphi) \rangle \right\}
   \equiv \frac{1}{2 \pi} \left\{ 1 + 2 \sum_{n=1}^{\infty} \cos(n \varphi) R_{n0} \right\} \; .
  \end{equation}
 On the one side, azimuthal distributions are very favorable observables to be compared with experimental analyses, even more than the azimuthal-correlation moments. This stems from the fact that, since measured distributions cannot cover the whole $(2 \pi)$ range of azimuthal angle due to detector limitations, observables differential on $\varphi$ permit us to dampen the descending accuracy loss. 
One the other side, numeric calculations of Eq.~(\ref{eq_azimuthal_distribution}) are not easy to perform, due to the large number of azimuthal coefficients required and due to instabilities in the $\nu$-integration that progressively increase with $n$. This requires sizeable effort from the numerical side (see Section~\ref{jethad}).
 
\end{enumerate}

\subsection{Prelude: Mueller--Navelet jets at 7 TeV LHC, theory versus experiment}
\label{th_vs_exp}

\begin{figure}[b]
\centering

   \includegraphics[scale=0.45,clip]{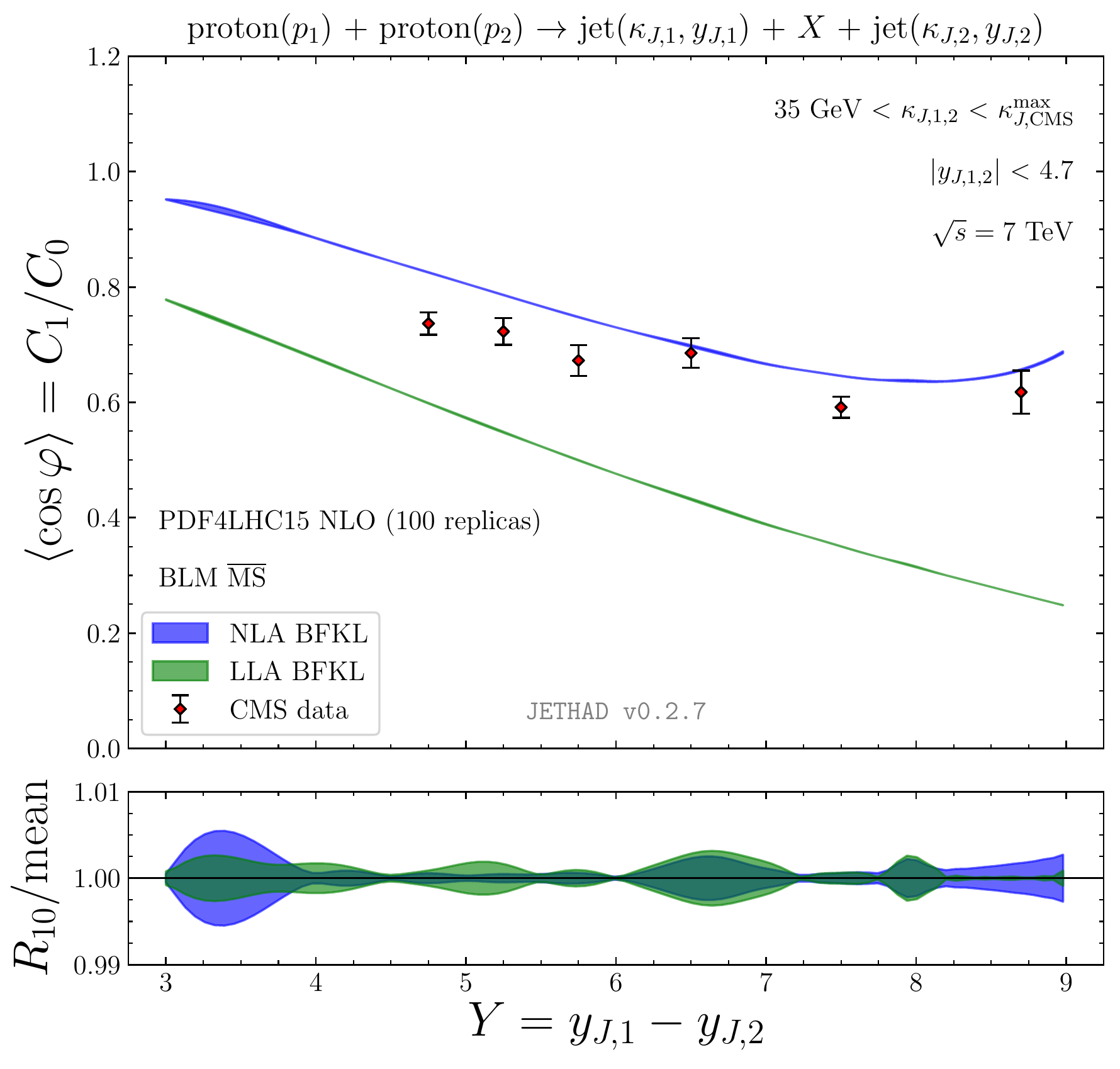}
   \hspace{0.25cm}
   \includegraphics[scale=0.45,clip]{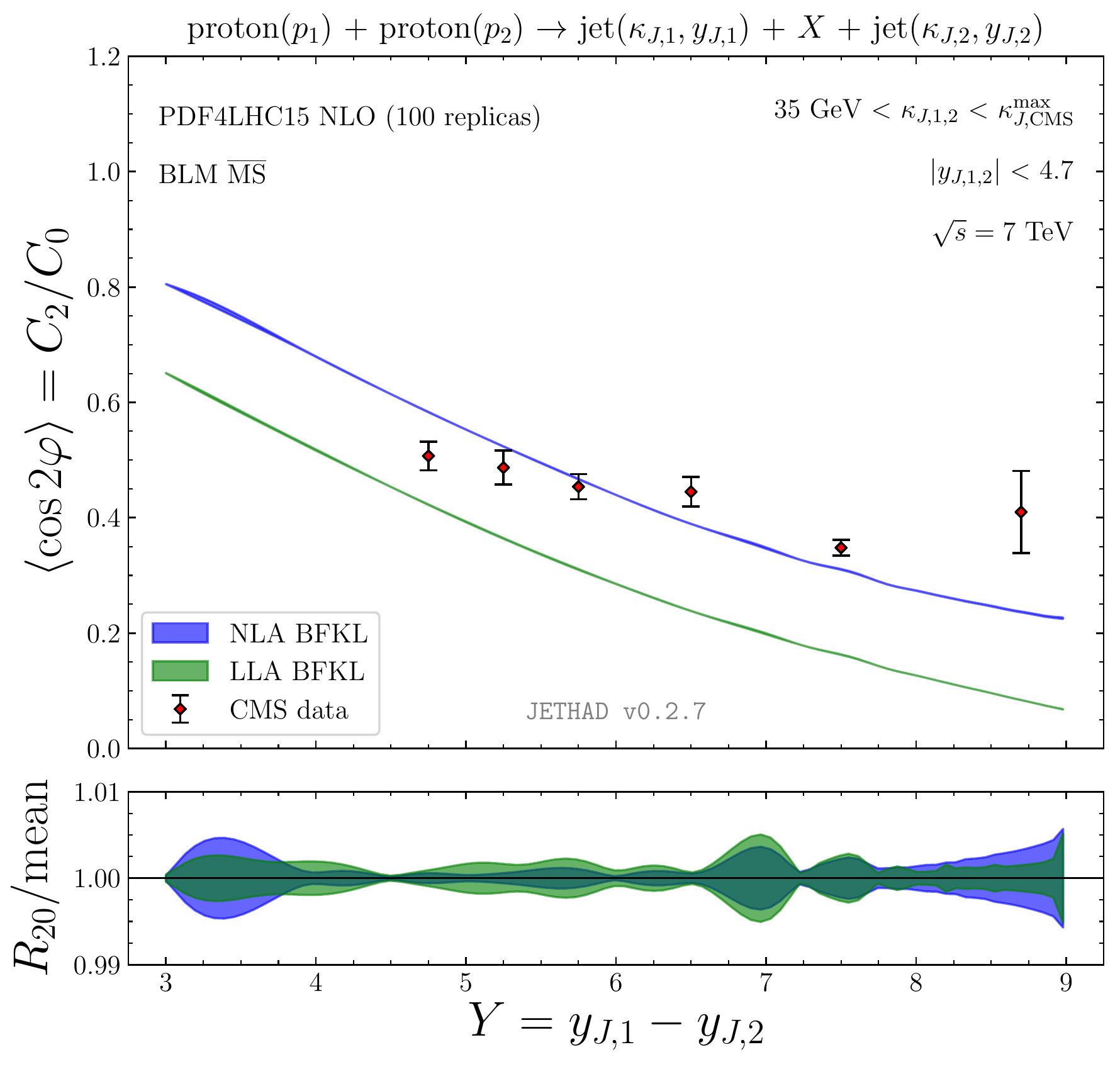}

\caption{$Y$-dependence of $R_{10} \equiv C_1/C_0$ and $R_{20} \equiv C_2/C_0$ of the Mueller--Navelet jet production (left panel of Fig.~\ref{fig:processes}) for 
$\mu_{F1,2} = \mu_R = \mu_R^{\rm BLM}$ and $\sqrt{s} = 7$ TeV (\textit{symmetric CMS} configuration). Calculations are done in the $\MSb$ renormalization scheme. Ancillary plots below primary panels show the reduced azimuthal correlations, \emph{i.e.} the envelope of $R_{nm}$ replicas' results divided by their mean value.}
\label{fig:MN-tve-CMS}
\end{figure}

In this Section predictions for different azimuthal correlations, $R_{nm} \equiv C_n/C_m$, are given for the Mueller--Navelet jet production channel (left panel of Fig.~\ref{fig:processes}) in the \textit{symmetric CMS} configuration (see Section~\ref{sCMS}) and compared with recent CMS data at $\sqrt{s} = 7$ TeV~\cite{Khachatryan:2016udy}. We actually improve the analysis conducted in Ref.~\cite{Caporale:2014gpa}, by implementing the ``exact", numeric BLM scale-optimization method instead of using the two approximated, analytic ones (we are advisedly anticipating details on the scale-optimization issue, discussed in detail in Section~\ref{blm}).
At variance with the previous work~\cite{Caporale:2014gpa}, where  the {\tt MSTW~2008}~\cite{Martin:2009iq} NLO PDF set (precursor of the {\tt MMHT~2014} one~\cite{Harland-Lang:2014zoa}) was employed, we adopt the NLO {\tt PDF4LHC15{\textunderscore}100} parametrization~\cite{Butterworth:2015oua} and we estimate the uncertainty of our predictions by calculating the standard deviation at the hand of the replica method (see Section~\ref{PDF_FF} for further details). 
As in Ref.~\cite{Caporale:2014gpa}, final calculations are done in the $\MSb$ renormalization scheme. We remind that it is possible to recover the analytic expression for the azimuthal coefficients, $C_n$, in the $\MSb$ scheme from the corresponding ones, given in the MOM one (Eq.~(\ref{Cn_NLA_int_blm})), by making the substitutions (note that the value of the renormalization scale is left unchanged)
\[
 T^{\rm conf} \to - T^\beta \;, \qquad \as^{(\rm MOM)} \to \as^{(\MSb)} \;,
\]
with $T^{\rm conf}$, $T^\beta$, $\as^{(\MSb)}$ and $\as^{(\rm MOM)}$ given in Section~\ref{strong_coupling}. The two factorization scales and the renormalization one are set equal to the one provided, as outcome for each value of $Y$, by the BLM procedure: $\mu_{F1,2} = \mu_R = \mu_R^{\rm BLM}$.

Results for $R_{10}$ and $R_{20}$, reported in Fig.~\ref{fig:MN-tve-CMS}, unambiguously highlight that the pure LLA calculations overestimate by far the decorrelation between the two tagged jets, whereas the agreement with experimental data definitely improves, when NLA BFKL corrections, together with the BLM prescription, are included. For this process and for the considered transverse-momentum kinematics, the uncertainty coming from the numerical multidimensional integration over the final-state phase space is very small. Therefore, error bands strictly refer to the uncertainty hailing from the use of different PDF members inside the {\tt PDF4LHC15{\textunderscore}100} collection of 100 replicas. As expected, its effect is very small when ratios of azimuthal coefficients are taken.
Ancillary plots below primary panels of Fig.~\ref{fig:MN-tve-CMS} clearly indicate that the relative standard deviation of the replicas' results is smaller than 1\% in all cases. Here, nodal points bring information on uncertainty oscillations inherited from the collinear-PDF parametrization.

\subsection{Need for scale optimization}
\label{scale_optimization}

It is widely recognized that calculations purely based on the BFKL resummation suffer from large higher-order corrections both in the kernel of the gluon Green's function and in the non-universal impact factors. 
\emph{Inter alia}, NLA BFKL contributions to the zeroth conformal spin come up with roughly the same order of the corresponding LLA predictions, but with opposite sign.
The primary manifestation of this instability lies in the large uncertainties arising from the renormalization (and factorization) scale choice, which hamper any genuine attempt at high-precision calculations.
All that calls for some ``stabilizing'' procedure of the perturbative series, which can consist in:
\begin{itemize}
 \item[(\emph{i})]
  the inclusion of next-to-NLA terms (\emph{a priori} unknown), such as those ones prescribed by the renormalization group, as in the all-order~\cite{Caporale:2013uva,Vera:2007kn,Salam:1998tj,Ciafaloni:1998iv,Ciafaloni:1999yw,Ciafaloni:1999au,Ciafaloni:2000cb,Ciafaloni:2002xk,Ciafaloni:2002xf,Ciafaloni:2003ek,Ciafaloni:2003rd,Ciafaloni:2003kd,Vera:2005jt,Caporale:2007vs} or in the consistency-condition~\cite{Kwiecinski:1996td} \emph{collinear-improvement} procedures, or by energy-momentum conservation~\cite{Kwiecinski:1999yx};
 \item[(\emph{ii})]
  the restriction of undetected-radiation activity to only allow final-state gluons separated by a minimum distance in rapidity (\emph{rapidity-veto} approach~\cite{Schmidt:1999mz,Forshaw:1999xm,Caporale:2018qnm});
 \item[(\emph{iii})]
  an ``optimal'' choice of the values of the energy and renormalization scales, which, though arbitrary within the NLA, can have a sizeable numerical impact through subleading terms. 
\end{itemize}
As for the case (\emph{iii}), the most common \emph{optimization} methods are those ones inspired by: the \emph{principle of minimum sensitivity} (PMS)~\cite{PMS,PMS_2}, the \emph{fast apparent convergence} (FAC)~\cite{FAC,FAC_2,FAC_3}, and the \emph{Brodsky--Lepage--Mackenzie method} (BLM)~\cite{BLM,BLM_2,BLM_3,BLM_4,BLM_5}. Even though the selection of one or the other optimization procedure should not consistently affect the prediction of physical observables, it produces, \emph{de facto}, a non-negligible impact. Thus, as pointed out in Ref.~\cite{Caporale:2014gpa}, preference to a particular method should be assigned by grading the agreement with experimental data in a given setup and, consequently, assumed to apply also in other configurations. Predictions derived with BLM turned to be in fair agreement~\cite{Ducloue:2013bva,Caporale:2014gpa} with the only kinematic configuration for which experimental analyses are to date performed, \emph{i.e.} \textit{symmetric CMS} cuts in the Mueller--Navelet channel (for a detailed comparison of the three listed methods, se the Discussion Section of Ref.~\cite{Caporale:2014gpa}). In view of this outcome, the use of the BLM scheme was extended in time to the study of other semi-hard reactions.

In Section~\ref{blm} we briefly introduce the key features of BLM, which relies on the removal of the renormalization scale ambiguity by absorbing the non-conformal $\beta_0$-terms into the running coupling. Then, we apply it to the cases of our interest and present results for BLM optimal scale values and cross sections.

\subsubsection{BLM scale setting and BFKL cross section}
\label{blm}

The \emph{BLM-optimized} renormalization scale, $\mu_R^{\rm BLM}$, is the value of $\mu_R$ that makes the non-conformal, $\beta_0$-dependent terms entering the expression of the observable of interest vanish.
The inspection of the analytic structure of semi-hard observables shows that two groups of non-conformal contribution exist\footnote{We refer the reader to Ref.~\cite{Caporale:2015uva} for a complete treatment of the application of the BLM procedure to semi-hard reactions.}. The first one originates from the NLA BFKL kernel, while the second one from the NLO impact-factor corrections. This makes $\mu_R^{\rm BLM}$ non-universal, but dependent on the energy of the process. 

In order to apply the BLM procedure, a finite renormalization from the $\overline{\rm MS}$ to
the physical MOM scheme needs to be executed (see Section~\ref{strong_coupling}).
Then, the condition for the BLM scale setting for a given azimuthal coefficient, $C_n$, is determined by solving the following integral equation:
\begin{equation}
\label{int_Cn_beta}
  C_n^{(\beta)}(s,Y) = 
  \int_{\kappa_1^{\rm min}}^{\kappa_1^{\rm max}} \drv \kappa_1
  \int_{\kappa_2^{\rm min}}^{\kappa_2^{\rm max}} \drv \kappa_2
  \int_{y_1^{\rm min}}^{y_1^{\rm max}} \drv y_1
  \int_{y_2^{\rm min}}^{y_2^{\rm max}} \drv y_2
  \, \delta \left( Y - (y_1 - y_2) \right)
  \, {\cal C}_n^{(\beta)}  = 0 \, ,
\end{equation}
where
\[
 {\cal C}^{(\beta)}_n
 \propto \!\!
 \int^{\infty}_{-\infty} \!\!\drv\nu\,e^{Y \bar \alpha^{\rm MOM}_s(\mu^{\rm BLM}_R)\chi(n,\nu)}
% \left(\alpha^{\rm MOM}_s (\mu^{\rm BLM}_R)\right)^3
 c_1(n,\nu,\kappa_1, x_1)[c_2(n,\nu,\kappa_2,x_2)]^*
% \frac{\beta_0}{2 N_c}
\]
\begin{equation}
\label{Cn_beta}
 \times \, \left[{\hat f}(\nu) + \bar \alpha^{\rm MOM}_s(\mu^{\rm BLM}_R) Y \: \frac{\chi(n,\nu)}{2} \left(- \frac{\chi(n,\nu)}{2} + {\hat f}(\nu) \right) \right] \, ,
\end{equation}

with
\begin{equation}
\label{fnu_hat}
{\hat f}(\nu) = f(\nu) + \frac{5}{3} + 2 \ln \frac{\mu^{\rm BLM}_R}{\mu_N} - 2 - \frac{4}{3} I \, .
\end{equation}

\begin{figure}[t]
\centering

   \includegraphics[scale=0.56,clip]{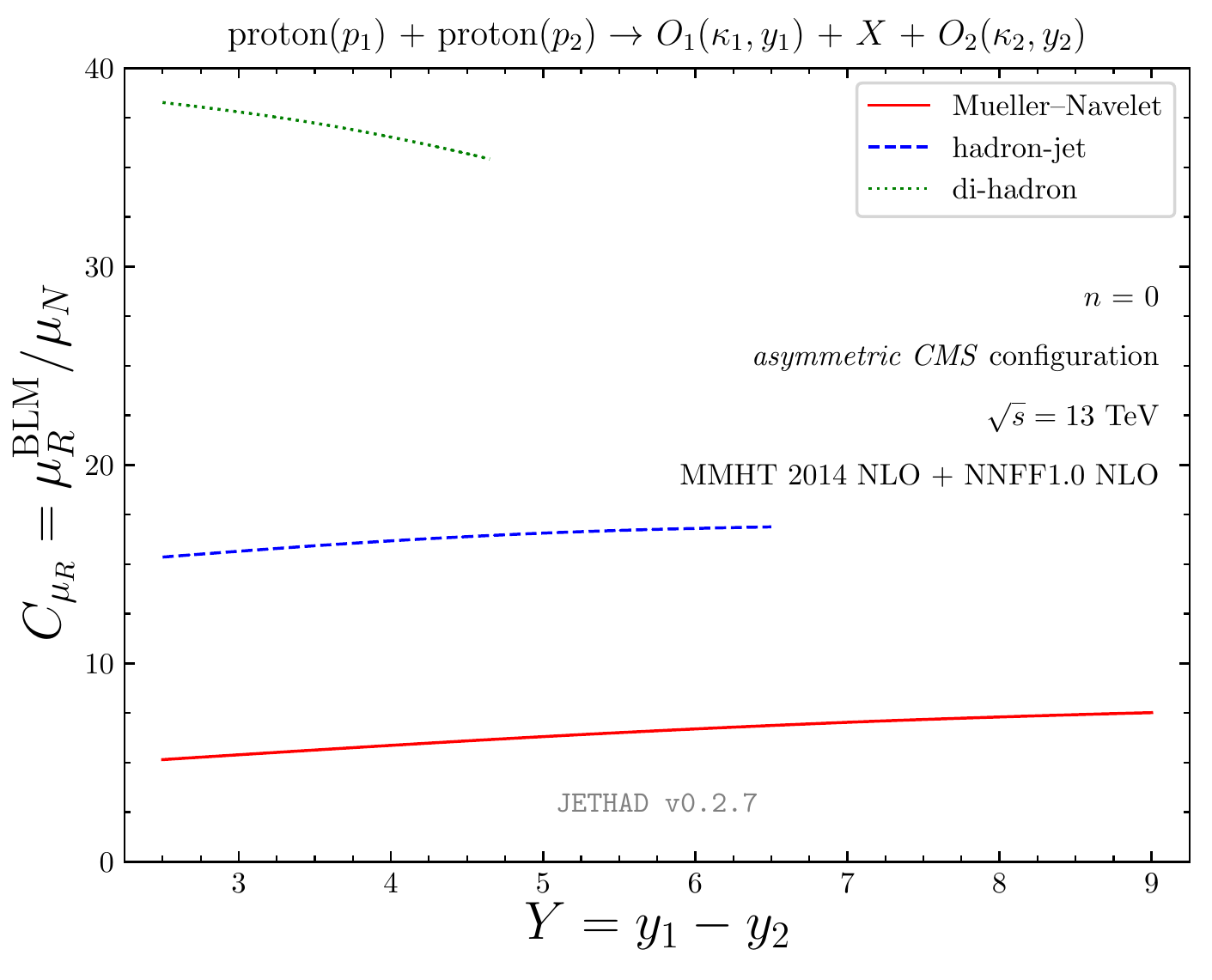}
   \hspace{0.25cm}
   \includegraphics[scale=0.56,clip]{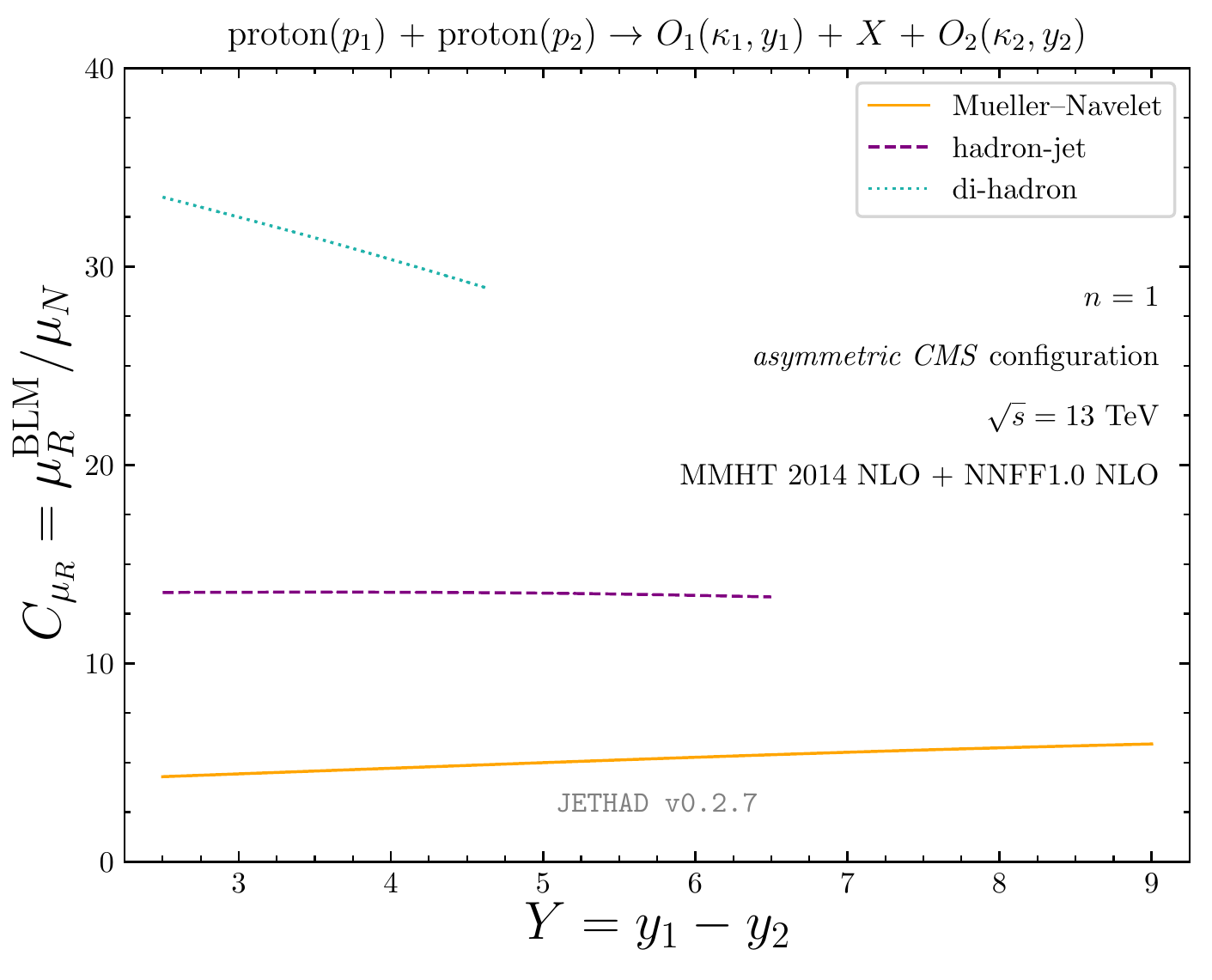}

   \includegraphics[scale=0.56,clip]{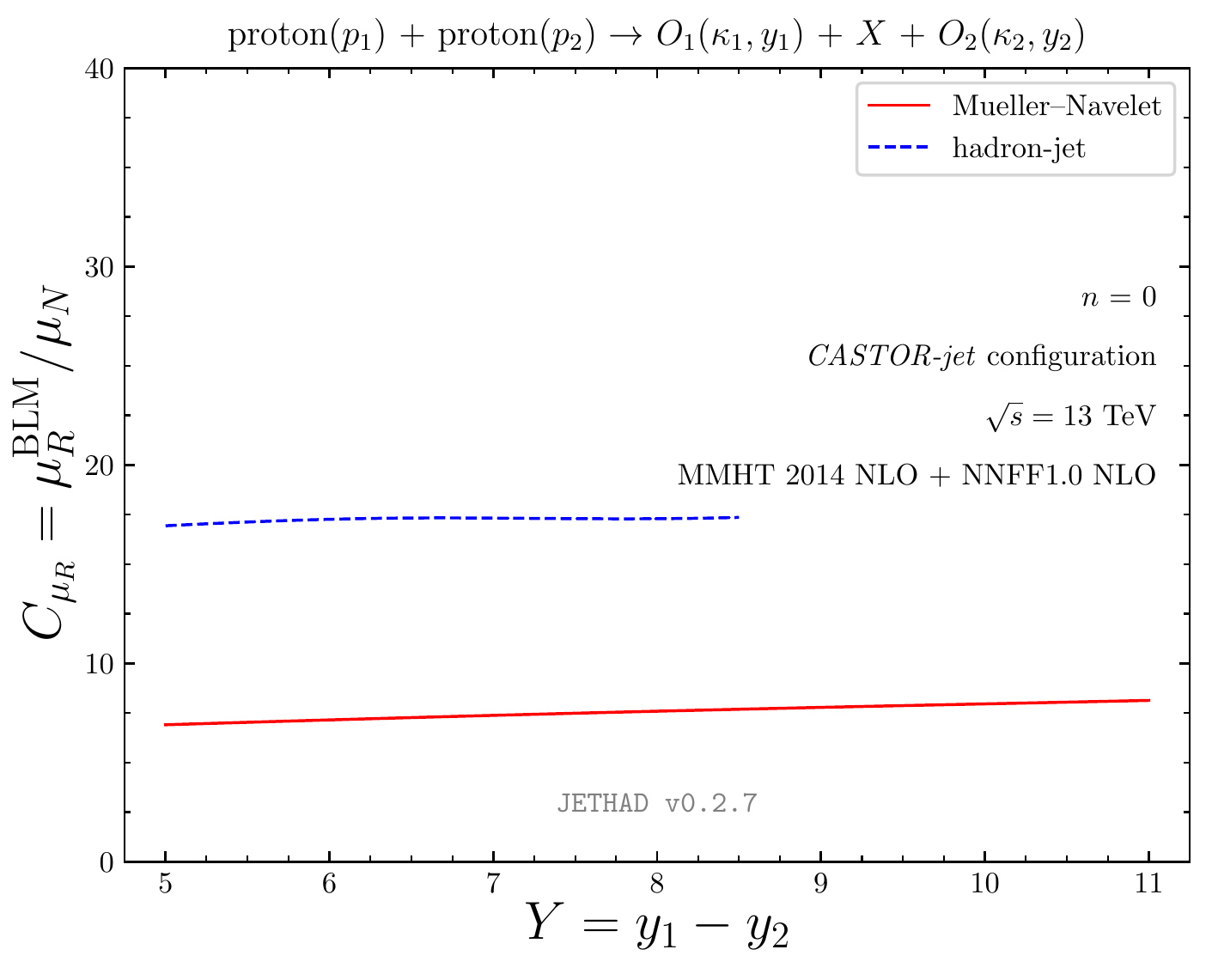}
   \hspace{0.25cm}
   \includegraphics[scale=0.56,clip]{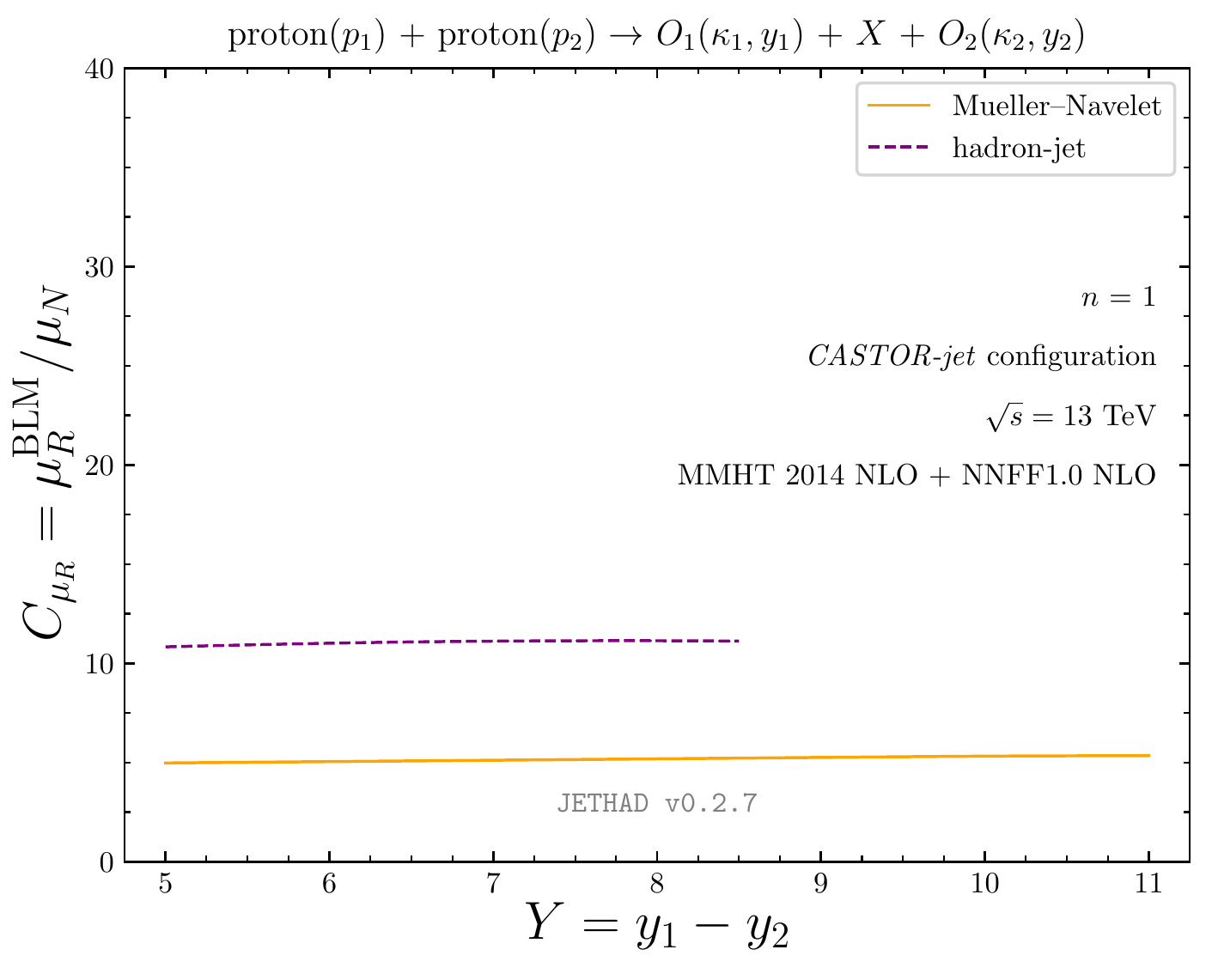}

\caption{BLM scales for the three considered reactions (Fig.~\ref{fig:processes}) versus the final-state rapidity interval, $Y$, for the $\varphi$-averaged cross section $C_0$ (left) and the azimuthal coefficient $C_1$ (right). Both the factorization scales, $\mu_{F1,2}$, have been set equal to $\mu_R^{\rm BLM} \equiv C_{\mu_R} \, \mu_N$. Results for $\sqrt{s} = 13$ TeV in the \textit{asymmetric CMS} (\textit{CASTOR-jet}) configuration are given in upper (lower) panels. Hadron emissions, when considered, are described in terms of {\tt NNFF1.0}~\cite{Bertone:2017tyb} NLO FF parametrizations (see Section~\ref{PDF_FF} for further details).}
\label{fig:SO-BLM-CMS-CST}
\end{figure}

\begin{figure}[t]
\centering

   \includegraphics[scale=0.56,clip]{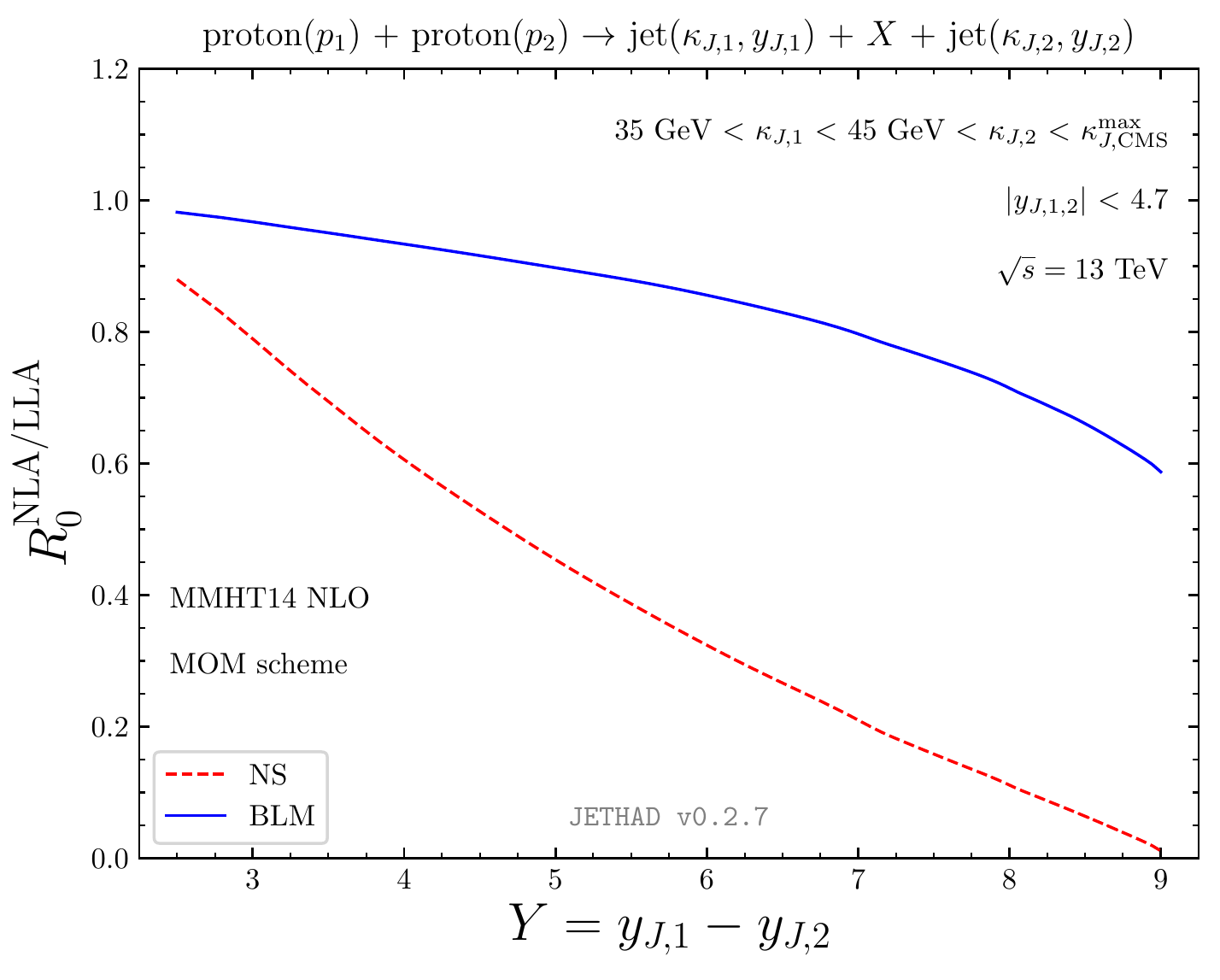}
   \hspace{0.25cm}
   \includegraphics[scale=0.56,clip]{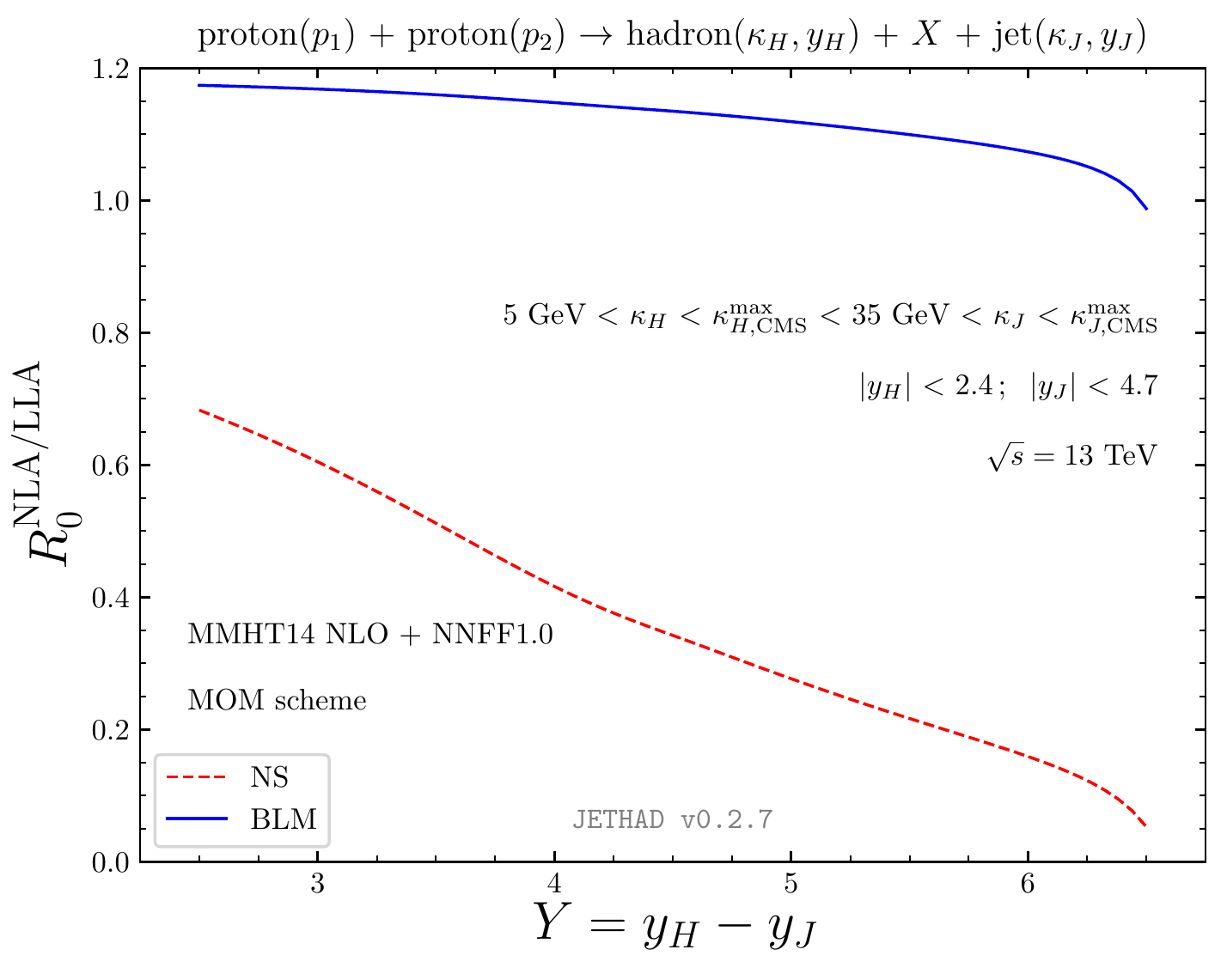}

   \includegraphics[scale=0.56,clip]{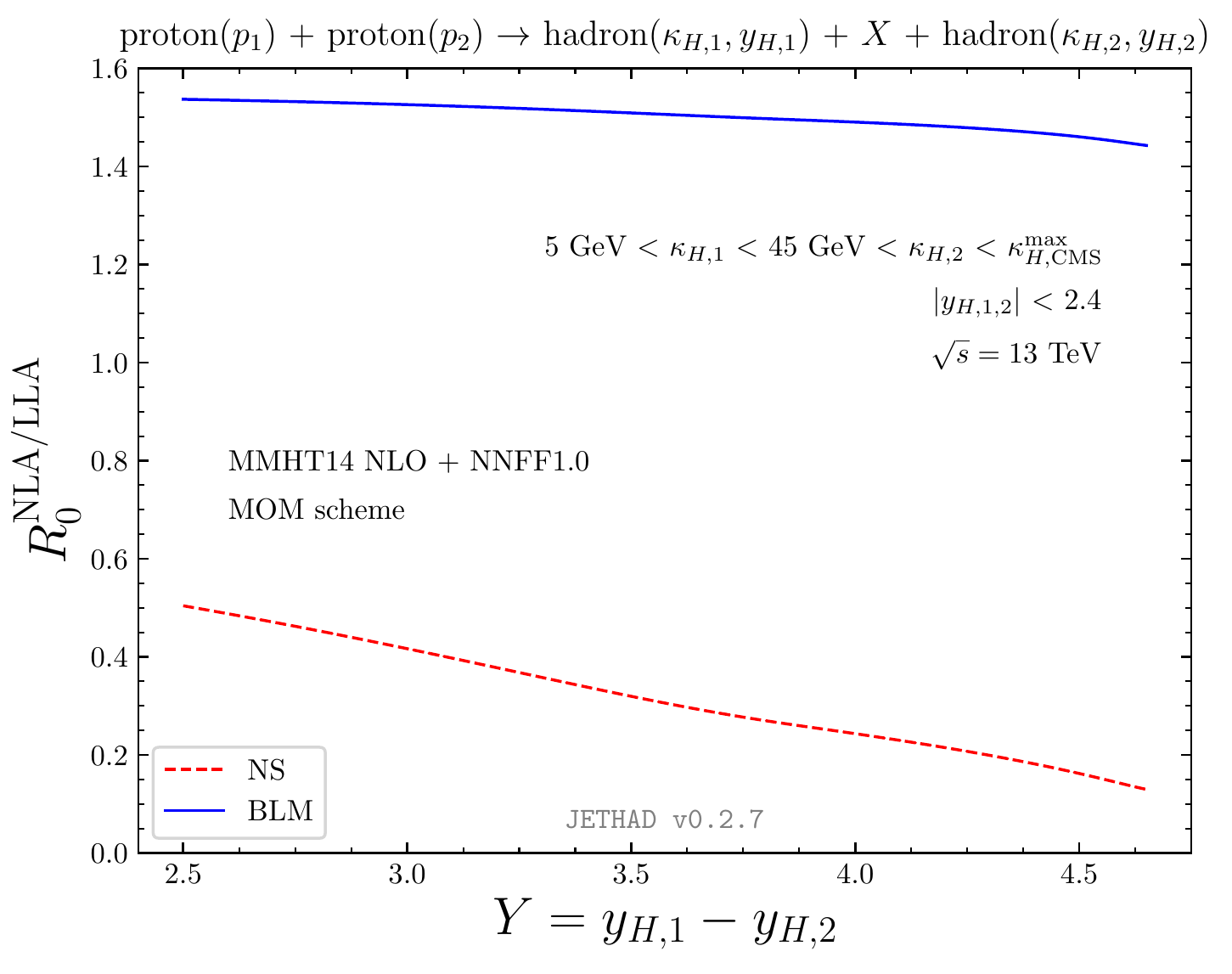}

\caption{$Y$-dependence of the $R_0^{\rm NLA/LLA}$ ratio for the three considered reactions (Fig.~\ref{fig:processes}) in the \textit{asymmetric CMS} configuration and for $\sqrt{s} = 13$ TeV. Hadron emissions, when considered, are described in terms of {\tt NNFF1.0}~\cite{Bertone:2017tyb} NLO FF parametrizations (see Section~\ref{PDF_FF} for further details).}
\label{fig:R0-CMS}
\end{figure}

\begin{figure}[t]
\centering

   \includegraphics[scale=0.56,clip]{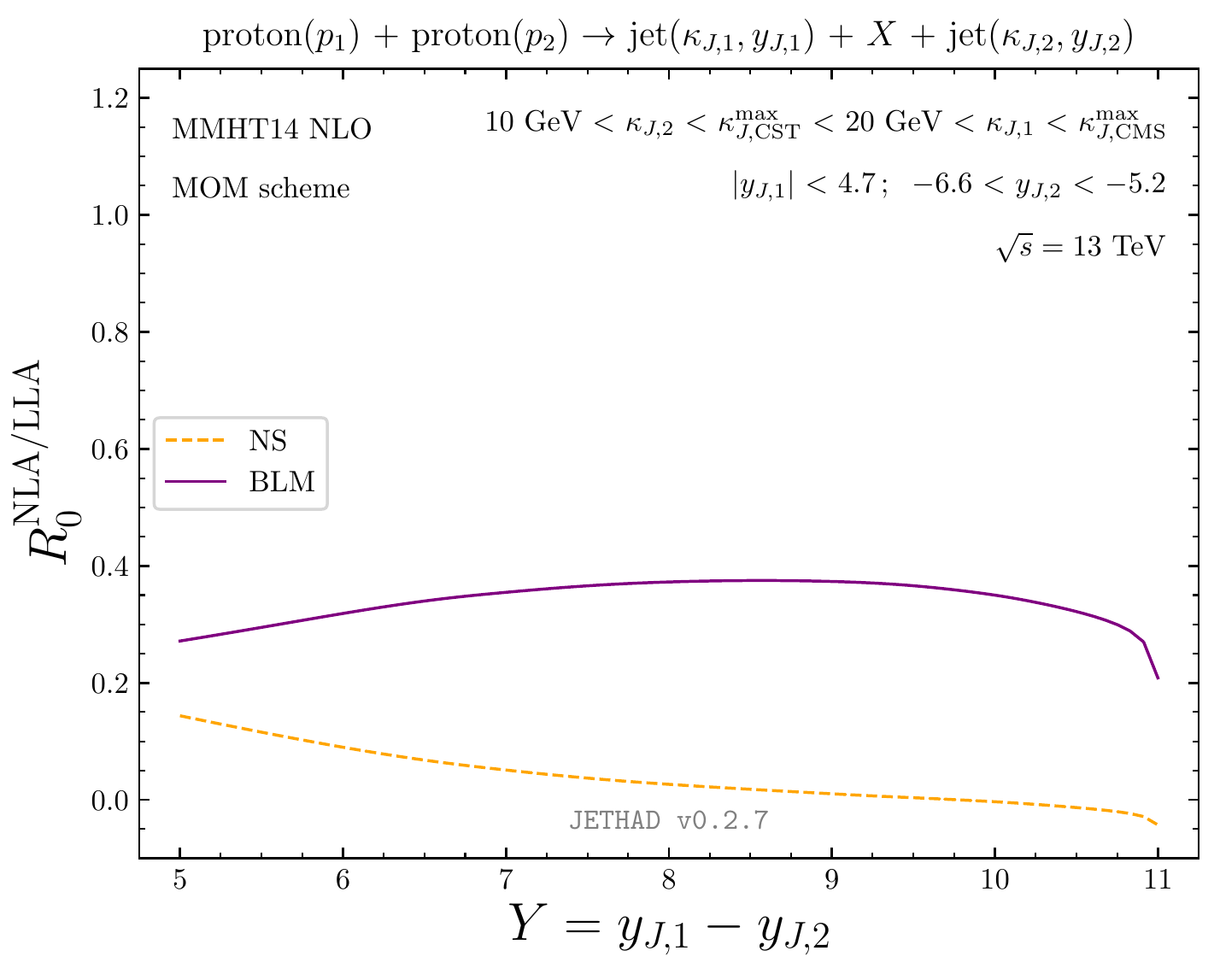}
   \hspace{0.25cm}
   \includegraphics[scale=0.56,clip]{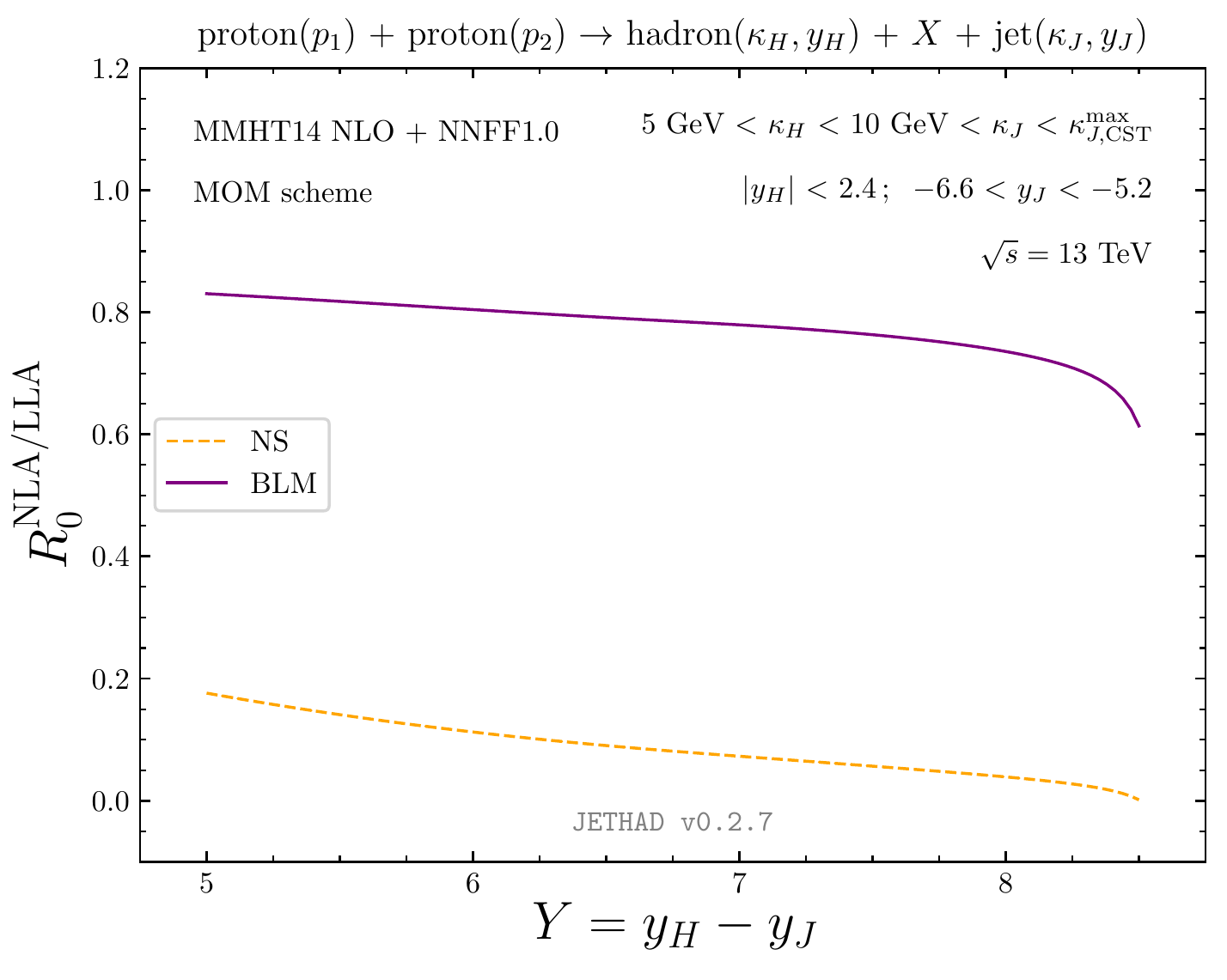}

\caption{$Y$-dependence of the $R_0^{\rm NLA/LLA}$ ratio for the three considered reactions (Fig.~\ref{fig:processes}) in the \textit{CASTOR-jet} configuration and for $\sqrt{s} = 13$ TeV. Hadron emissions, when considered, are described in terms of {\tt NNFF1.0}~\cite{Bertone:2017tyb} NLO FF parametrizations (see Section~\ref{PDF_FF} for further details).}
\label{fig:R0-CST}
\end{figure}

\begin{figure}[t]
\centering

   \includegraphics[scale=0.56,clip]{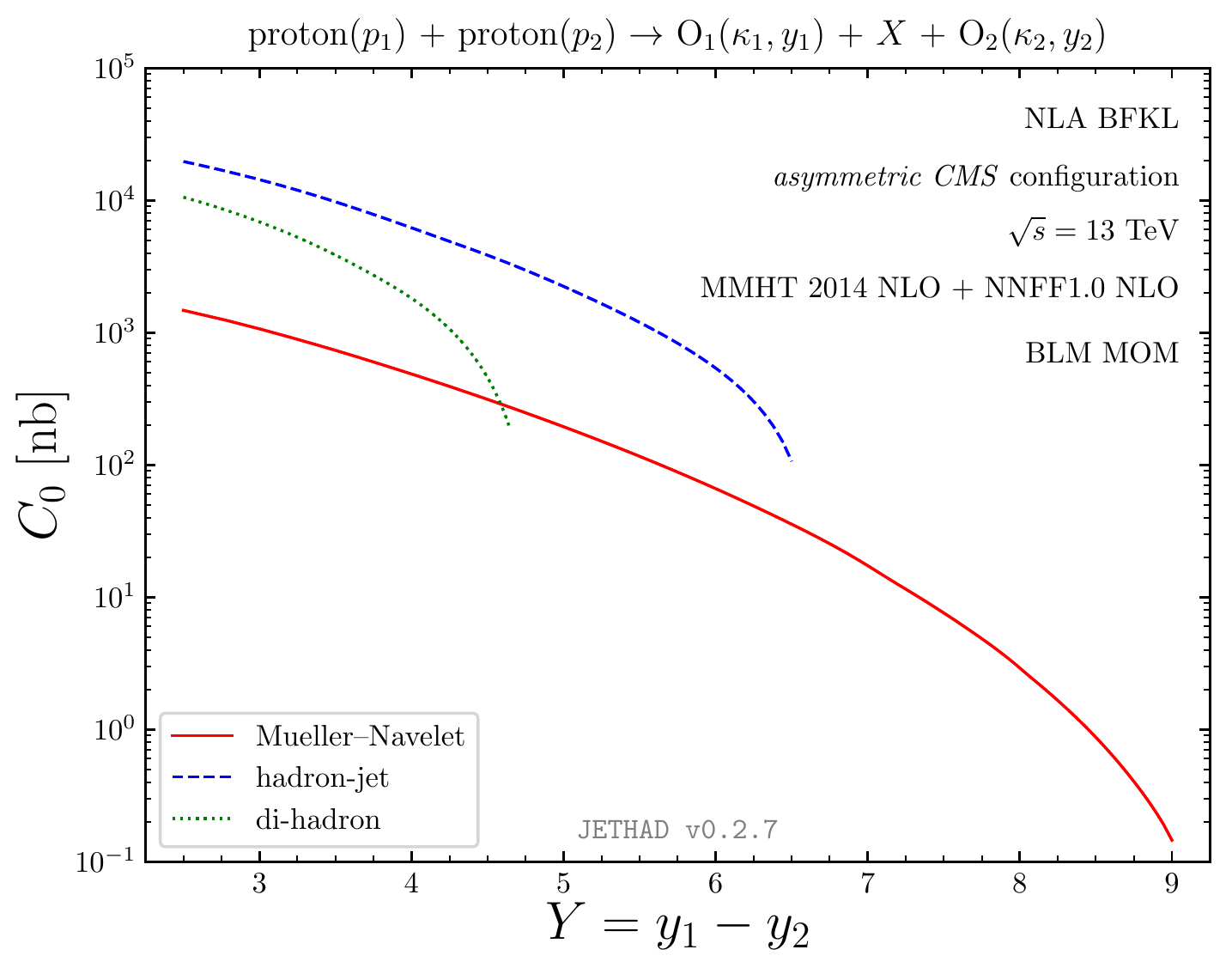}
   \hspace{0.25cm}
   \includegraphics[scale=0.56,clip]{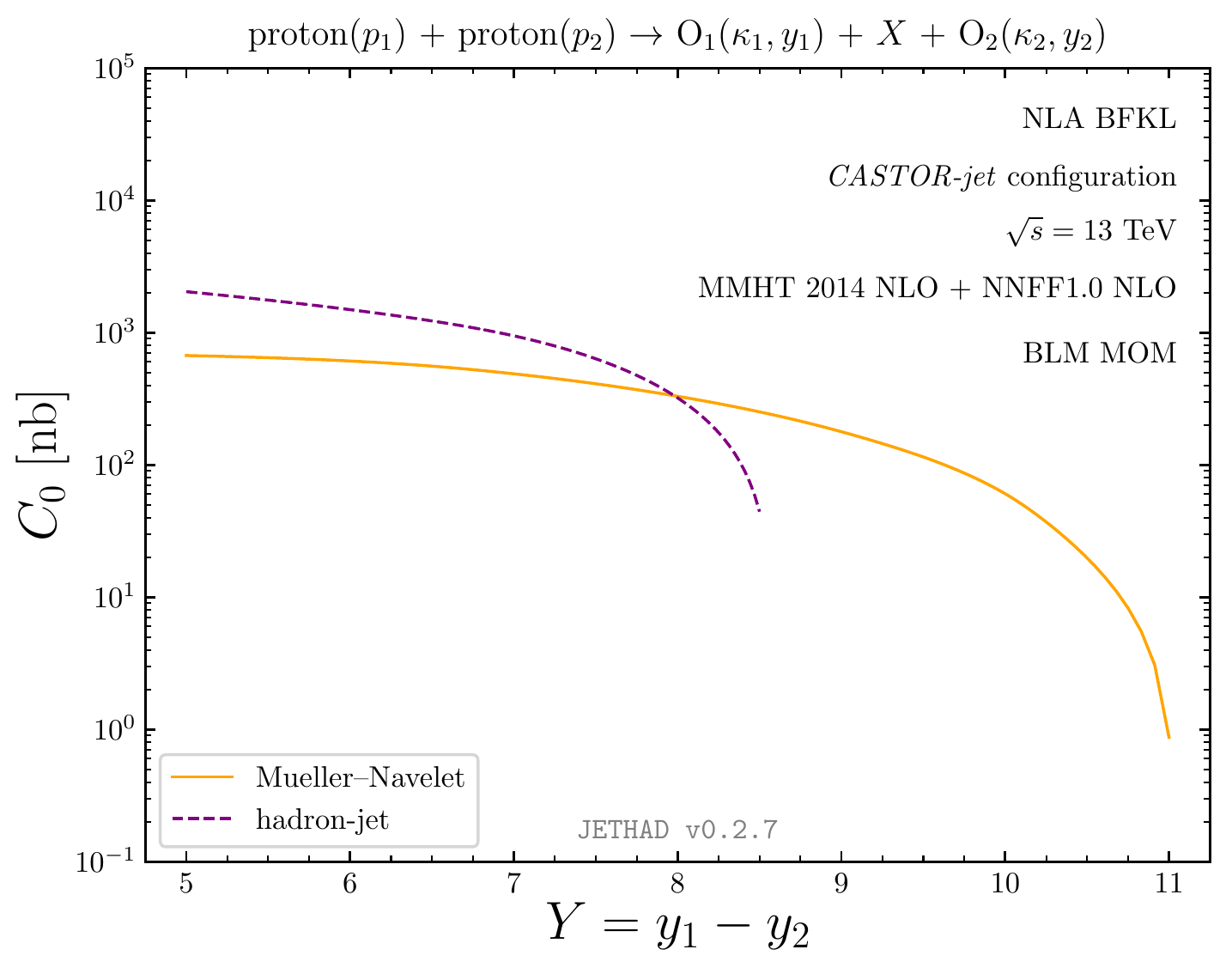}

\caption{$Y$-dependence of the $\varphi$-averaged cross section, $C_0$, for the three considered reactions (Fig.~\ref{fig:processes}) in the NLA BFKL accuracy. Results for $\sqrt{s} = 13$ TeV in the \textit{asymmetric CMS} (\textit{CASTOR-jet}) configuration are given in the left (right) panel. Hadron emissions, when considered, are described in terms of {\tt NNFF1.0}~\cite{Bertone:2017tyb} NLO FF parametrizations (see Section~\ref{PDF_FF} for further details).}
\label{fig:C0-BLM-CMS-CST}
\end{figure}

It is convenient to introduce the ratio of the 
BLM to the \emph{natural} scale suggested by the kinematic of the process, \emph{i.e.}
$\mu_N \equiv \sqrt{\kappa_1\kappa_2}$, so that $ C_{\mu_R} \equiv \mu_R^{\rm BLM}/\mu_N$, and hunt for those values of $C_{\mu_R}$ which solve Eq.~(\ref{int_Cn_beta}). 

As a last step, the value of the scale given by the procedure is plugged into expressions for the integrated coefficients, getting the following definition, valid in the NLA BFKL accuracy and in the MOM scheme:
\[
C_n^{\rm NLA} = 
 \int_{\kappa_1^{\rm min}}^{\kappa_1^{\rm max}} \drv \kappa_1
 \int_{\kappa_2^{\rm min}}^{\kappa_2^{\rm max}} \drv \kappa_2
 \int_{y_1^{\rm min}}^{y_1^{\rm max}} \drv y_1
 \int_{y_2^{\rm min}}^{y_2^{\rm max}} \drv y_2
 \,\; \delta \left( Y - (y_1 - y_2) \right)
\]
\[
 \times \,
 \int_{-\infty}^{+\infty} \drv \nu \,
 \frac{e^Y}{s}\,
 e^{Y \bar \alpha^{\rm MOM}_s(\mu^{\rm BLM}_R)\left[\chi(n,\nu)
 +\bar \alpha^{\rm MOM}_s(\mu^{\rm BLM}_R)\left(\bar \chi(n,\nu) +\frac{T^{\rm conf}}
 {3}\chi(n,\nu)\right)\right]}
 c_1(n,\nu,\kappa_1, x_1)[c_2(n,\nu,\kappa_2,x_2)]^*
\]
\begin{equation}
\label{Cn_NLA_int_blm}
\times \,
\left(\alpha^{\rm MOM}_s (\mu^{\rm BLM}_R)\right)^2
 \left\{1 + \alpha^{\rm MOM}_s(\mu^{\rm BLM}_R)\left[\frac{\bar c_1(n,\nu,\kappa_1,x_1)}{c_1(n,\nu,\kappa_1, x_1)}
 +\left[\frac{\bar c_2(n,\nu,\kappa_2, x_2)}{c_2(n,\nu,\kappa_2,x_2)}\right]^*
 +\frac{2T^{\rm conf}}{\pi} \right] \right\} \, ,
\end{equation}
where $\bar c_{i}(n,\nu,\kappa_i,x_i)$ are the NLO impact-factor corrections after the \emph{subtraction} 
of those terms, entering their expressions, which are proportional to $\beta_0$ and can be universally
expressed through the LO impact factors:
\begin{equation}
\label{IF_LO_sub}
\bar c_{i}(n,\nu,\kappa_i,x_i) = \hat c_{i}(n,\nu,\kappa_i,x_i) - \frac{\beta_0}{4 N_c} \left[ i \frac{\drv}{\drv \nu} \ln c_{i}(n,\nu,\kappa_i,x_i) + \left( \ln \mu_R^2 + \frac{5}{3} \right) c_{i}(n,\nu,\kappa_i,x_i) \right] \, .
\end{equation}

The analogous BLM-MOM expression for the DGLAP case in the high-energy limit reads:
\[
C_n^{\rm DGLAP} = 
 \int_{\kappa_1^{\rm min}}^{\kappa_1^{\rm max}} \drv \kappa_1
 \int_{\kappa_2^{\rm min}}^{\kappa_2^{\rm max}} \drv \kappa_2
 \int_{y_1^{\rm min}}^{y_1^{\rm max}} \drv y_1
 \int_{y_2^{\rm min}}^{y_2^{\rm max}} \drv y_2
 \,\; \delta \left( Y - (y_1 - y_2) \right)
\]
\[
 \times \,
 \int_{-\infty}^{+\infty} \drv \nu \,
\frac{e^Y}{s}\,
c_1(n,\nu,\kappa_1, x_1)[c_2(n,\nu,\kappa_2,x_2)]^*
\left(\alpha^{\rm MOM}_s (\mu^{\rm BLM}_R)\right)^2
\]
\begin{equation}
\label{Cn_DGLAP_int_blm}
\times \,
 \left\{1 + \alpha^{\rm MOM}_s(\mu^{\rm BLM}_R)\left[Y \frac{C_A}{\pi} \chi(n,\nu) + \frac{\bar c_1(n,\nu,\kappa_1,x_1)}{c_1(n,\nu,\kappa_1, x_1)}
 +\left[\frac{\bar c_2(n,\nu,\kappa_2, x_2)}{c_2(n,\nu,\kappa_2,x_2)}\right]^*
 +\frac{2T^{\rm conf}}{\pi} \right] \right\} \, .
\end{equation}

Finally, a valid formula for the azimuthal coefficients in the pure LLA BFKL accuracy can be obtained by making in Eq.~(\ref{LLA_Cn}) the substitution (note that the functional form of the strong coupling and the value of the renormalization scale are simultaneously varied)
\[
 \as^{(\MSb)}(\mu_R) \to \as^{(\rm MOM)}(\mu_R^{\rm BLM}) \; .
\]

Eq.~(\ref{int_Cn_beta}) requires a numerical solution, whose algorithm is implemented in {\Jethad} (see Section~\ref{jethad} for further details) and universally holds for any semi-hard final state. Before the advent of {\Jethad}, some analytic, approximate BLM methods were developed. In those cases (see Eqs.~(42)-(46) of Ref.~\cite{Caporale:2015uva}), the BLM scale is treated as a function of the variable $\nu$ and is fixed in order to make vanish either the NLO impact-factor $\beta_0$-dependent terms, the NLA BFKL kernel ones, or the whole integrand of Eq.~(\ref{int_Cn_beta}).

In all the calculations of this work the two factorization scales, $\mu_{F1,2}$, are set equal to the renormalization scale $\mu_R$, as assumed by most of the PDF parametrizations. All results are obtained in the MOM renormalization scheme (except for the ones presented in Section~\ref{th_vs_exp}, where the $\MSb$ scheme is selected). 

Values of the BLM scales for $C_0$ and $C_1$, for all the three considered production channels (Fig.~\ref{fig:processes}), are presented in Fig.~\ref{fig:SO-BLM-CMS-CST}. At fixed rapidity distance, $Y$, the $C_{\mu_R}$ ratio definitely increases with the number of charged-light hadrons emitted in the final state, up to around $30 \div 40$ units of natural scales. 
This phenomenon, already observed in previous analyses on the inclusive di-hadron production~\cite{Celiberto:2016hae,Celiberto:2017ptm}, could be controlled by the behavior of collinear FFs as functions of $\mu_F$.
More in particular, when FFs are convoluted with PDFs in our LO impact factors~(Eq.~(\ref{hadron_IF_LO})), the dominant contribution to PDFs in the kinematic sector of our interest is given by the gluon, and thus only the behavior of the gluon FF plays a role. Preliminary studies on FF parametrizations depicting the emissions of different hadron species, including heavy-flavored ones~\cite{Celiberto:2021dzy}, have highlighted how the $\mu_F$-dependence of (gluon) FFs can enhance or worsen the stability of our azimuthal coefficients. Smooth-behaved, non-decreasing gluon FF functions, characteristic of heavy-flavored bound states (see, \emph{e.g.}, parametrizations for $\Lambda_c$ baryons~\cite{Kniehl:2020szu}, charged $D^*$ mesons~\cite{Kneesch:2007ey,Anderle:2017cgl,Soleymaninia:2017xhc} and $b$-flavored hadrons~\cite{Kniehl:2008zza,Kramer:2018vde}), have a stabilizing effect. This situation is in some aspects closer to jet emissions, where no FFs are employed. Conversely, light-flavored FF sets, typical of the study presented in this work, lead to an increased sensitivity of predictions on energy scales, as well as to a stronger discrepancy between LLA and NLA. Thus, in our case, the larger scales prescribed by the BLM method come out as a technical consequence of the stability requirement.
In addition, the function $f(\nu)$ entering the expression given in Eq.~(\ref{fnu_hat}), is zero in the Mueller--Navelet channel and non-zero in the hadron-tag case. This radically changes the analytic structure of the BLM equation~(\ref{Cn_beta}). Finally, a side study (Section~\ref{had_species}) on single hadron-species emission(s) addresses the potential presence of intrinsic contributions to the BLM-scale values coming from the FFs.

In Figs.~\ref{fig:R0-CMS} and~\ref{fig:R0-CST} we compare predictions for the ratio, $R_0^{\rm NLA/LLA} \equiv C_0^{\rm NLA}/C_0^{\rm LLA}$, obtained with BLM optimization to the corresponding ones calculated at \emph{natural} scales, \emph{i.e.} $\mu_R = \mu_N \equiv \sqrt{\kappa_1 \kappa_2}$ and $\mu_{F1,2} = \kappa_{1,2}$, after having checked that the alternative choice for $\mu_{F1,2}$, $\mu_{F1} = \mu_{F2} = \mu_N$, produces almost the same results with respect to the previous one.
In particular, Fig.~\ref{fig:R0-CMS} shows the $Y$-dependence of the $R_0^{\rm NLA/LLA}$ ratio for the three considered reactions (Fig.~\ref{fig:processes}) in the \textit{asymmetric CMS} configuration. We notice that 
NLA corrections, generally with opposite sign with respect LLA results, become larger and larger in absolute value at increasing rapidity interval, thus making the $R_0^{\rm NLA/LLA}$ smaller and smaller (we stress that the label ``NLA'' in our plots \emph{always} stands for LLA plus higher-order terms, and not just the latter ones). This is an expected phenomenon in the BFKL approach which, however, becomes milder when scales are optimized. The adoption of the BLM method leads to a scale choice that permits to mimic the most relevant subleading terms, thus stabilizing the perturbative series. From the operational point of view, this results in a reduction of the distance between the LLA and the NLA, namely it raises the $R_0^{\rm NLA/LLA}$ ratio in our plots, making it ideally close to one. This is exactly what happens in the Mueller--Navelet case (upper left panel of Fig.~\ref{fig:R0-CMS}), while its value at BLM scales exceeds one in the case of hadron detection (remaining panels). The explanation for this, apparently odd behavior, has not to be hunted in the use of BLM, but rather in a combination of two distinct effects, already present in the expression for $C_0$ given in Eq.~(\ref{Cn_NLA_int_blm}) and therefore independent from the scale choice. On the one side, going from the $\MSb$ to the MOM renormalization scheme generates a non-exponentiated, positive extra factor proportional to $T^{\rm conf}$. On the other side, it is easy to prove that the $C_{gg}$ coefficient in Eq.~(\ref{Cgg_hadron}) gives a large and positive contribution to the hadron NLO impact-factor correction. All that makes $C_0$ at the NLA larger than the corresponding LLA one \emph{before} switching any eventual scale-optimization procedure on. 
For the sake of completeness, the $Y$-dependence of the $R_0^{\rm NLA/LLA}$ ratio for the di-jet and the hadron jet case (first two panels of Fig.~\ref{fig:processes}) in the \textit{CASTOR-jet} configuration is presented in Fig.~\ref{fig:R0-CST}.

Ultimately, the $Y$-dependence of the NLA $\varphi$-averaged cross section, $C_0$, for all the considered reactions is examined. Final-state configurations molded on the \textit{asymmetric CMS} (\textit{CASTOR-jet}) event selection are presented in the left (right) panel of Fig.~\ref{fig:C0-BLM-CMS-CST}. Here, a definite hierarchy among them is stringently respected, except for values of $Y$ close to the upper bound given by the kinematics: hadron detections dominate in statistics with respect to jet tags.

\begin{figure}[h]
\centering

   \includegraphics[scale=0.56,clip]{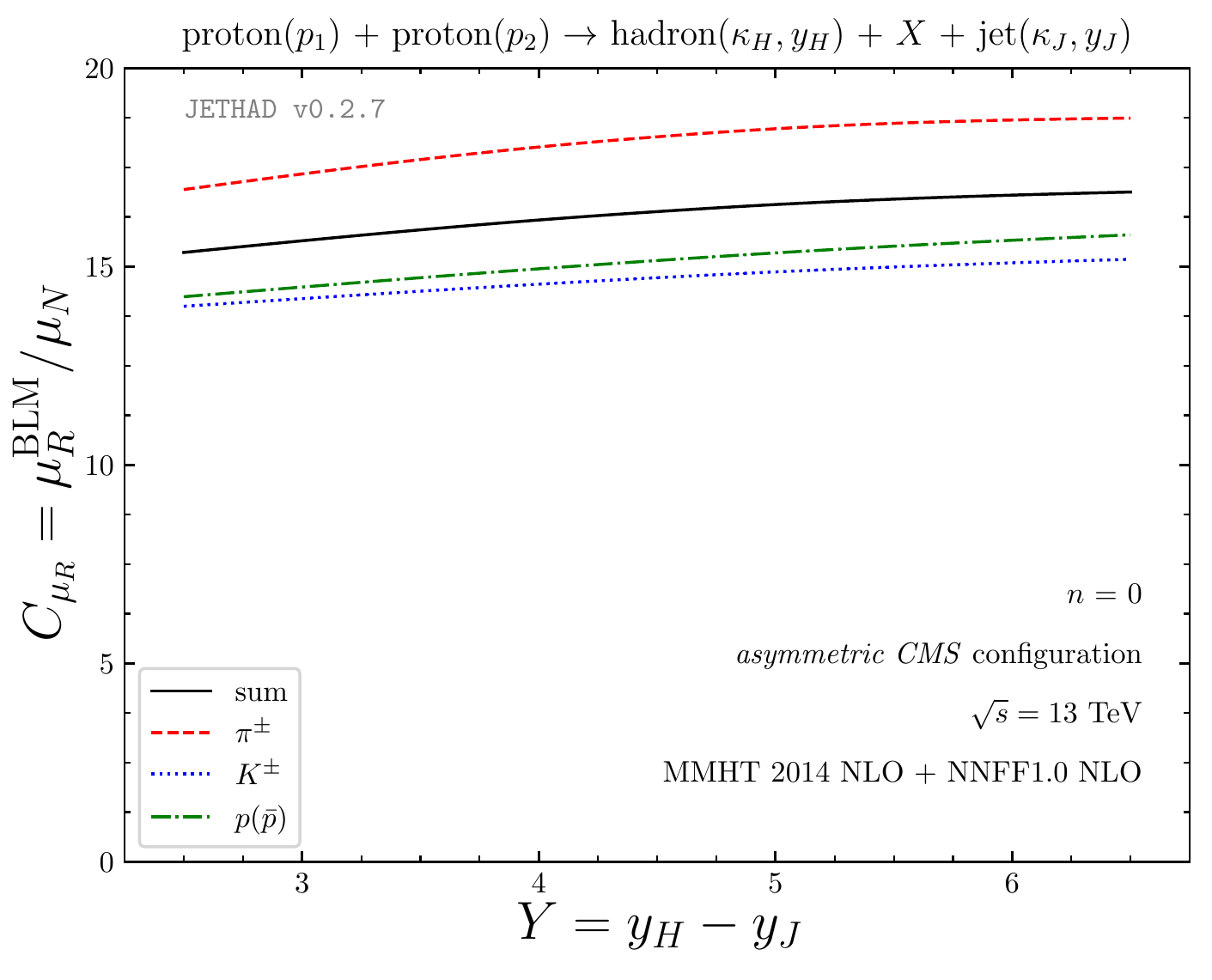}
   \hspace{0.25cm}
   \includegraphics[scale=0.56,clip]{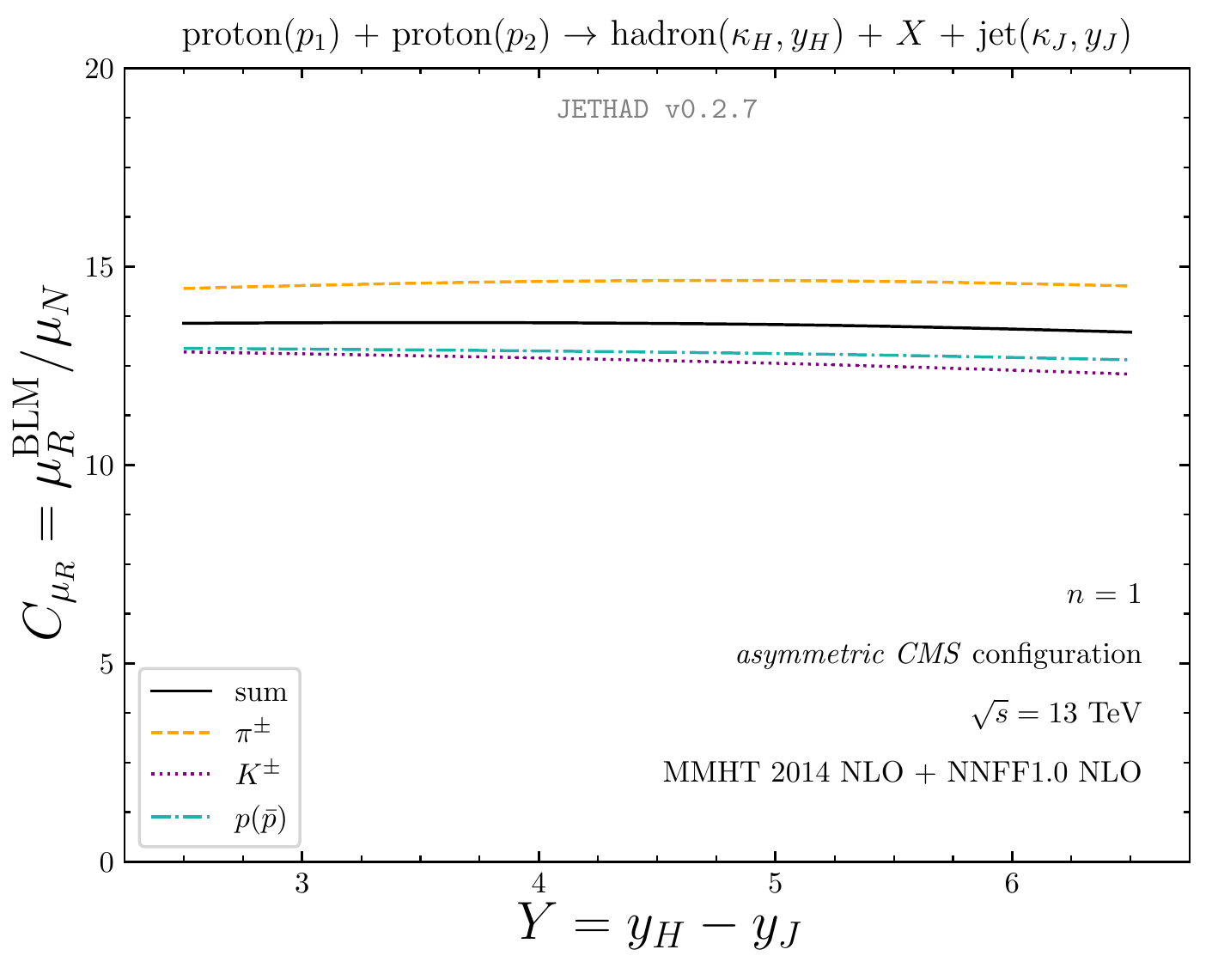}

   \includegraphics[scale=0.56,clip]{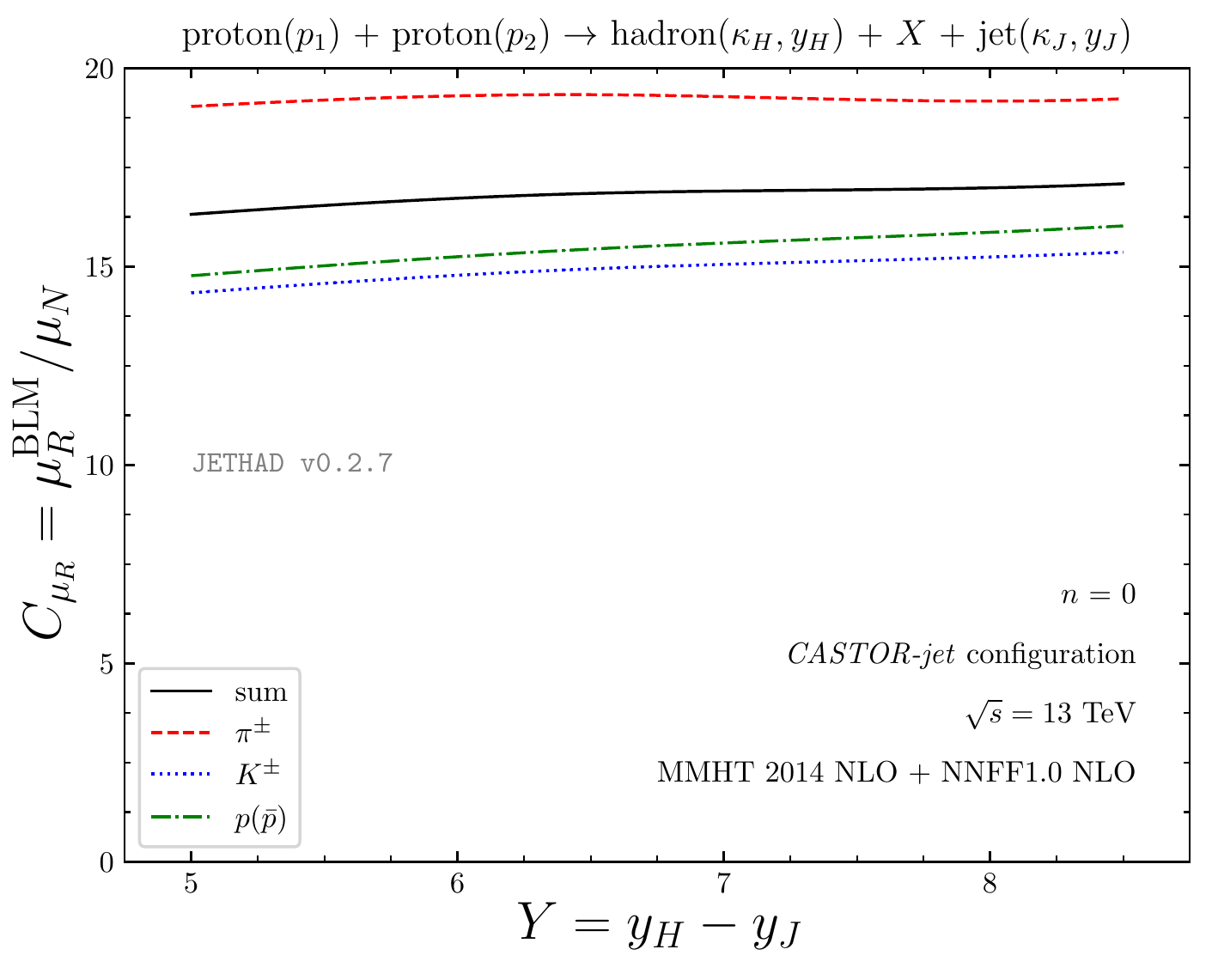}
   \hspace{0.25cm}
   \includegraphics[scale=0.56,clip]{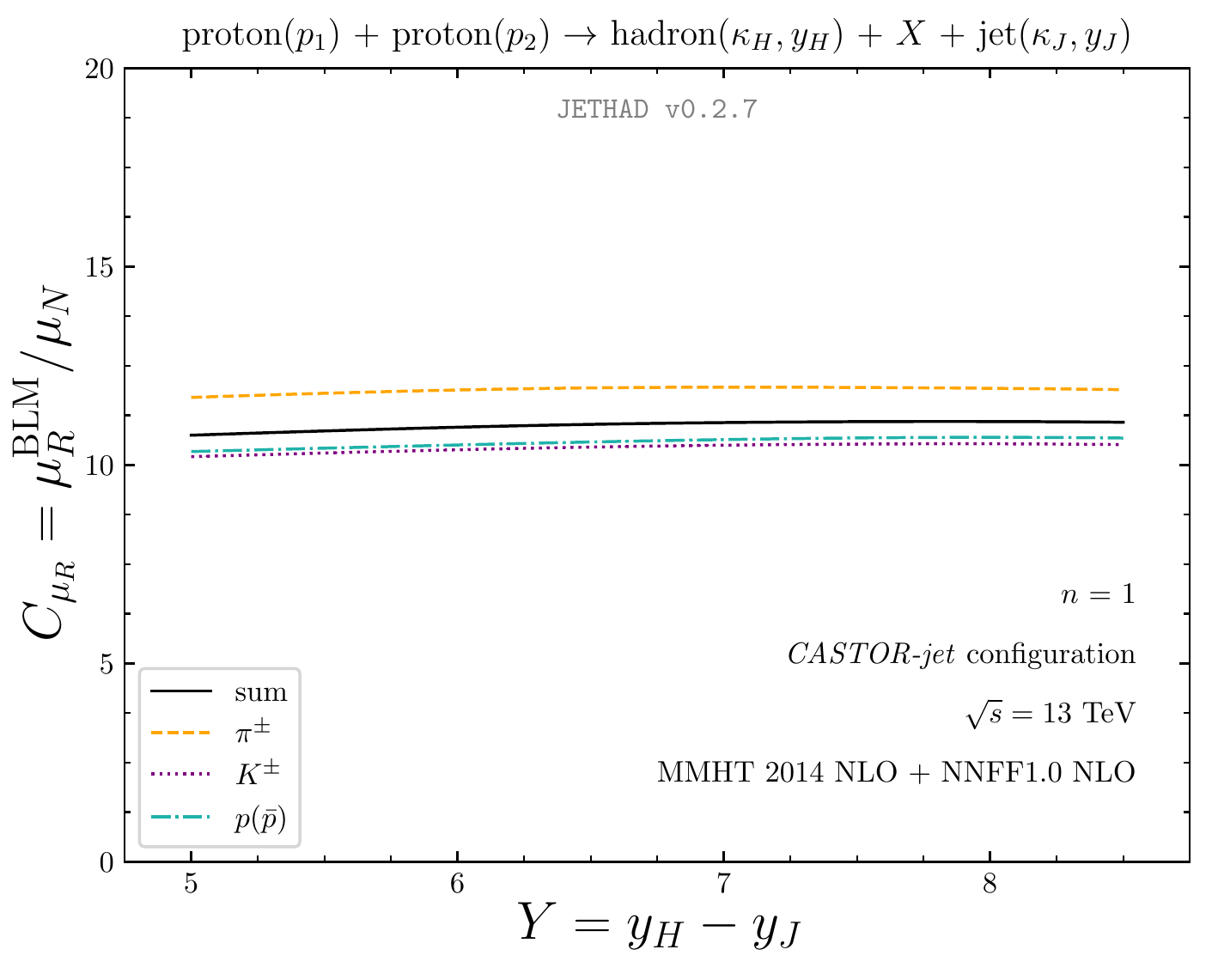}

\caption{BLM scales for the inclusive hadron-jet production (central panel in Fig.~\ref{fig:processes}) versus the final-state rapidity interval, $Y$, for the $\varphi$-averaged cross section $C_0$ (left) and the azimuthal coefficient $C_1$ (right), for $\sqrt{s} = 13$ TeV in the \textit{asymmetric CMS} configuration.
Predictions for single hadron-species emissions ($\pi^\pm$, $K^\pm$, $p(\bar{p})$) are compared with the standard case, where the \emph{sum} over all species is taken. Both the factorization scales, $\mu_{F1,2}$, have been set equal to $\mu_R^{\rm BLM} \equiv C_{\mu_R} \, \mu_N$. Results for $\sqrt{s} = 13$ TeV in the \textit{asymmetric CMS} (\textit{CASTOR-jet}) configuration are given in upper (lower) panels.}
\label{fig:HSA-SO-BLM-CMS-CST}
\end{figure}

\begin{figure}[h]
\centering

   \includegraphics[scale=0.56,clip]{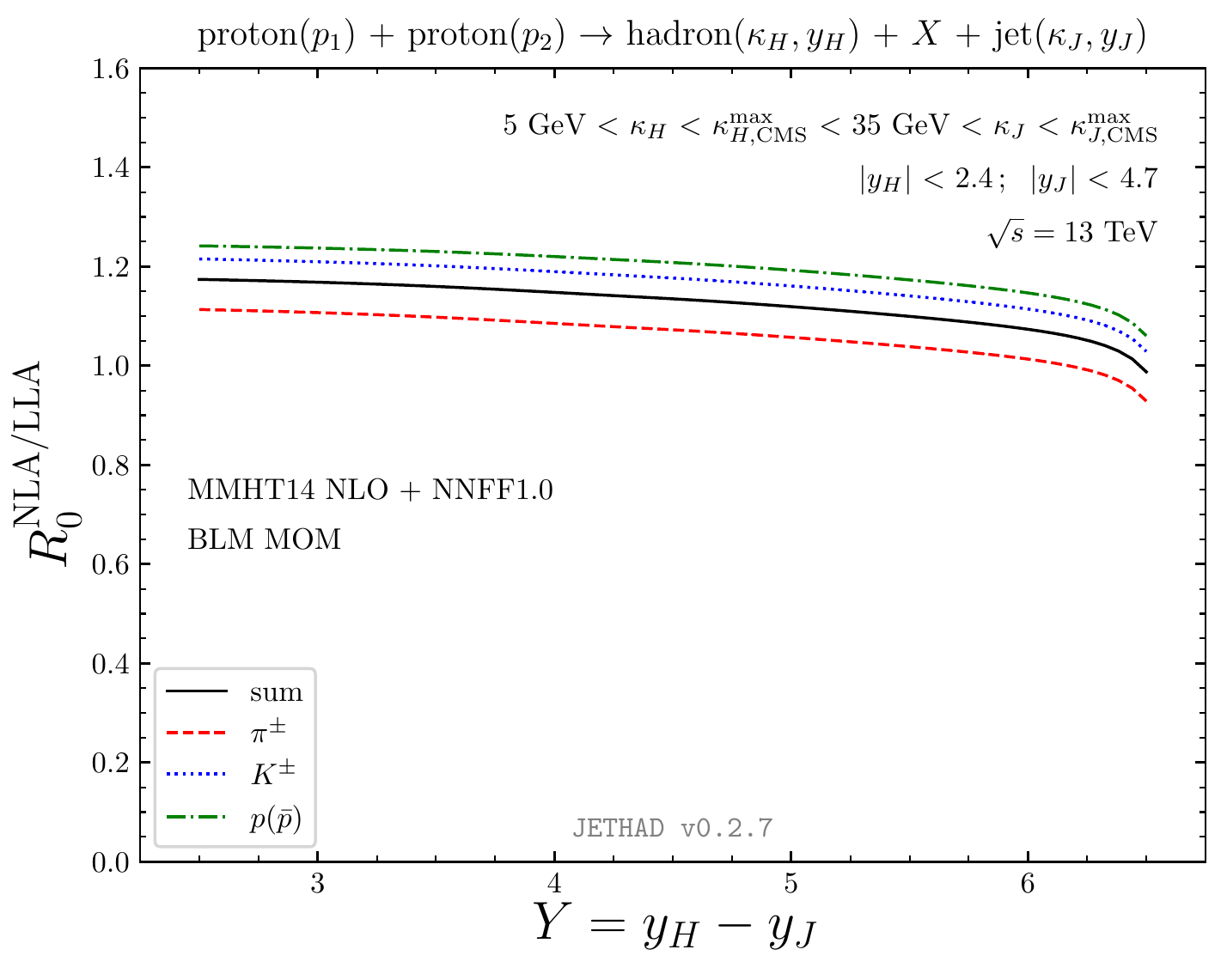}
   \hspace{0.25cm}
   \includegraphics[scale=0.56,clip]{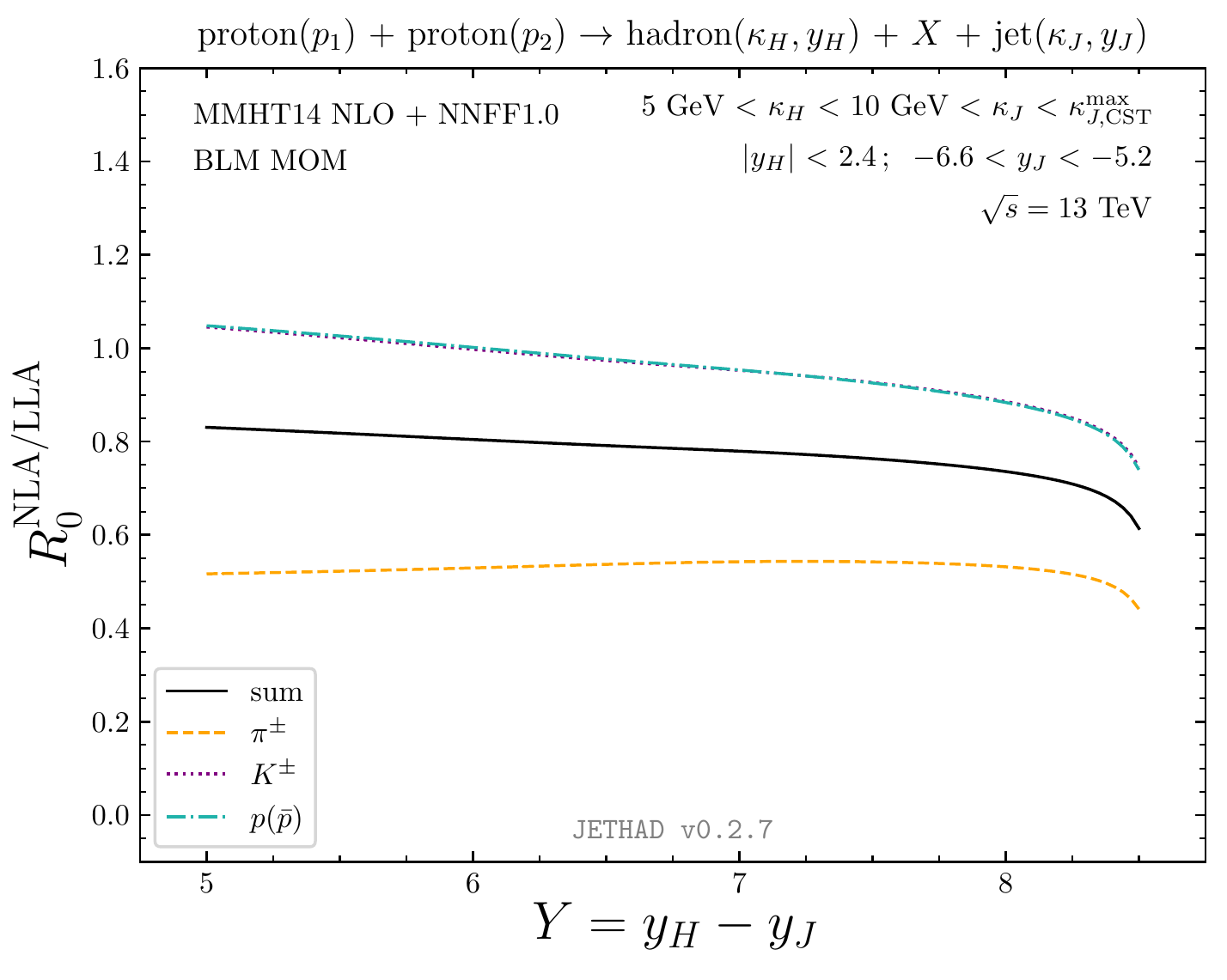}

\caption{$Y$-dependence of the $R_0^{\rm NLA/LLA}$ ratio for the inclusive hadron-jet production (central panel in Fig.~\ref{fig:processes}) in the \textit{asymmetric CMS} (left) and in the \textit{CASTOR-jet} (right) configuration, for $\sqrt{s} = 13$ TeV. Predictions for single hadron-species emissions ($\pi^\pm$, $K^\pm$, $p(\bar{p})$) are compared with the standard case, where the \emph{sum} over all species is taken.}
\label{fig:HSA-R0-BLM-CMS-CST}
\end{figure}

\begin{figure}[h]
\centering

   \includegraphics[scale=0.56,clip]{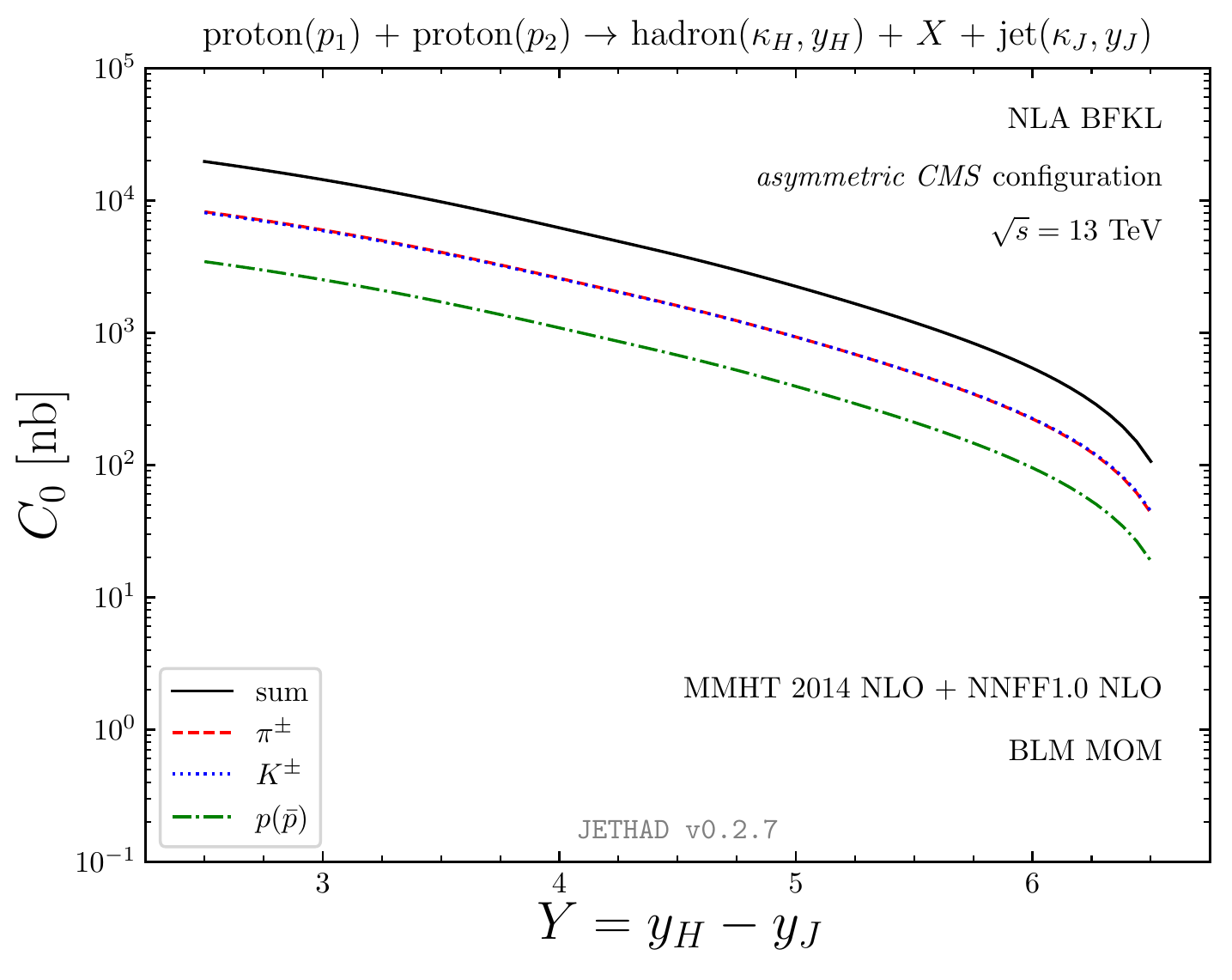}
   \hspace{0.25cm}
   \includegraphics[scale=0.56,clip]{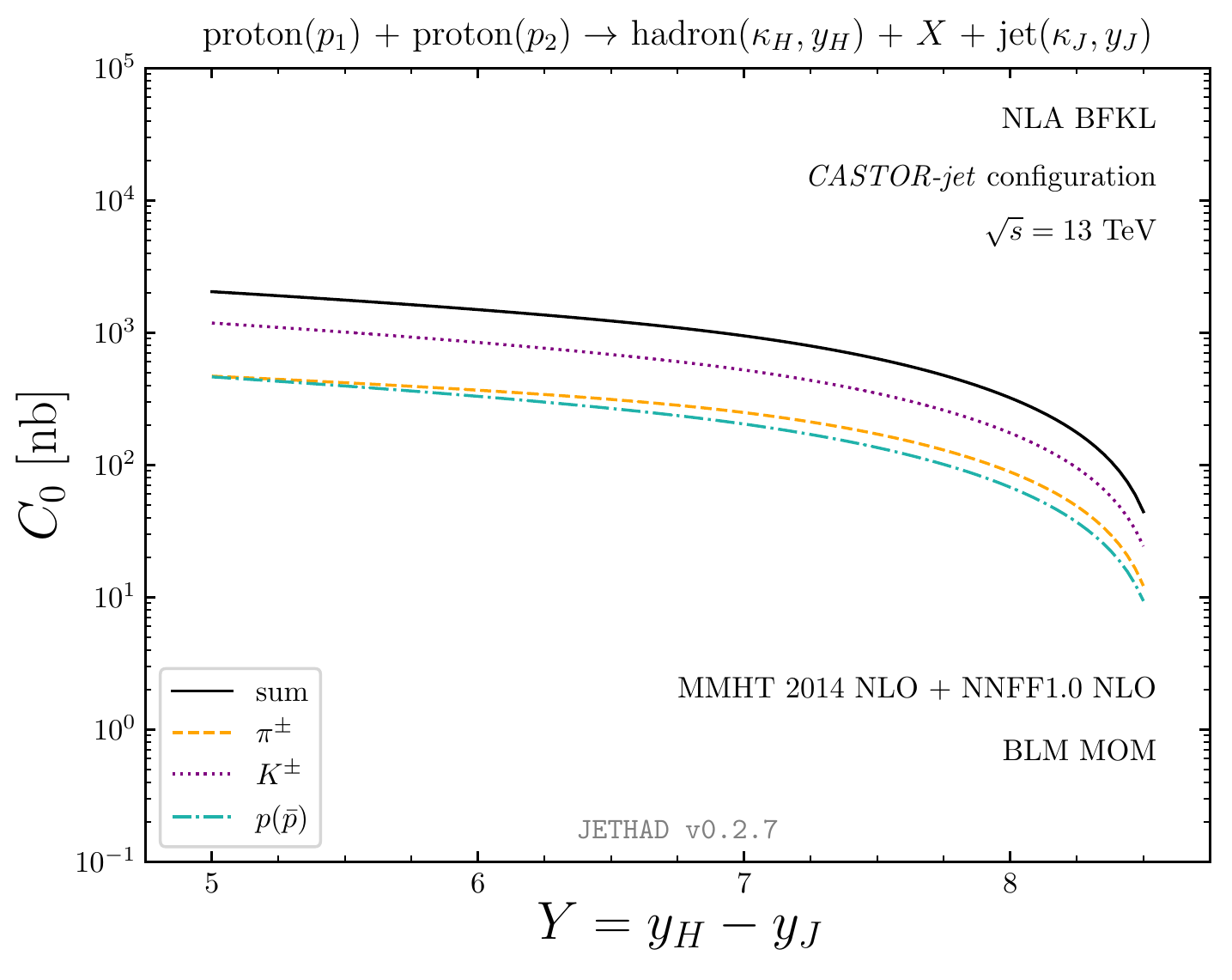}

\caption{$Y$-dependence of the $\varphi$-averaged cross section, $C_0$, for inclusive hadron-jet production (central panel in Fig.~\ref{fig:processes}) in the NLA BFKL accuracy, in the \textit{asymmetric CMS} (left) and in the \textit{CASTOR-jet} (right) configuration, and for $\sqrt{s} = 13$ TeV. Predictions for single hadron-species emissions ($\pi^\pm$, $K^\pm$, $p(\bar{p})$) are compared with the standard case, where the \emph{sum} over all species is taken.}
\label{fig:HSA-C0-BLM-CMS-CST}
\end{figure}

\subsubsection{Hadron-species analysis}
\label{had_species}

In this Section we hunt for possible contributions to the optimal scale values intrinsically coming from the description of the parton-to-hadron fragmentation. Predictions for single hadron-species detections ($\pi^\pm$, $K^\pm$, $p(\bar{p})$) are matched to the standard case, where the \emph{sum} over all species is intended. For the sake of simplicity, we take the inclusive hadron-jet production (central panel of Fig.~\ref{fig:processes}) as reference reaction and consider both the \textit{asymmetric CMS} and the \textit{CASTOR-jet} final-state configurations.

The inspection of results of the BLM scales (Fig.~\ref{fig:HSA-SO-BLM-CMS-CST}) for values of the conformal spin, $n = 0$ and $n = 1$, indicates that $\pi^\pm$ emissions lead to larger scales which are, however, of the same order of the other single-species cases and of the \emph{sum} one. Therefore, no clues of FF intrinsic effects emerge from our analysis. This can be due to the fact that, while FFs depend on the mass, $M_H$, of the corresponding light-flavored hadron, $H$, the hard subprocess is mass-independent. Thus, residual logarithmic contributions in $M_H$ can survive, generating a mild sensitivity of the scale-choice procedure to the given hadron species.

For completeness of presentation, the $Y$-dependence of the $R_0^{\rm NLA/LLA}$ ratio with BLM optimization is shown in Fig.~\ref{fig:HSA-R0-BLM-CMS-CST}, whereas the $\varphi$-averaged NLA BLM cross section is given in Fig.~\ref{fig:HSA-C0-BLM-CMS-CST}.

\subsection{BFKL versus DGLAP}
\label{bfkl_vs_dglap}

In this Section we present and discuss features of the core results of our work. The examination of azimuthal-correlation moments, given in Section~\ref{core}, is extended and accompanied by the analysis of azimuthal distributions, portrayed in Section~\ref{azimuthal_distribution}.

\subsubsection{Azimuthal correlations}
\label{core}

\begin{figure}[H]
\centering

   \includegraphics[scale=0.56,clip]{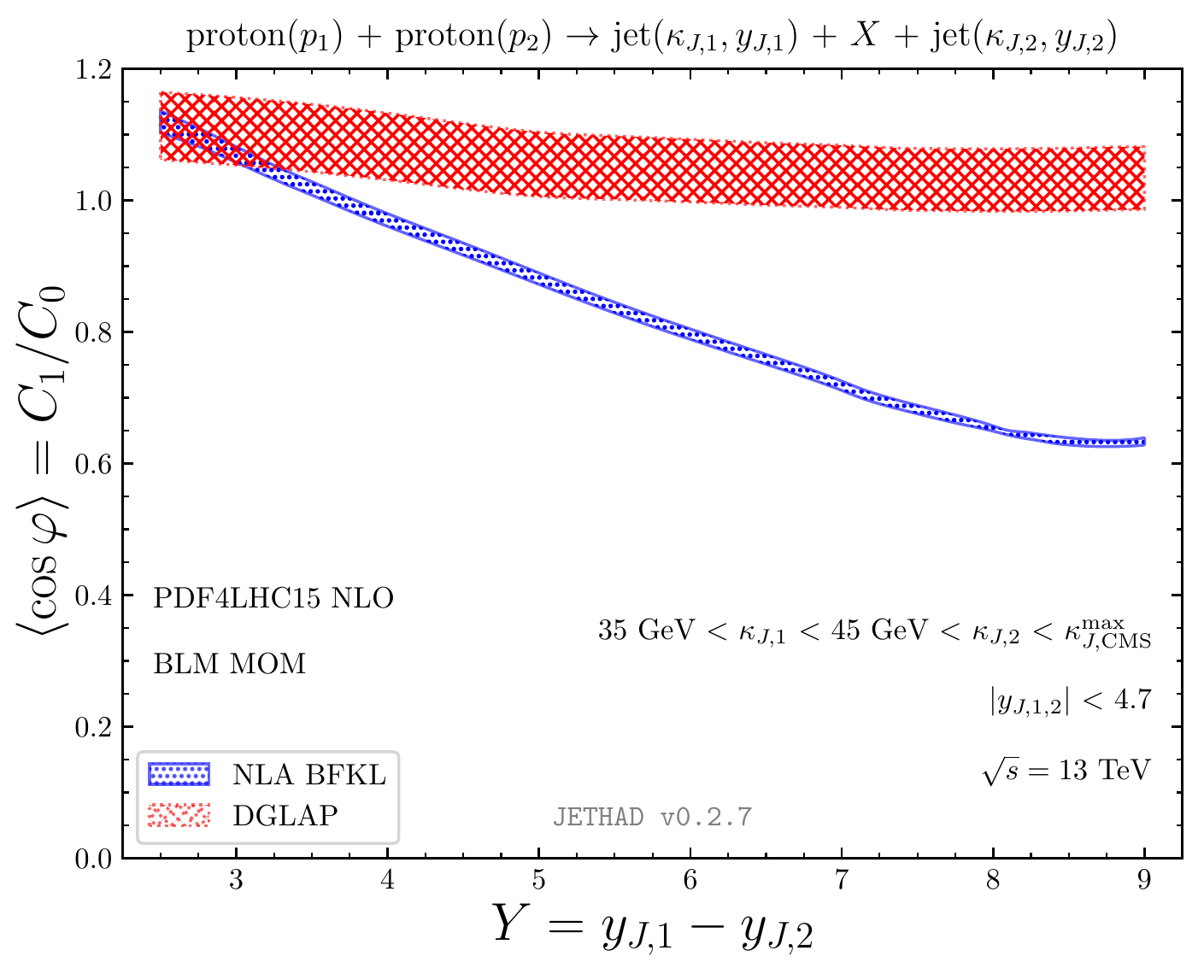}
   \hspace{0.25cm}
   \includegraphics[scale=0.56,clip]{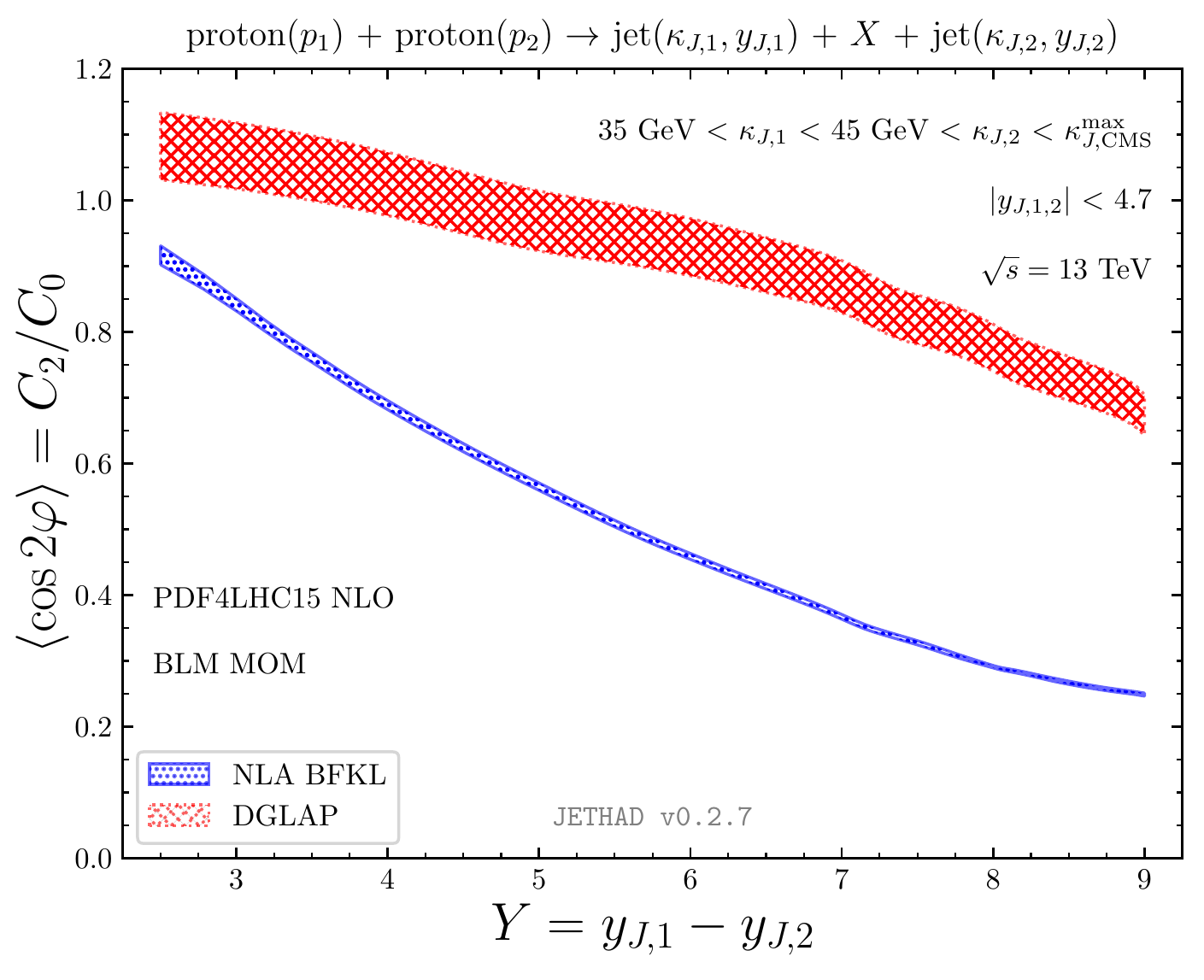}

   \includegraphics[scale=0.56,clip]{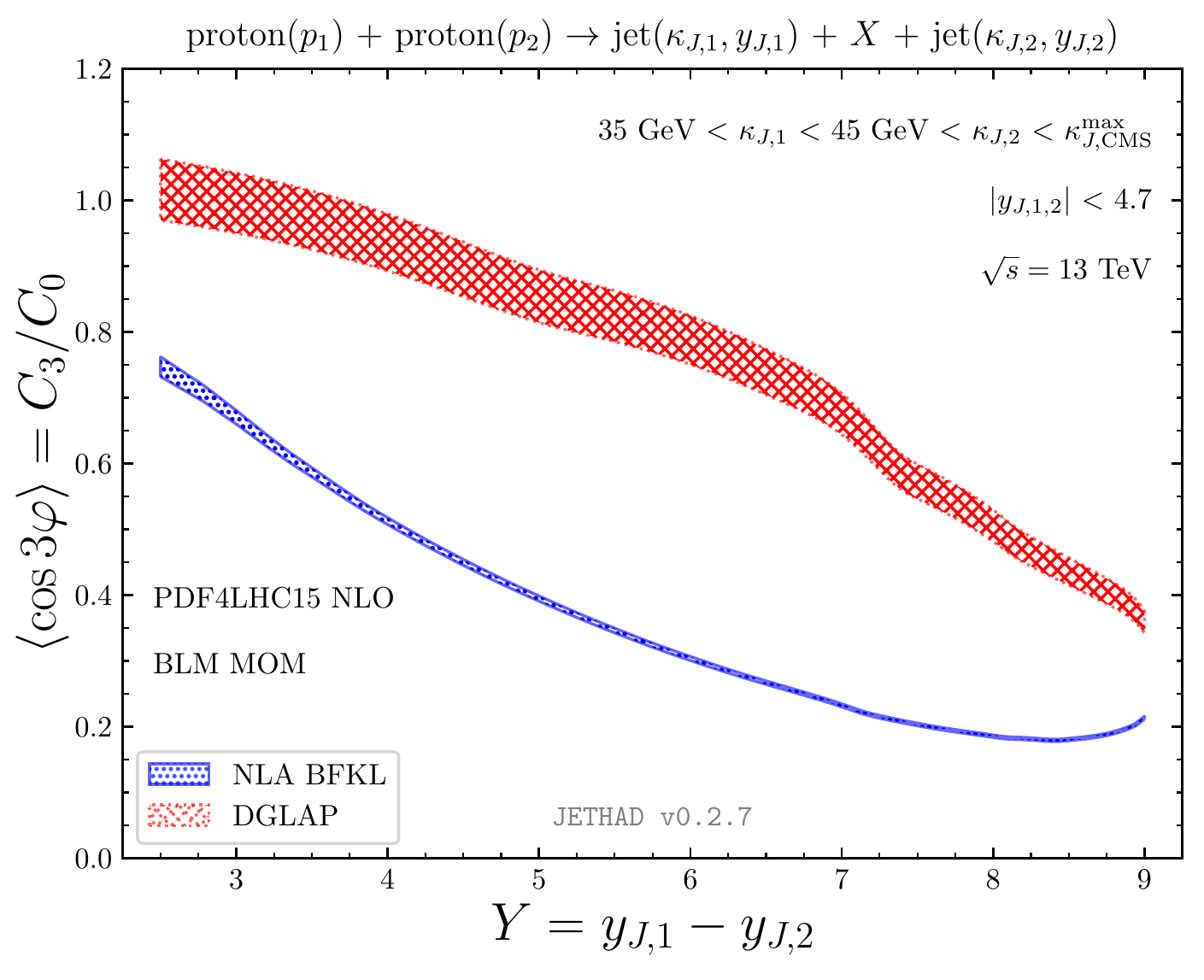}
   \hspace{0.25cm}
   \includegraphics[scale=0.56,clip]{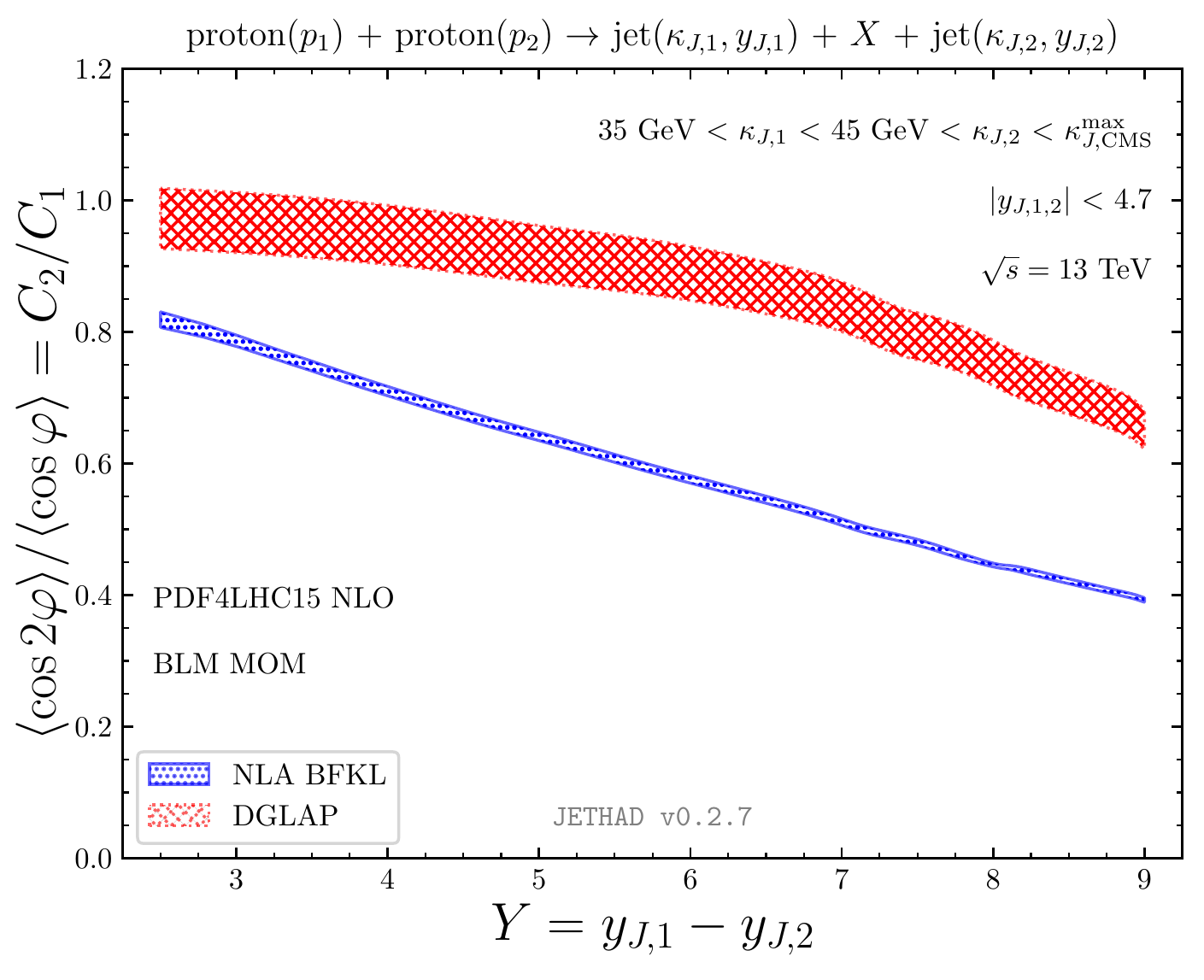}

   \includegraphics[scale=0.56,clip]{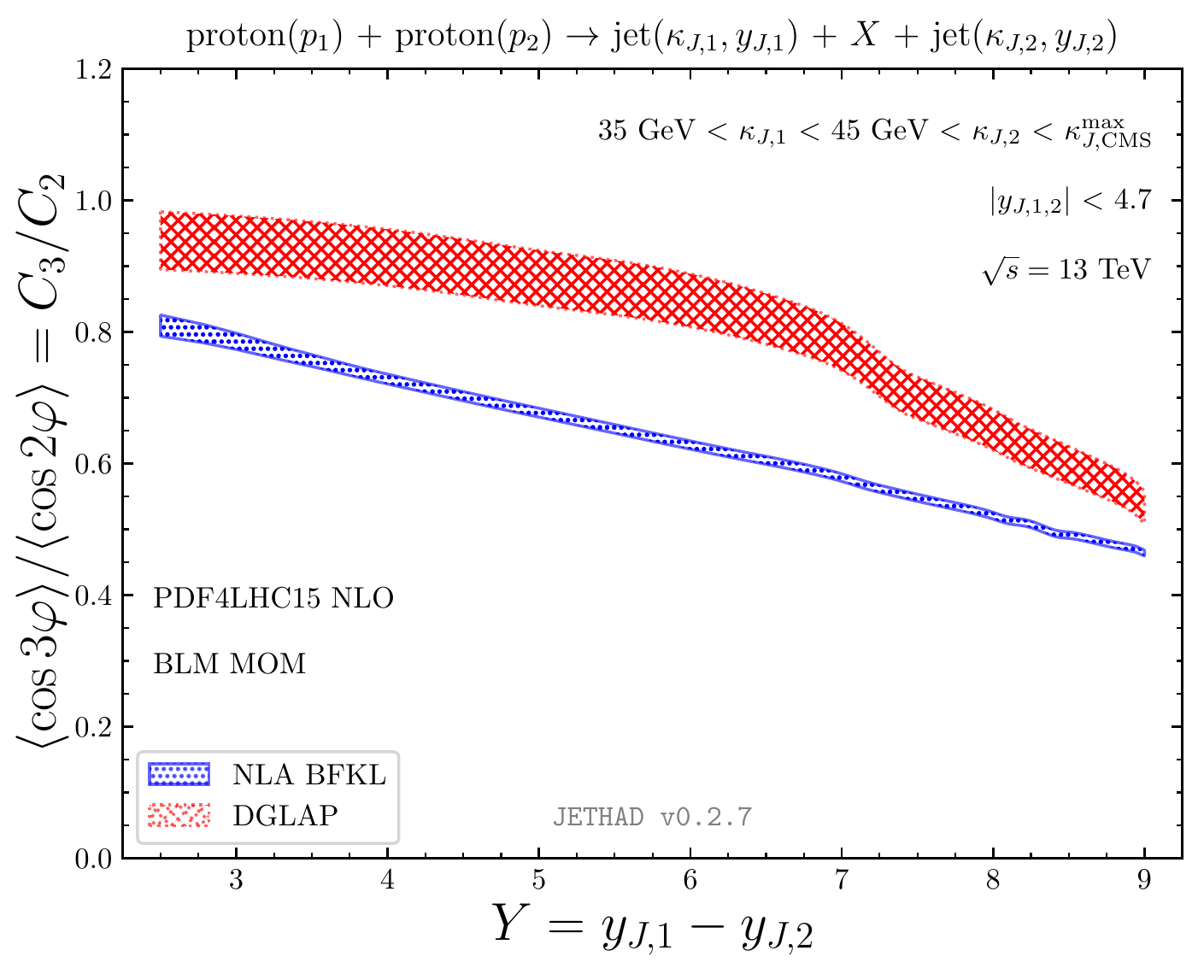}

\caption{$Y$-dependence of several azimuthal correlations, $R_{nm} \equiv C_n/C_m$, of the Mueller--Navelet jet production (left panel of Fig.~\ref{fig:processes}) for
$\mu_{F1,2} = \mu_R = \mu_R^{\rm BLM}$ and $\sqrt{s} = 13$ TeV (\textit{asymmetric CMS} configuration). Full NLA BFKL predictions are compared with the respective ones in the high-energy DGLAP limit.}
\label{fig:MN-BvD-CMS}
\end{figure}

\begin{figure}[H]
\centering

   \includegraphics[scale=0.56,clip]{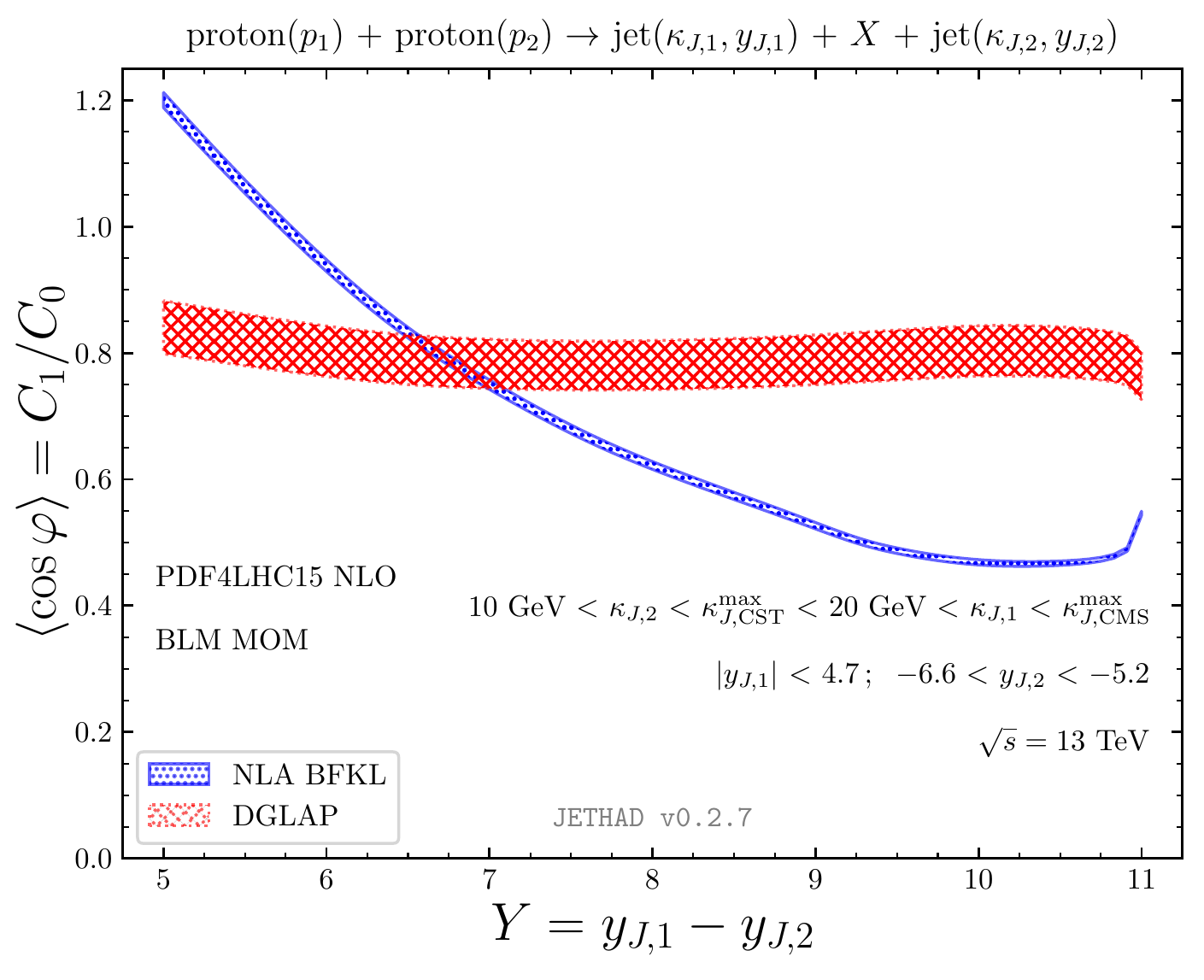}
   \hspace{0.25cm}
   \includegraphics[scale=0.56,clip]{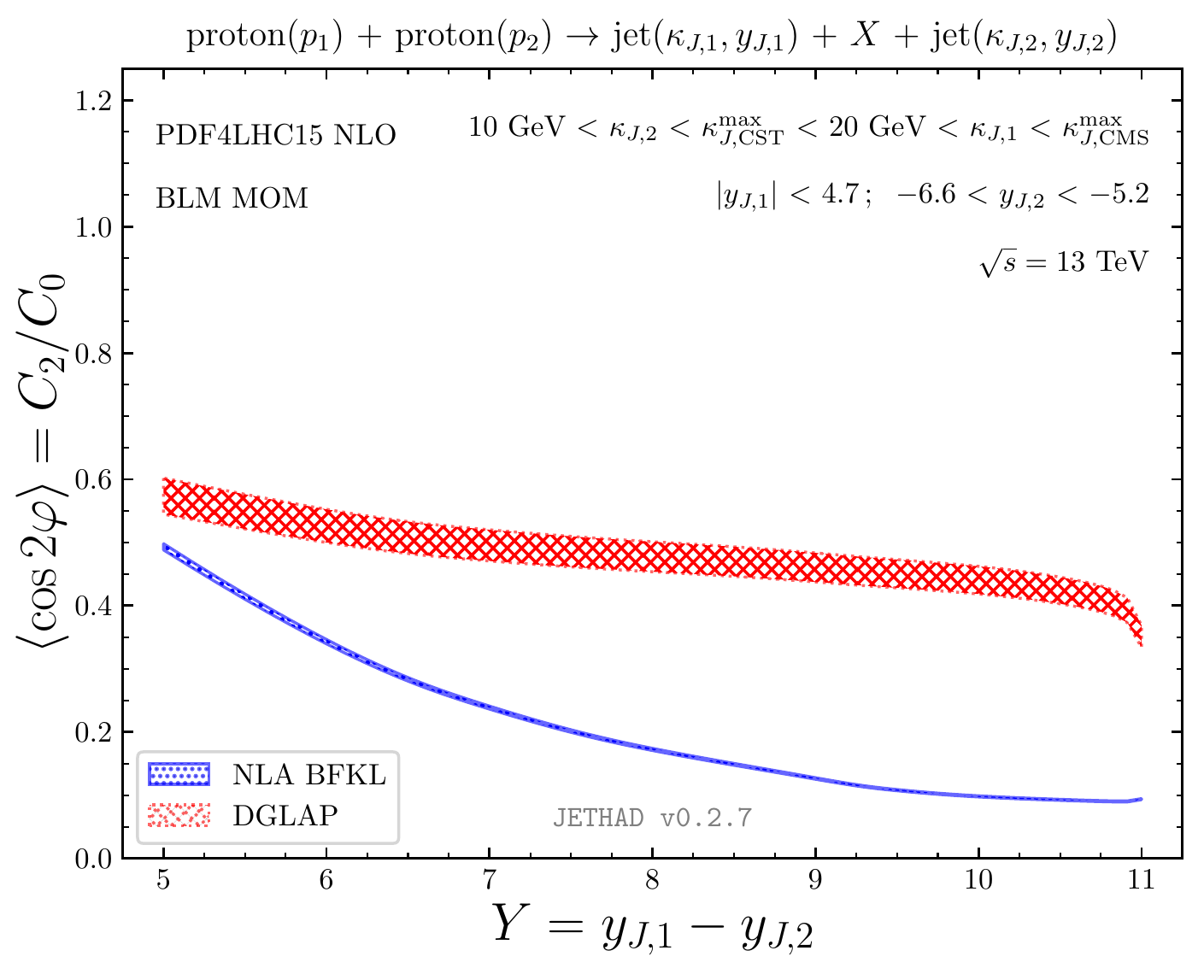}

   \includegraphics[scale=0.56,clip]{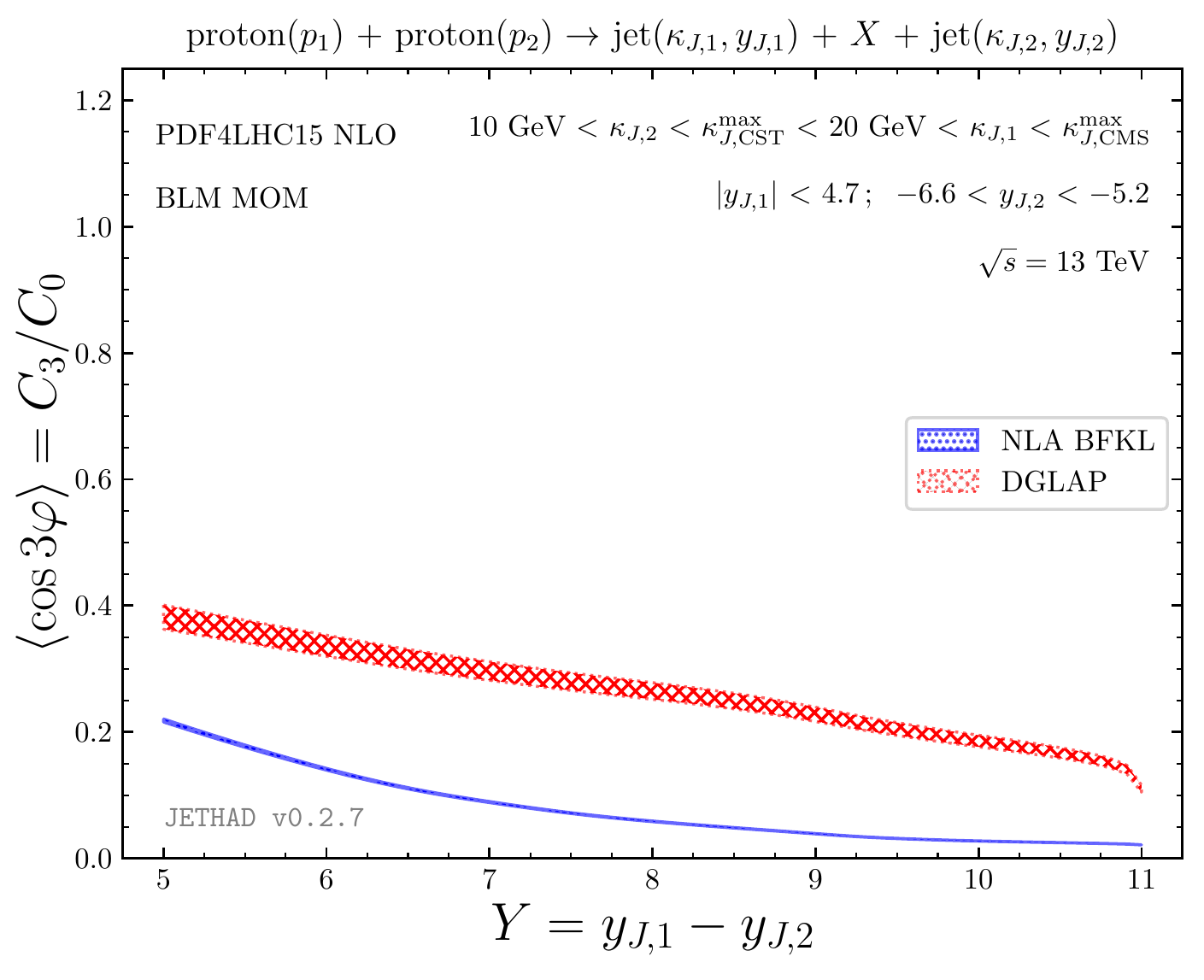}
   \hspace{0.25cm}
   \includegraphics[scale=0.56,clip]{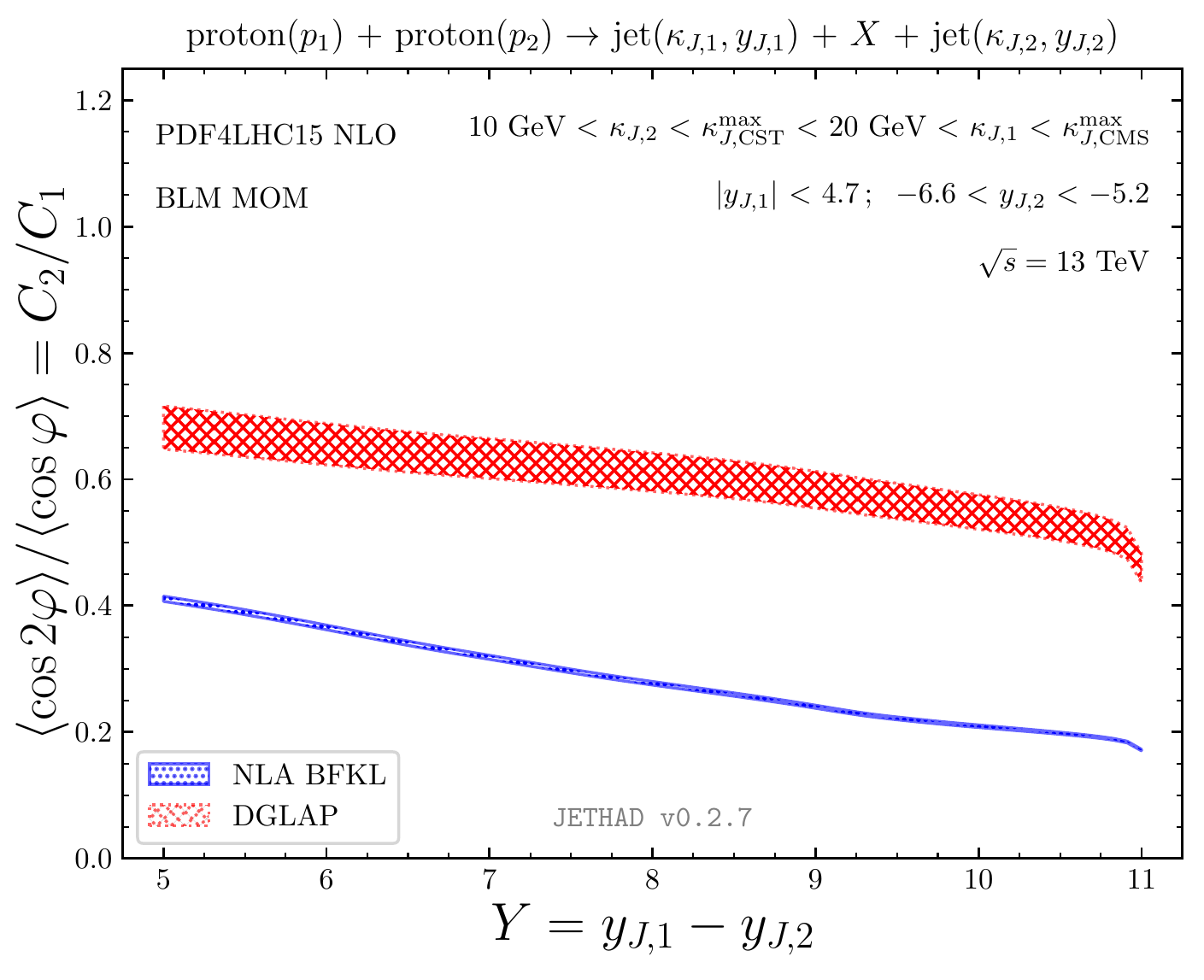}

   \includegraphics[scale=0.56,clip]{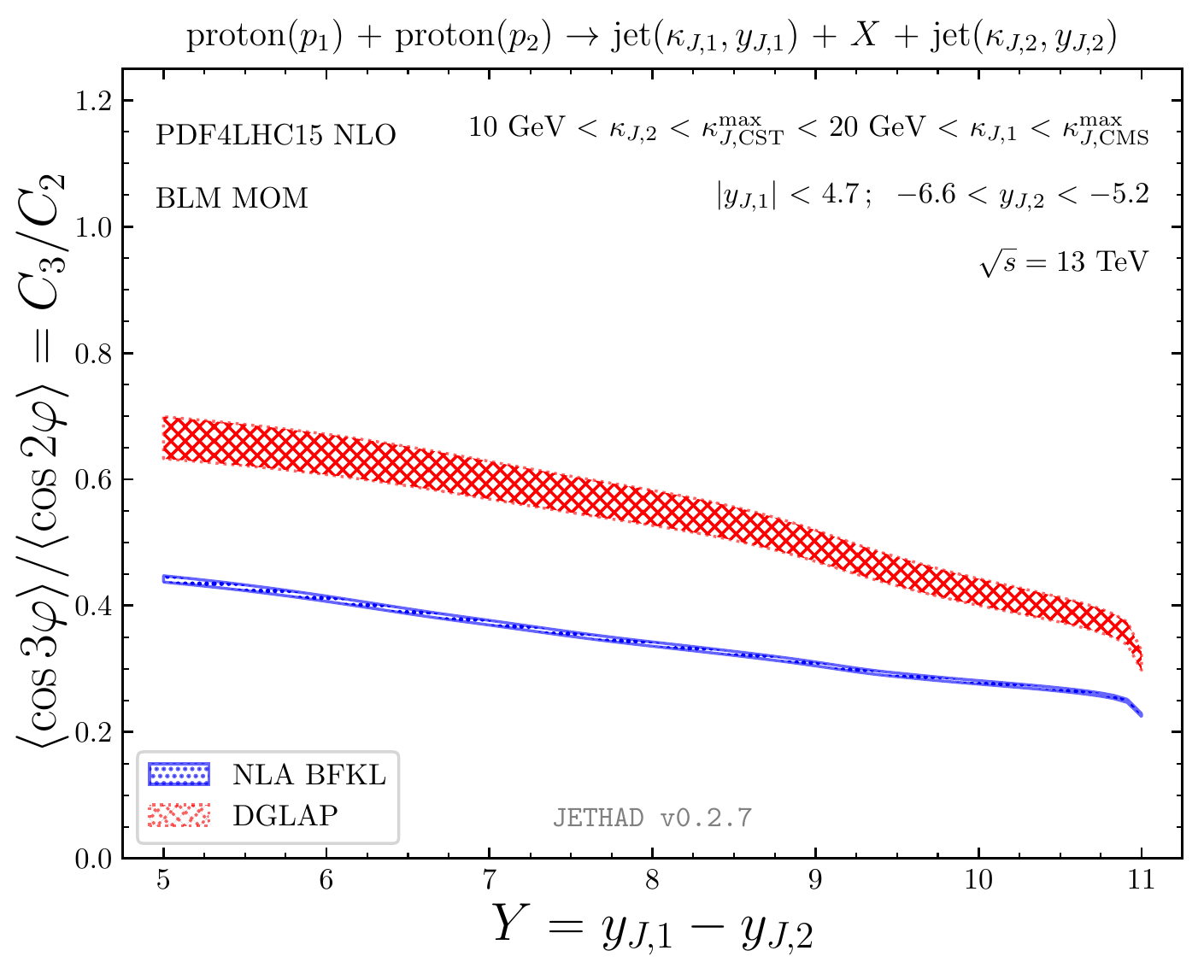}

\caption{$Y$-dependence of several azimuthal correlations, $R_{nm} \equiv C_n/C_m$, of the Mueller--Navelet jet production (left panel of Fig.~\ref{fig:processes}) for
$\mu_{F1,2} = \mu_R = \mu_R^{\rm BLM}$ and $\sqrt{s} = 13$ TeV (\textit{CASTOR-jet} configuration). Full NLA BFKL predictions are compared with the respective ones in the high-energy DGLAP limit.}
\label{fig:MN-BvD-CST}
\end{figure}

NLA BFKL predictions for the azimuthal ratios, $R_{nm} \equiv C_n/C_m$, in the Mueller--Navelet channel (left panel of Fig.~\ref{fig:processes}) are compared in Fig.~\ref{fig:MN-BvD-CMS} with the corresponding high-energy DGLAP ones in the (\textit{asymmetric CMS} configuration) for $\sqrt{s} = 13$ TeV. This extends and completes the study conducted in Refs.~\cite{Celiberto:2015yba,Celiberto:2015mpa}, where a sharp distance between BFKL-resummed and fixed-order calculations came out in the inclusive di-jet hadroproduction with \emph{partially} asymmetric configurations in the $\kappa$-plane, namely 35~GeV~$<$~$\kappa_{J,1}$~$<$~60~GeV and~$45, 50$~GeV~$<$~$\kappa_{J,2}$~$<$~60~GeV, and for $|y_{J,1,2}| < 4.7$, at $\sqrt{s} = 7$ TeV. Here, besides the use of \emph{fully} asymmetric transverse-momentum cuts (disjoint $\kappa$\textit{-windows}), the main improvement (panels of Fig.~\ref{fig:MN-BvD-CMS} correspond to the respective ones in Figs.~2~and~3 of Ref.~\cite{Celiberto:2015yba}) stands in the adoption of the ``exact" BLM scale-optimization procedure (see Section~\ref{blm}) instead of approximated, semi-analytic BLM choices, and in a fair reduction of the numeric uncertainty. Analogous results for the di-jet correlations in the more exclusive, \textit{CASTOR-jet} range, are displayed in Fig.~\ref{fig:MN-BvD-CST}. In both the cases, the pattern of the NLA BFKL $R_{n0}$ series presents a \emph{plateau} at large rapidity distance, $Y$, more emphasized in the \textit{CASTOR-jet} configuration, which visibly evolves into a \emph{turn-up} for $R_{10}$ (upper left panel of Fig.~\ref{fig:MN-BvD-CMS} and of Fig.~\ref{fig:MN-BvD-CST}).
The main reason for this behavior is that the increase of $Y$ values moves the parton longitudinal fractions towards the so-called \emph{threshold} region, where the energy of the di-jet system approaches the value of the center-of-mass energy, $\sqrt{s}$. 
Hence, PDFs are sounded in ranges close to the end-points of their definitions, where they exhibit large scaling violations and uncertainties\footnote{When the struck-parton longitudinal-momentum fraction approaches one, the effect of \emph{target-mass} corrections~\cite{Nachtmann:1973mr,Georgi:1976ve,Barbieri:1976bj,Barbieri:1976rd,Ellis:1982cd,Ellis:1982wd,DeRujula:1976baf,DeRujula:1976ih,Schienbein:2007gr,Accardi:2008ne,Accardi:2008pc} becomes relevant and must be necessarily taken into account.}. In these configurations, our formalism misses the sizeable effect of threshold
double logarithms which enter the perturbative series and have to be resummed to all orders. We note that error bands in the DGLAP case are larger with respect to the resummed ones. This originates from the highly-oscillatory behavior of the $\nu$-integrand in Eq.~(\ref{DGLAP_Cn}), not anymore faded by the exponential factor as in the NLA BFKL case, in Eq.~(\ref{BFKL_Cn}) (we refer the interested reader to Section~\ref{uncertainty} for more details on the numerical-uncertainty estimate of all the presented results).

\begin{figure}[H]
\centering

   \includegraphics[scale=0.56,clip]{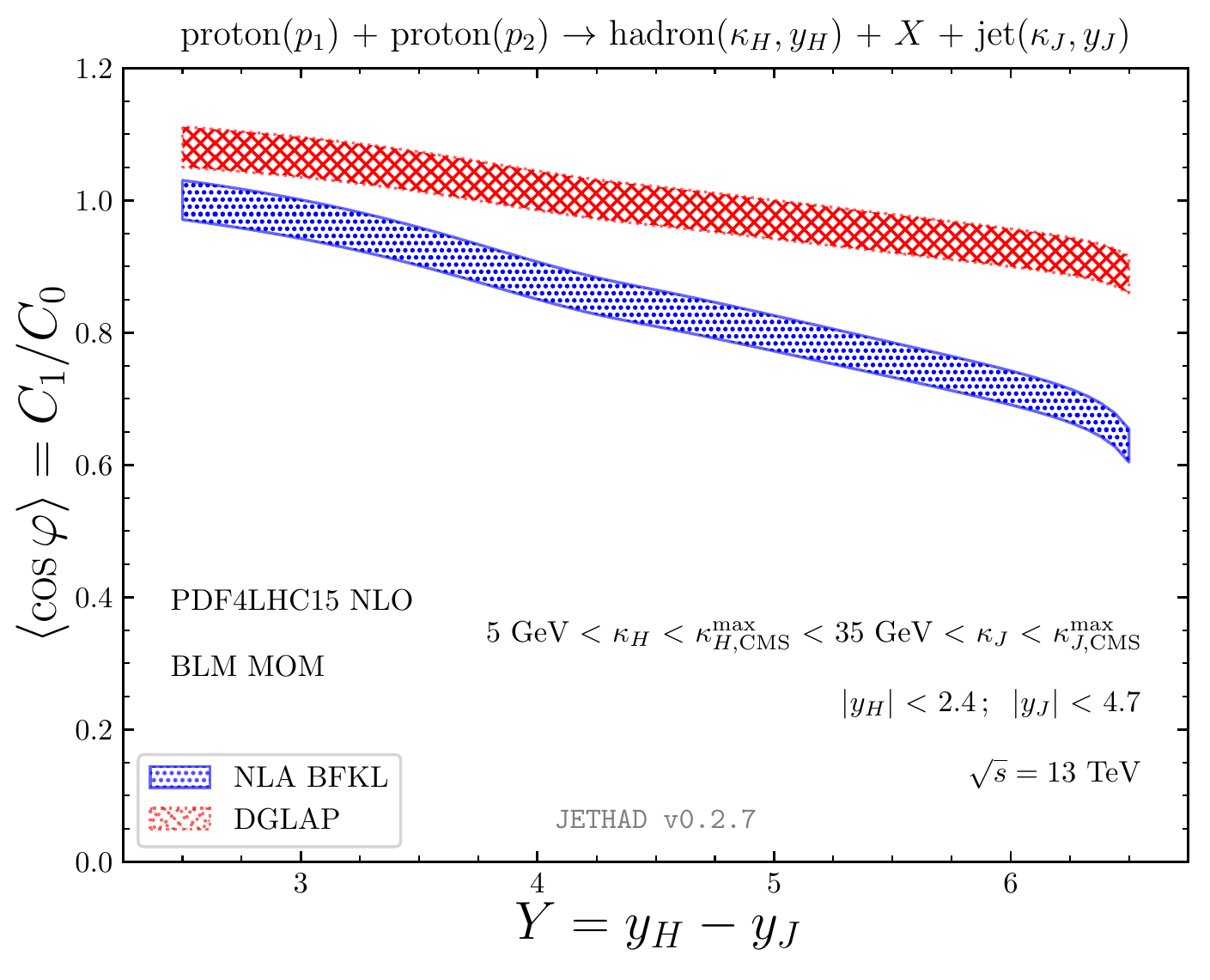}
   \hspace{0.25cm}
   \includegraphics[scale=0.56,clip]{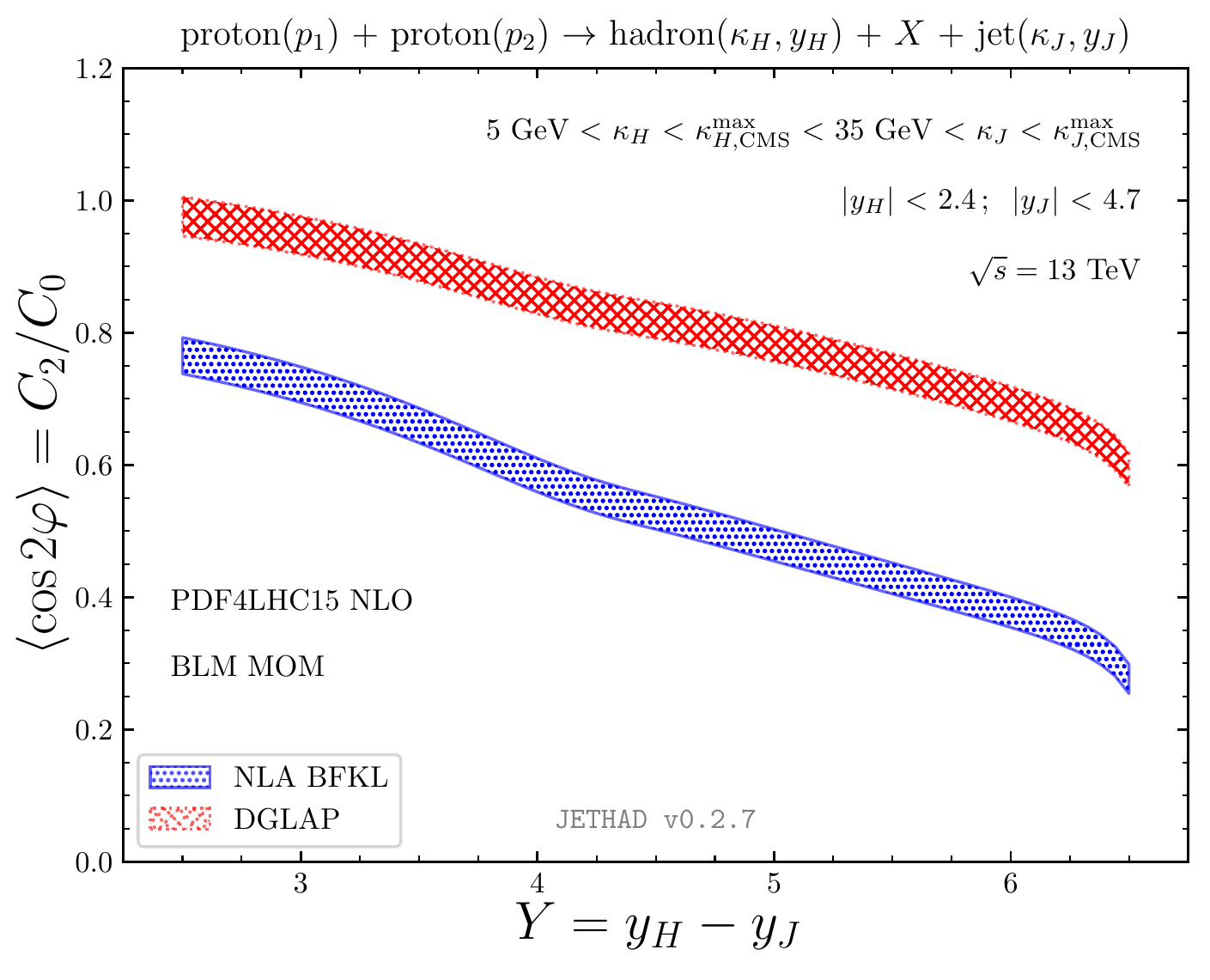}

   \includegraphics[scale=0.56,clip]{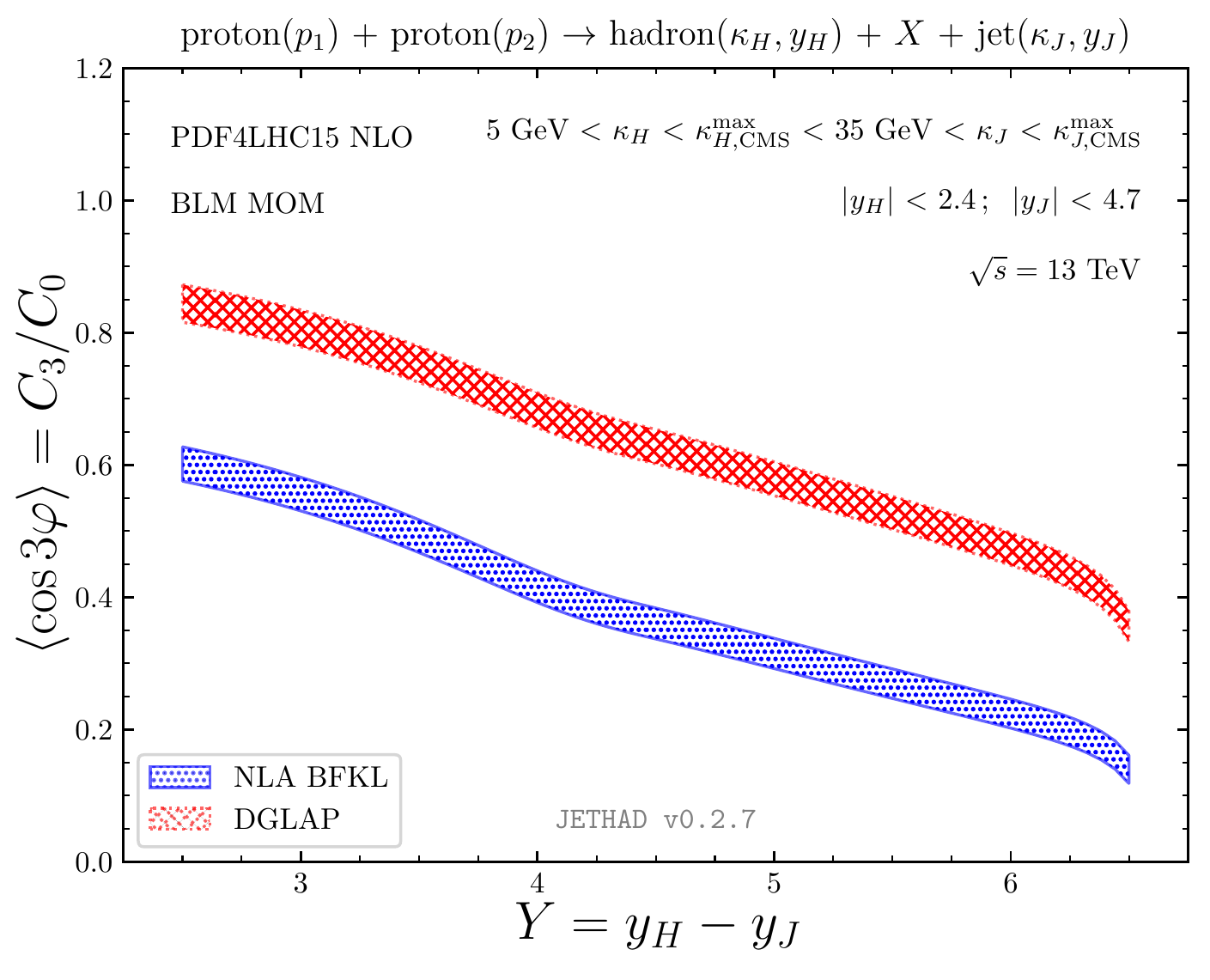}
   \hspace{0.25cm}
   \includegraphics[scale=0.56,clip]{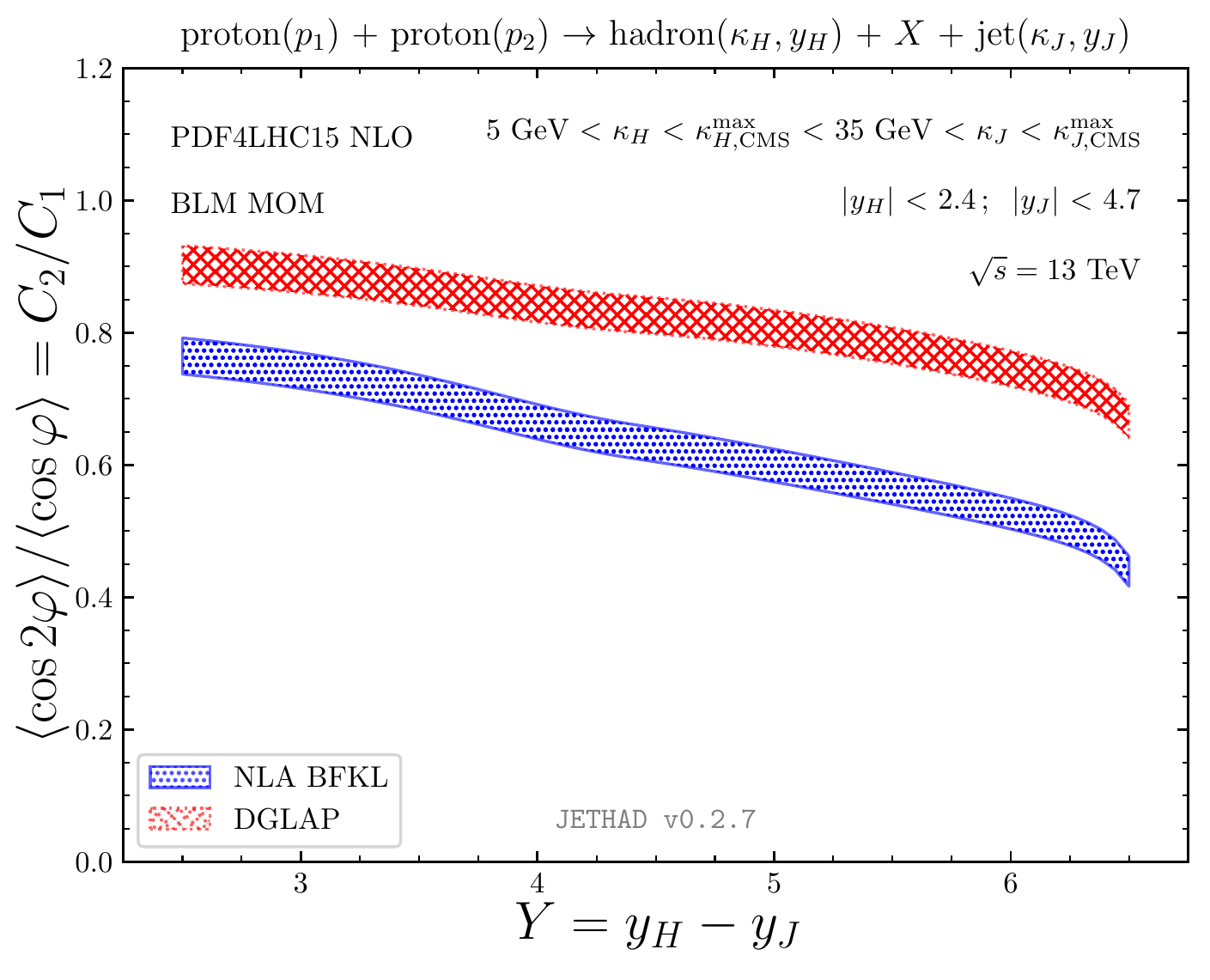}

   \includegraphics[scale=0.56,clip]{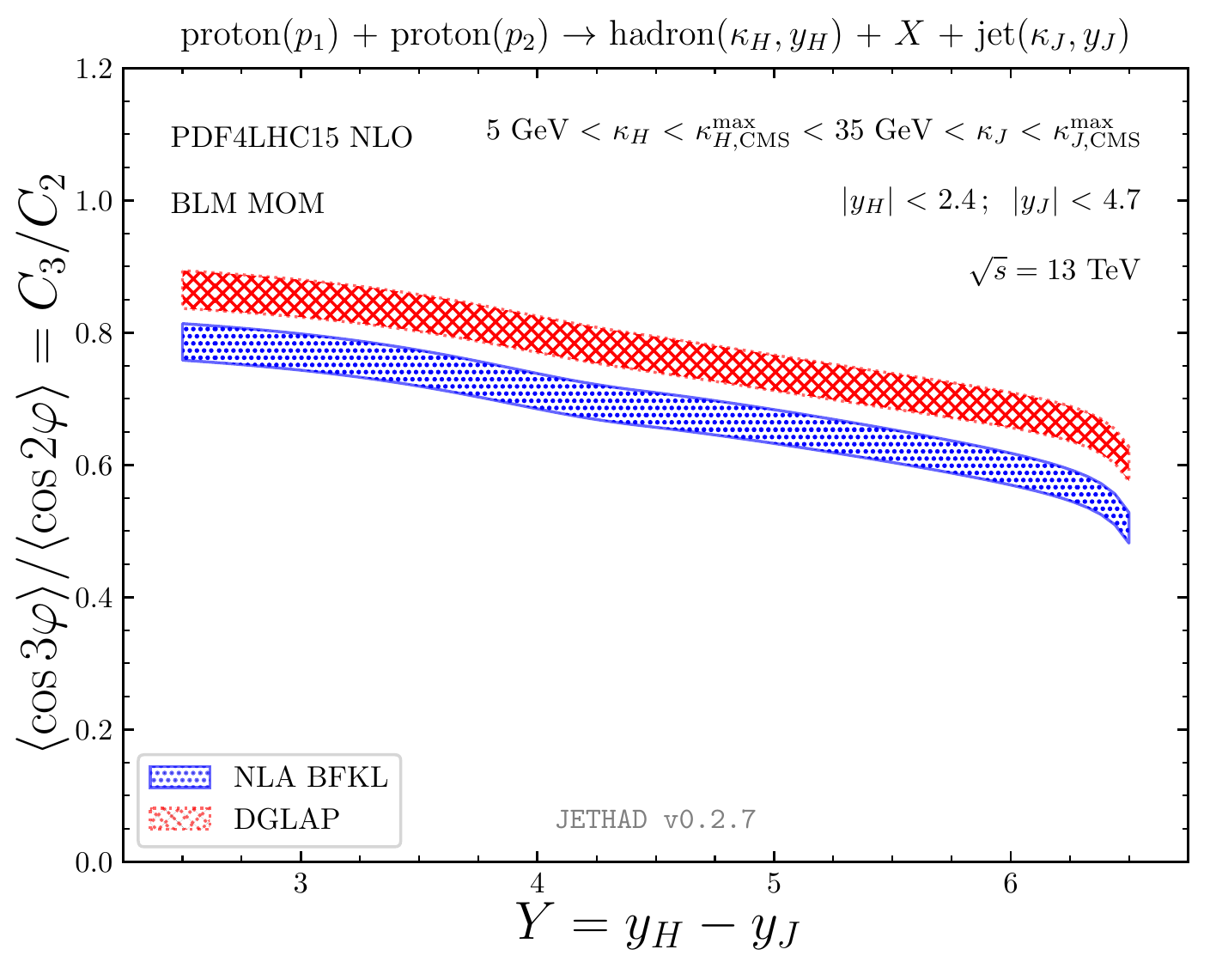}

\caption{$Y$-dependence of several azimuthal correlations, $R_{nm} \equiv C_n/C_m$, of the inclusive hadron-jet production (central panel of Fig.~\ref{fig:processes}) for
$\mu_{F1,2} = \mu_R = \mu_R^{\rm BLM}$ and $\sqrt{s} = 13$ TeV (\textit{asymmetric CMS} configuration). Full NLA BFKL predictions are compared with the respective ones in the high-energy DGLAP limit.}
\label{fig:HJ-BvD-CMS}
\end{figure}
\begin{figure}[H]
\centering

   \includegraphics[scale=0.56,clip]{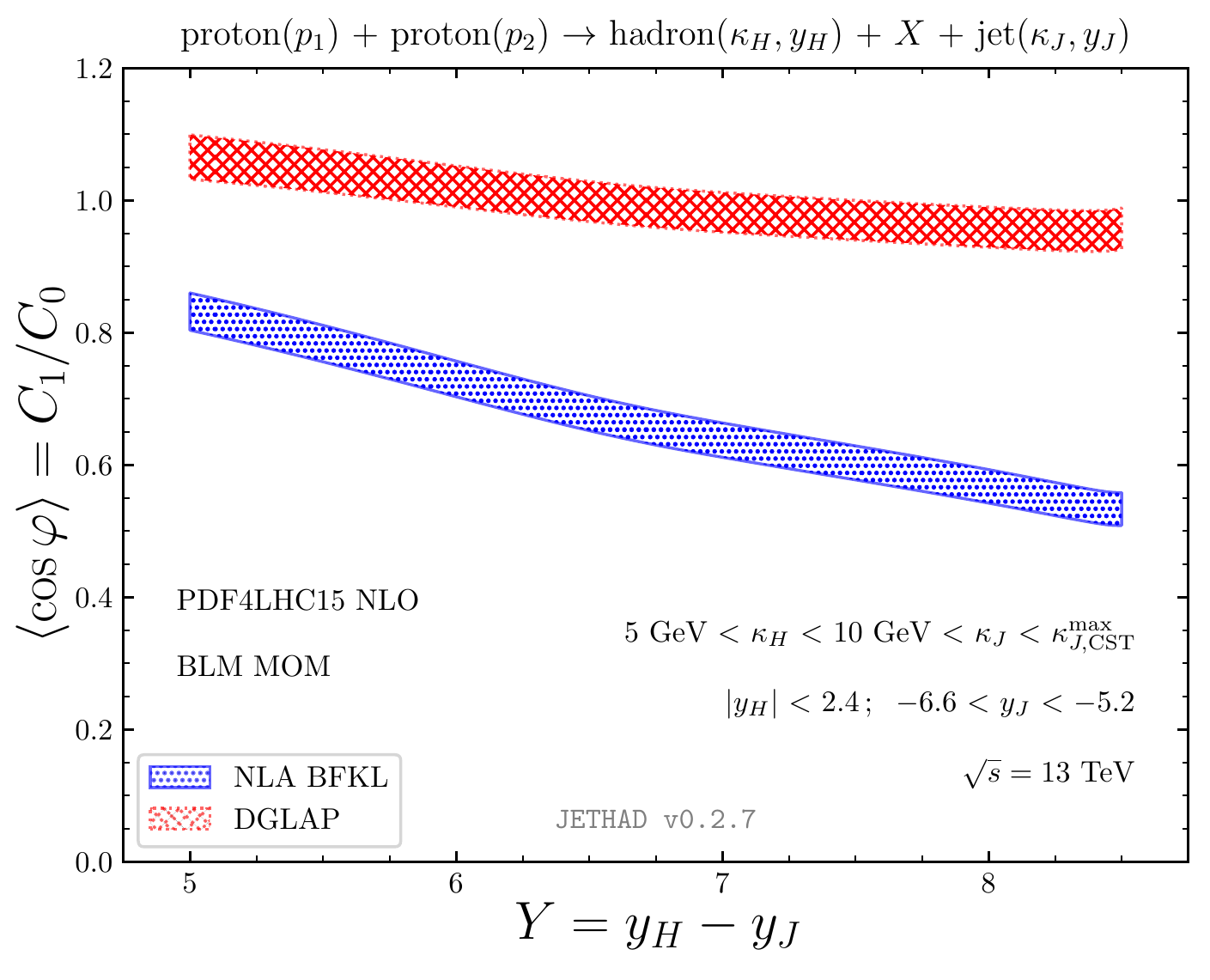}
   \hspace{0.25cm}
   \includegraphics[scale=0.56,clip]{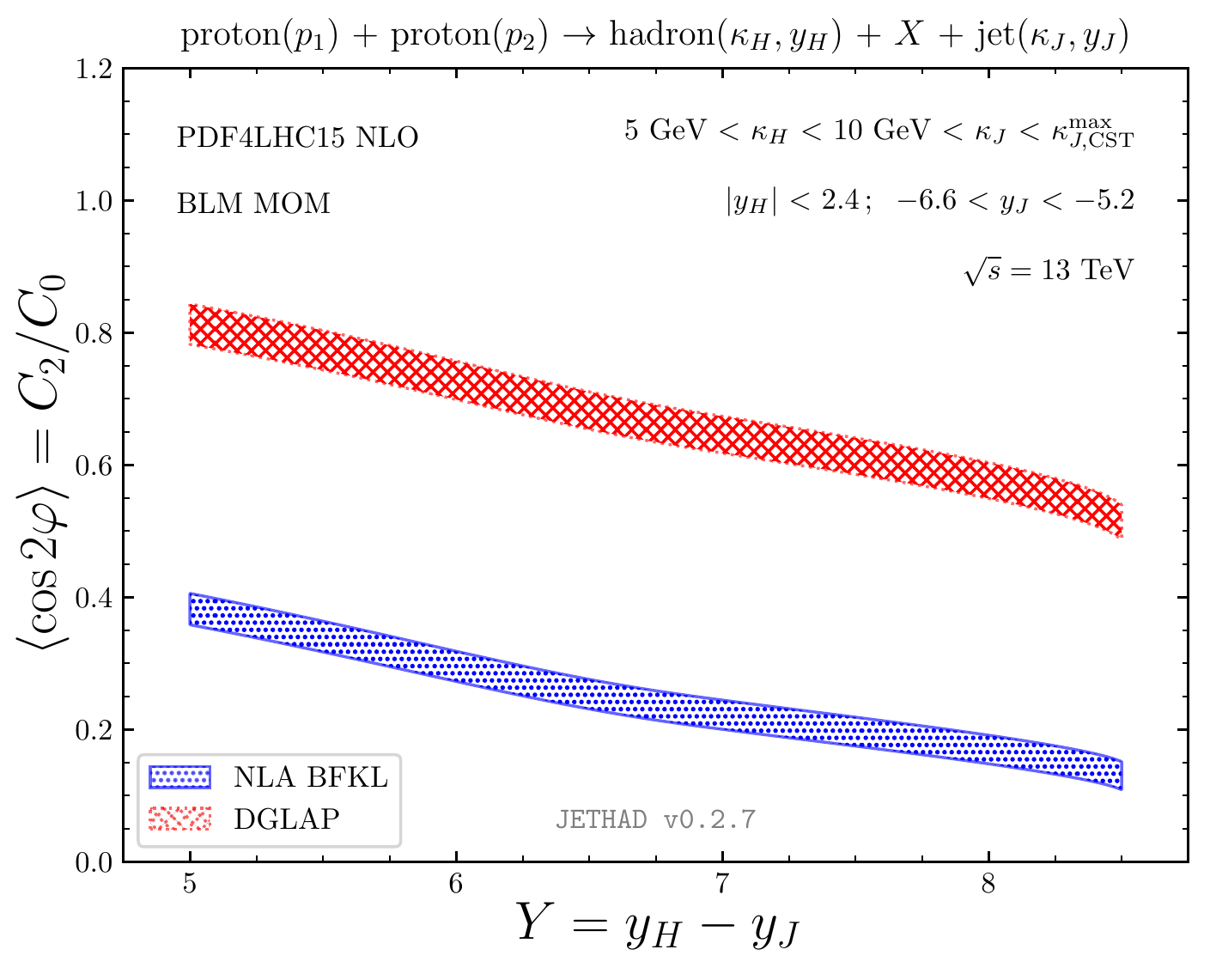}

   \includegraphics[scale=0.56,clip]{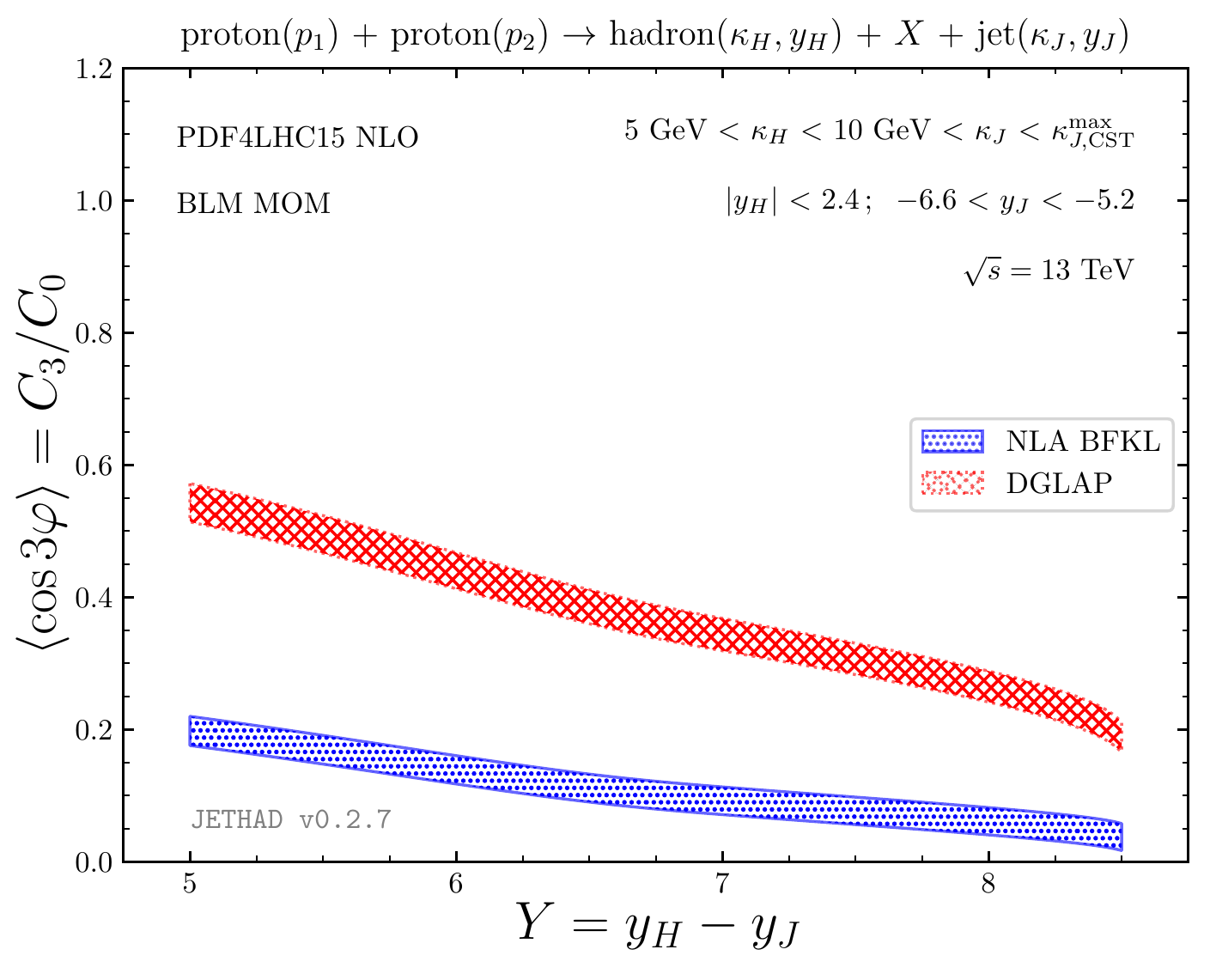}
   \hspace{0.25cm}
   \includegraphics[scale=0.56,clip]{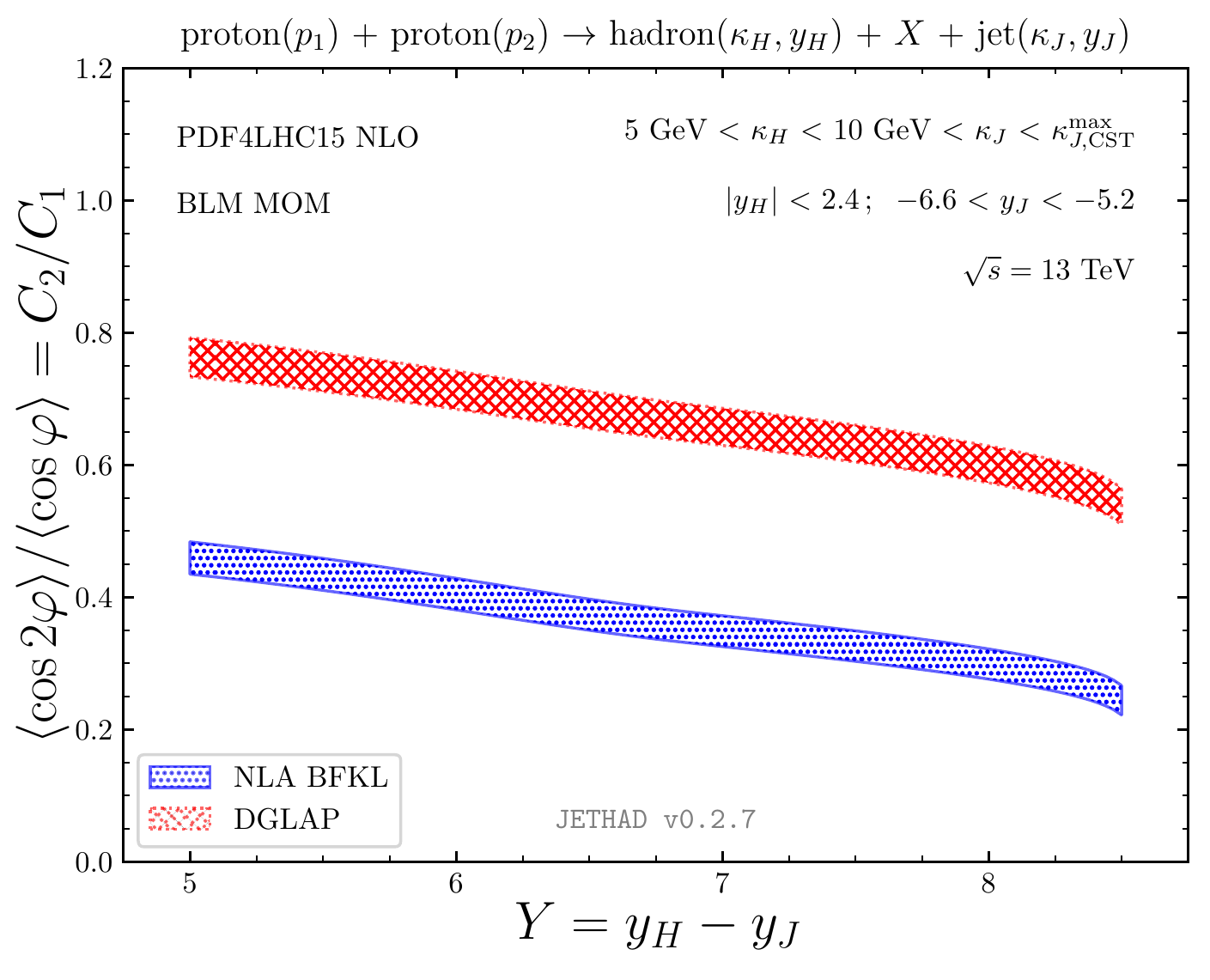}

   \includegraphics[scale=0.56,clip]{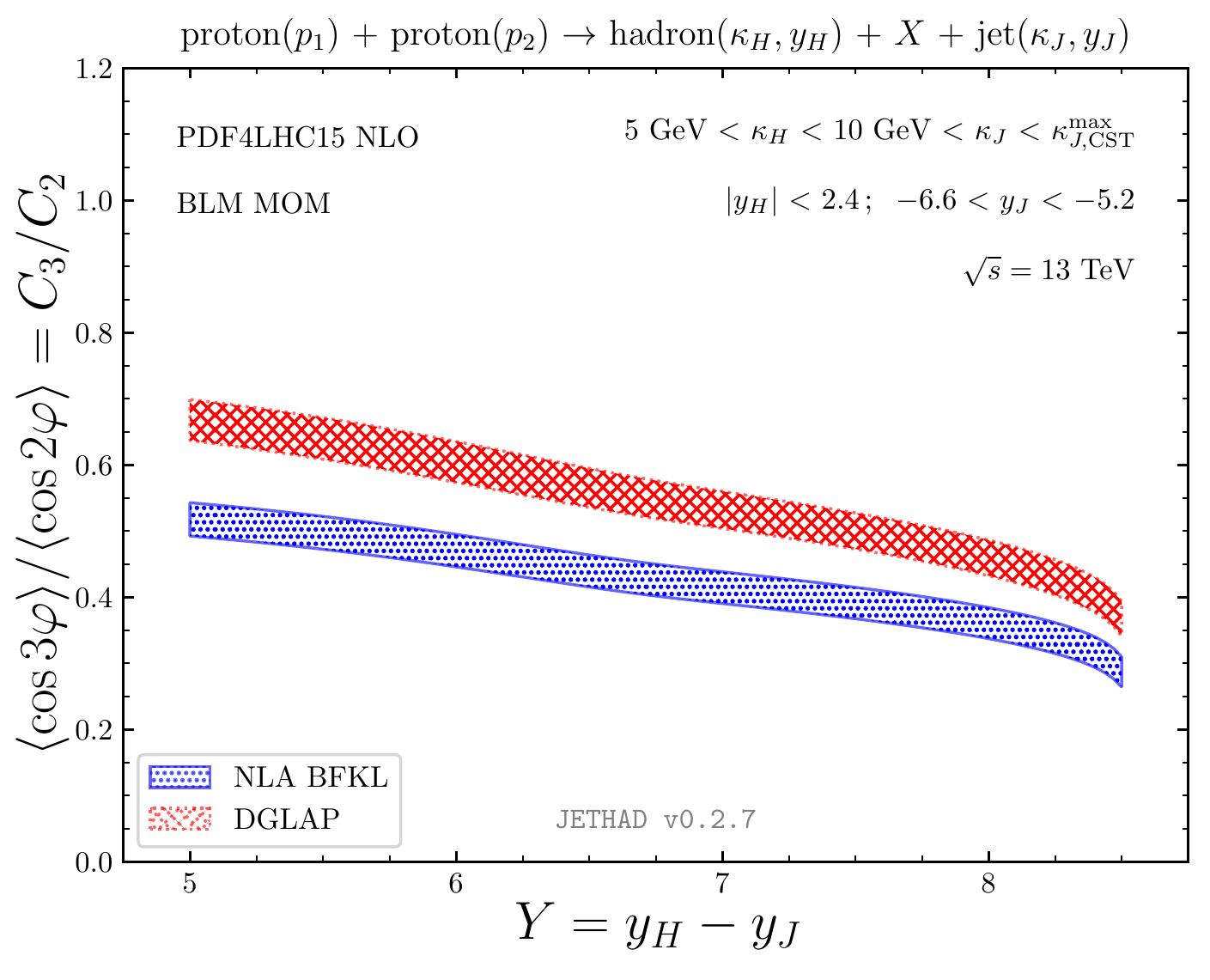}

\caption{$Y$-dependence of several azimuthal correlations, $R_{nm} \equiv C_n/C_m$, of the inclusive hadron-jet production (central panel of Fig.~\ref{fig:processes}) for
$\mu_{F1,2} = \mu_R = \mu_R^{\rm BLM}$ and $\sqrt{s} = 13$ TeV (\textit{CASTOR-jet} configuration). Full NLA BFKL predictions are compared with the respective ones in the high-energy DGLAP limit.}
\label{fig:HJ-BvD-CST}
\end{figure}
\begin{figure}[H]
\centering

   \includegraphics[scale=0.56,clip]{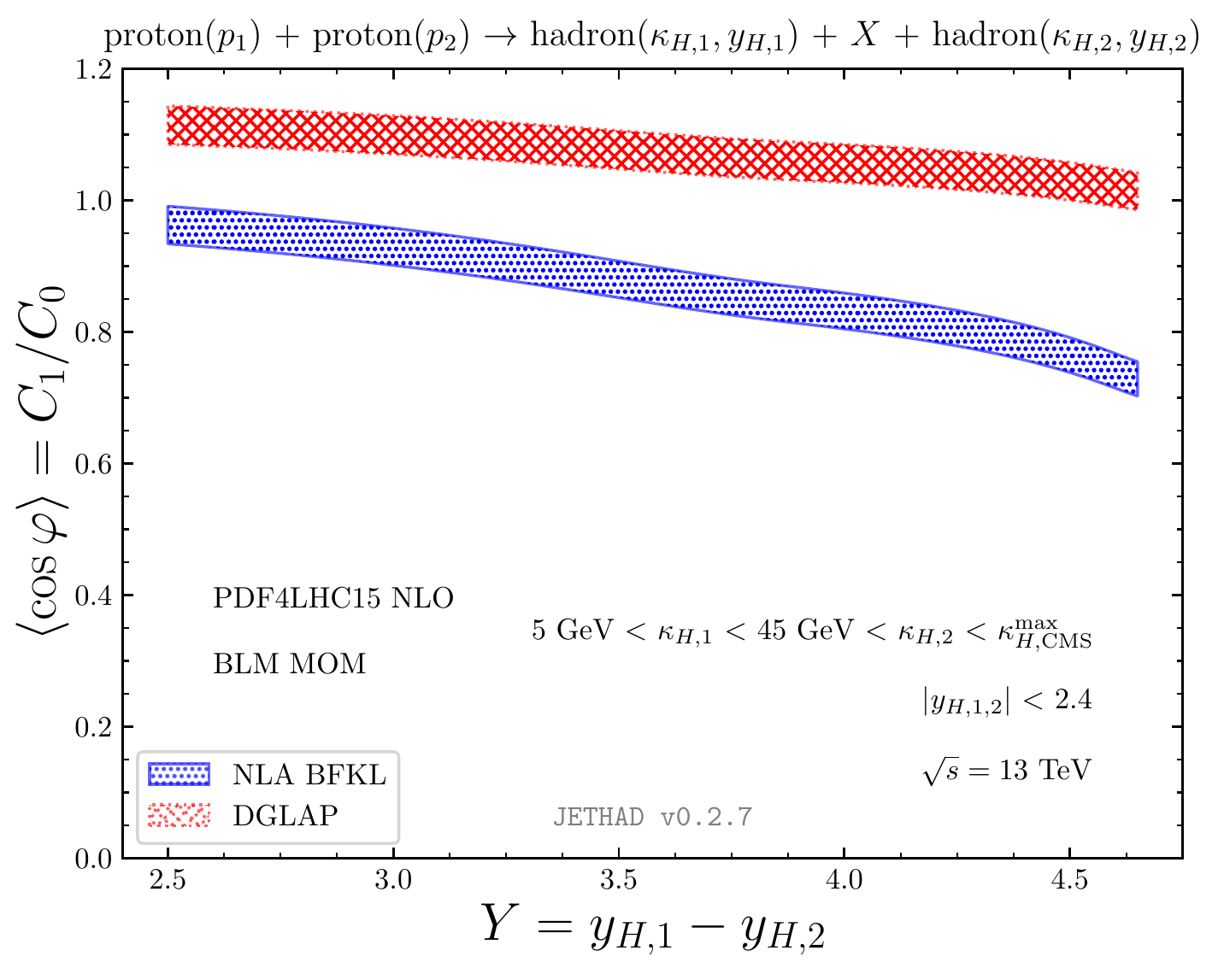}
   \hspace{0.25cm}
   \includegraphics[scale=0.56,clip]{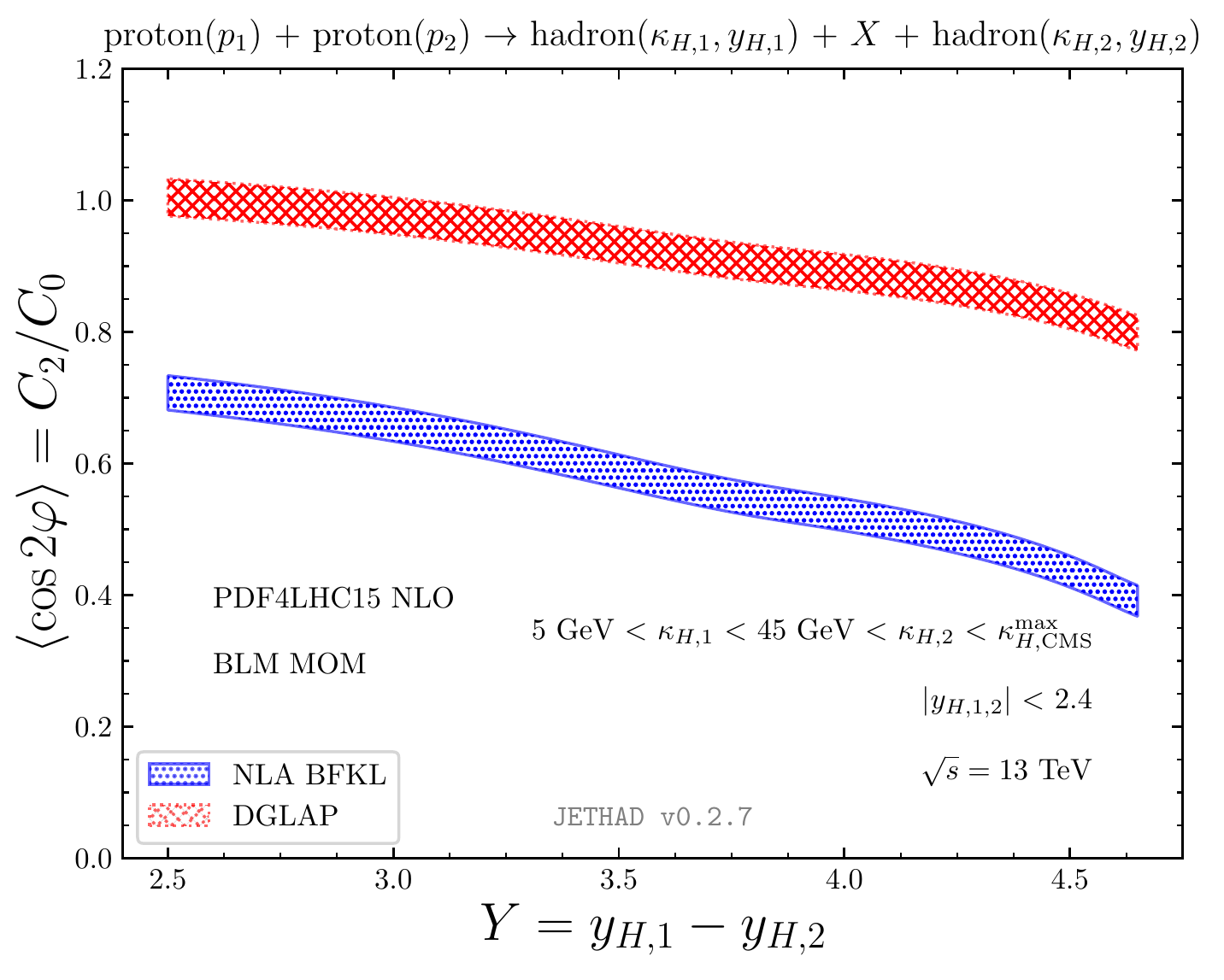}

   \includegraphics[scale=0.56,clip]{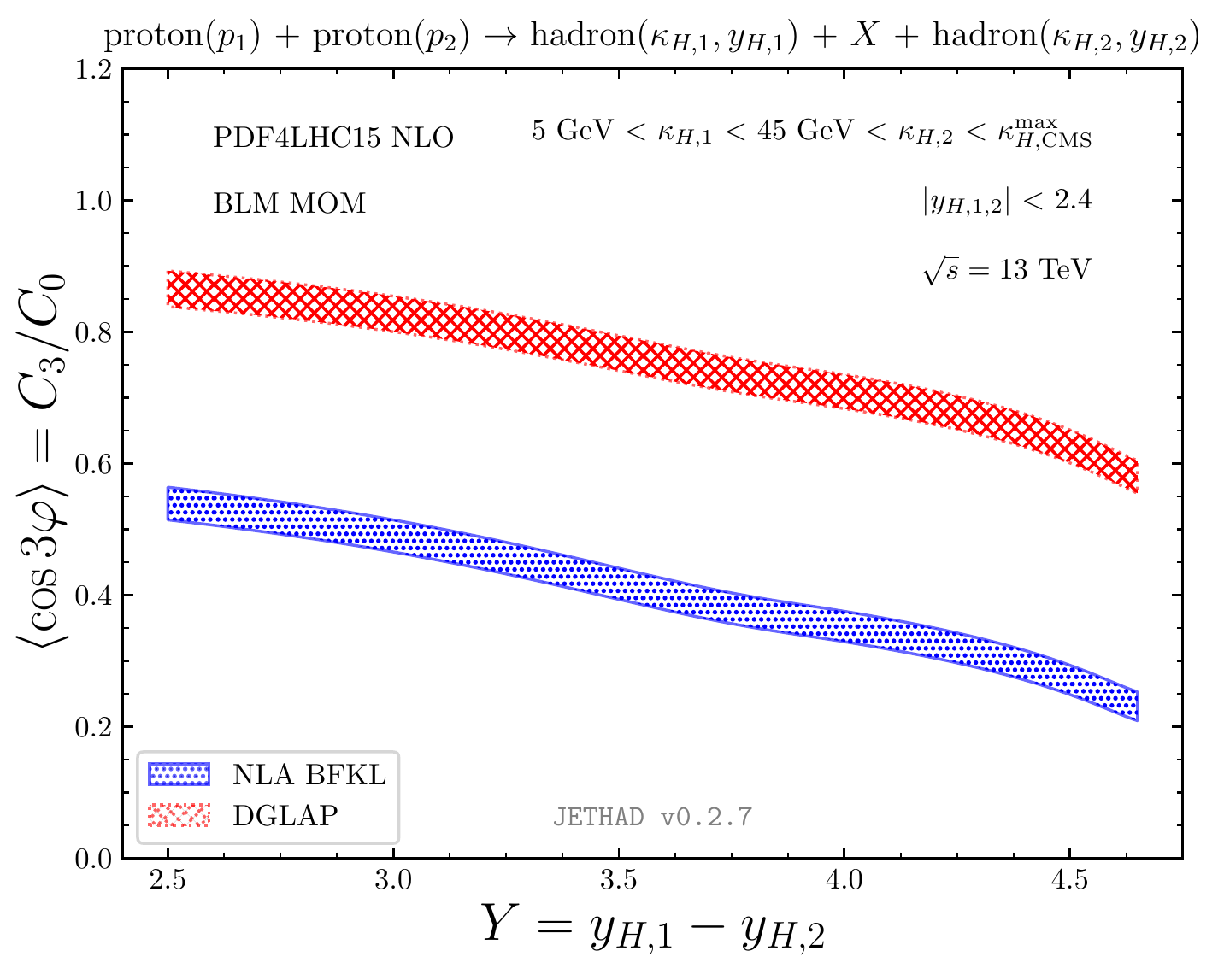}
   \hspace{0.25cm}
   \includegraphics[scale=0.56,clip]{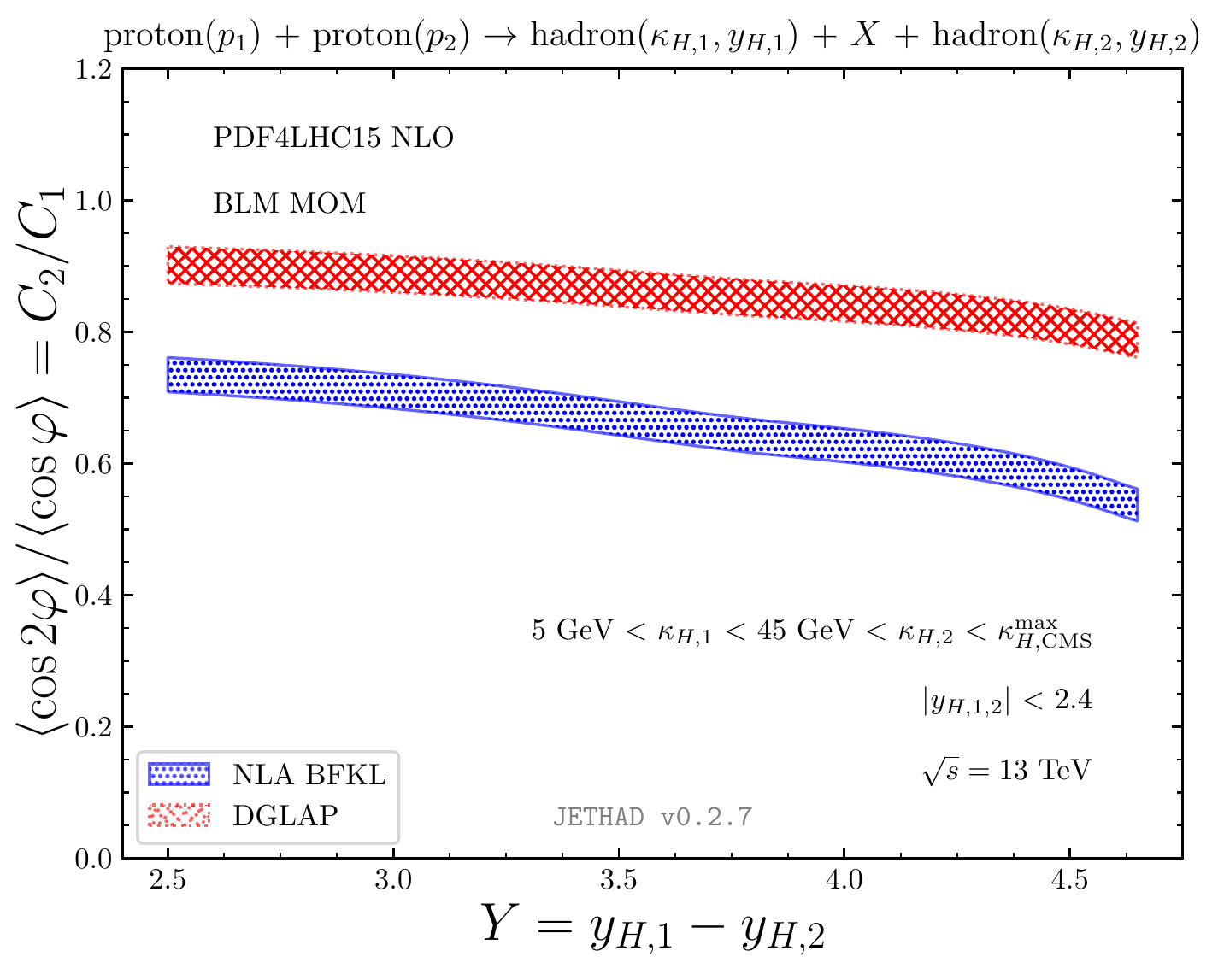}

   \includegraphics[scale=0.56,clip]{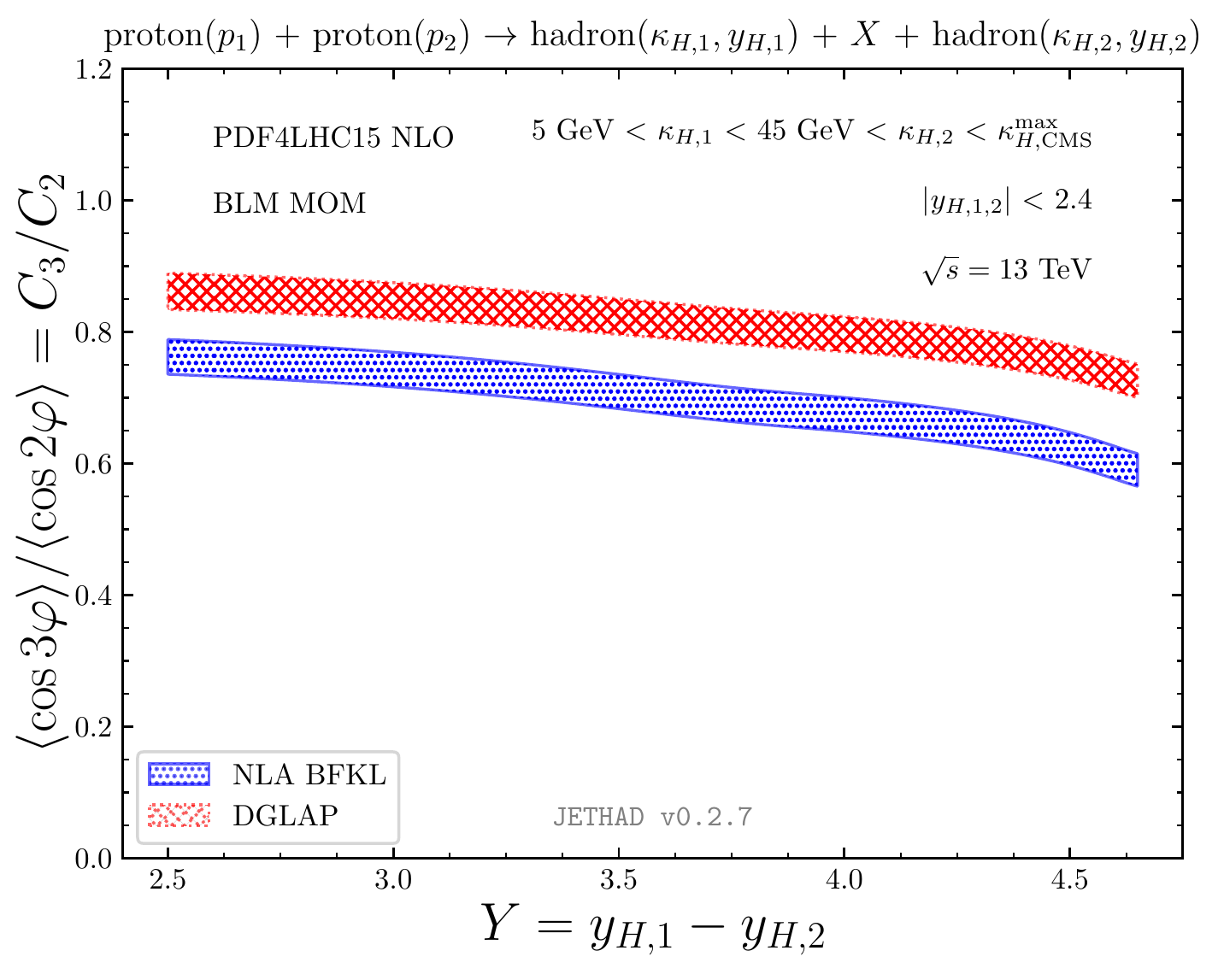}

\caption{$Y$-dependence of several azimuthal correlations, $R_{nm} \equiv C_n/C_m$, of the inclusive di-hadron production (right panel of Fig.~\ref{fig:processes}) for
$\mu_{F1,2} = \mu_R = \mu_R^{\rm BLM}$ and $\sqrt{s} = 13$ TeV (\textit{asymmetric CMS} configuration). Full NLA BFKL predictions are compared with the respective ones in the high-energy DGLAP limit.}
\label{fig:HH-BvD-CMS}
\end{figure}
\begin{figure}[H]
\centering

   \includegraphics[scale=0.56,clip]{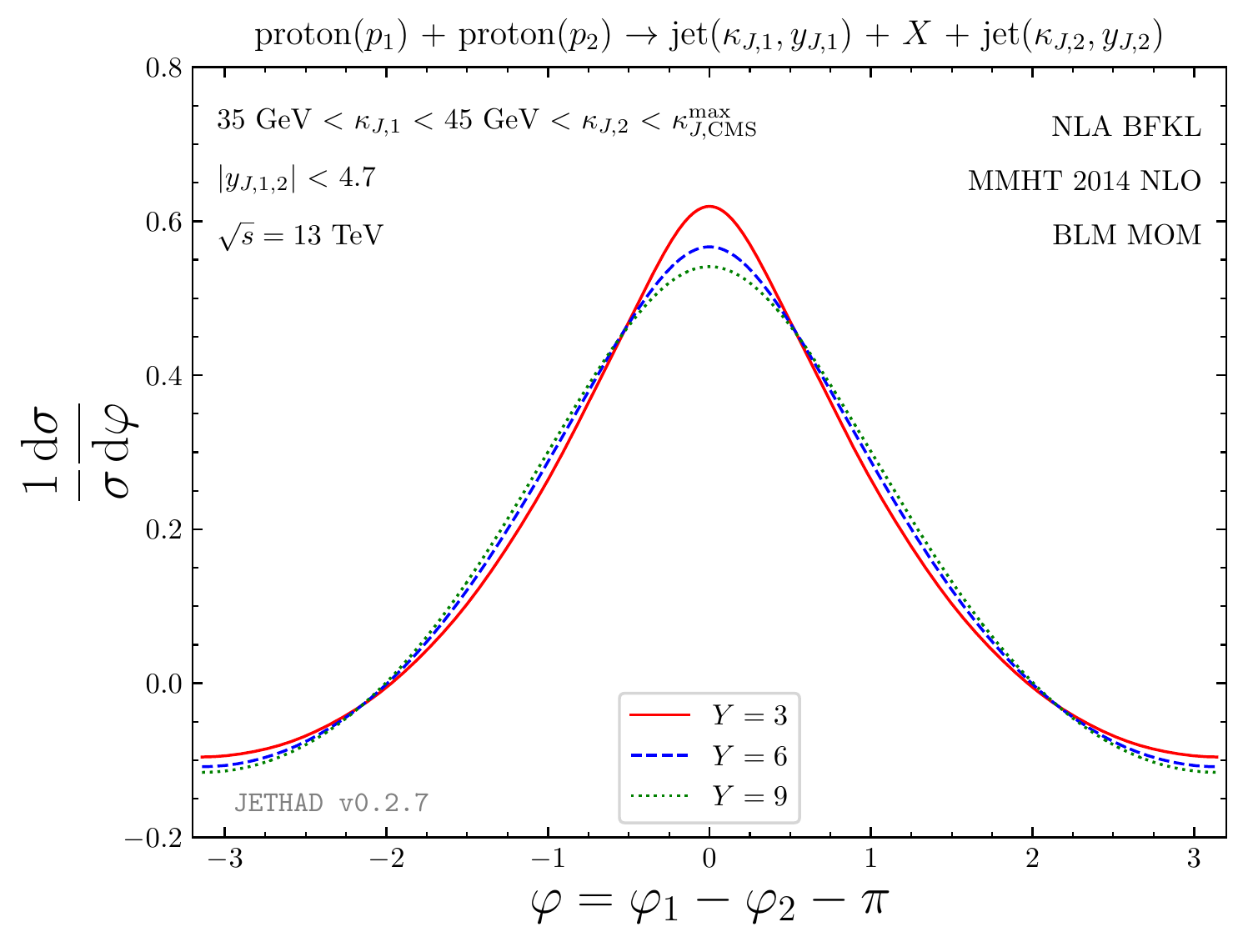}
   \hspace{-0.25cm}
   \includegraphics[scale=0.56,clip]{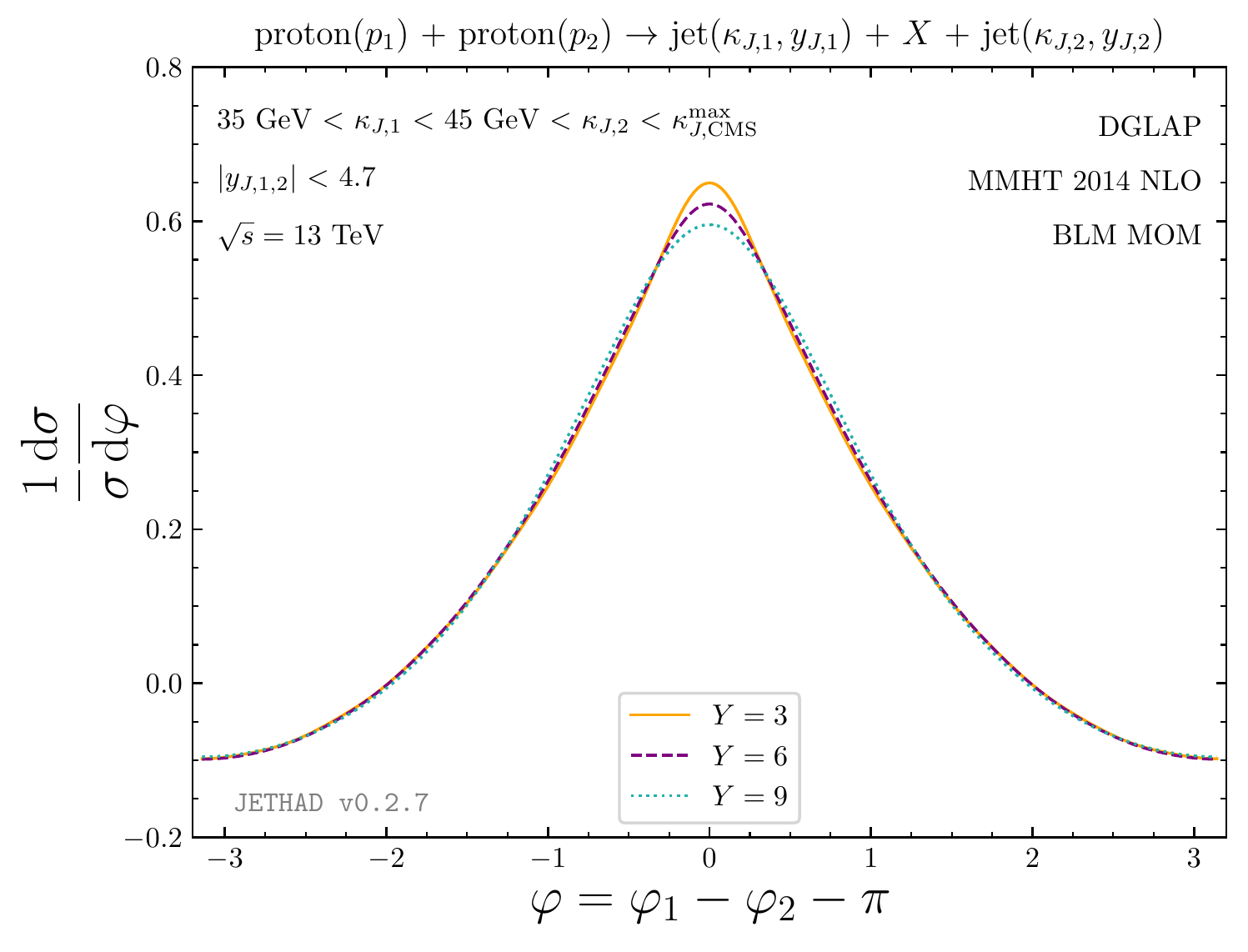}

   \includegraphics[scale=0.56,clip]{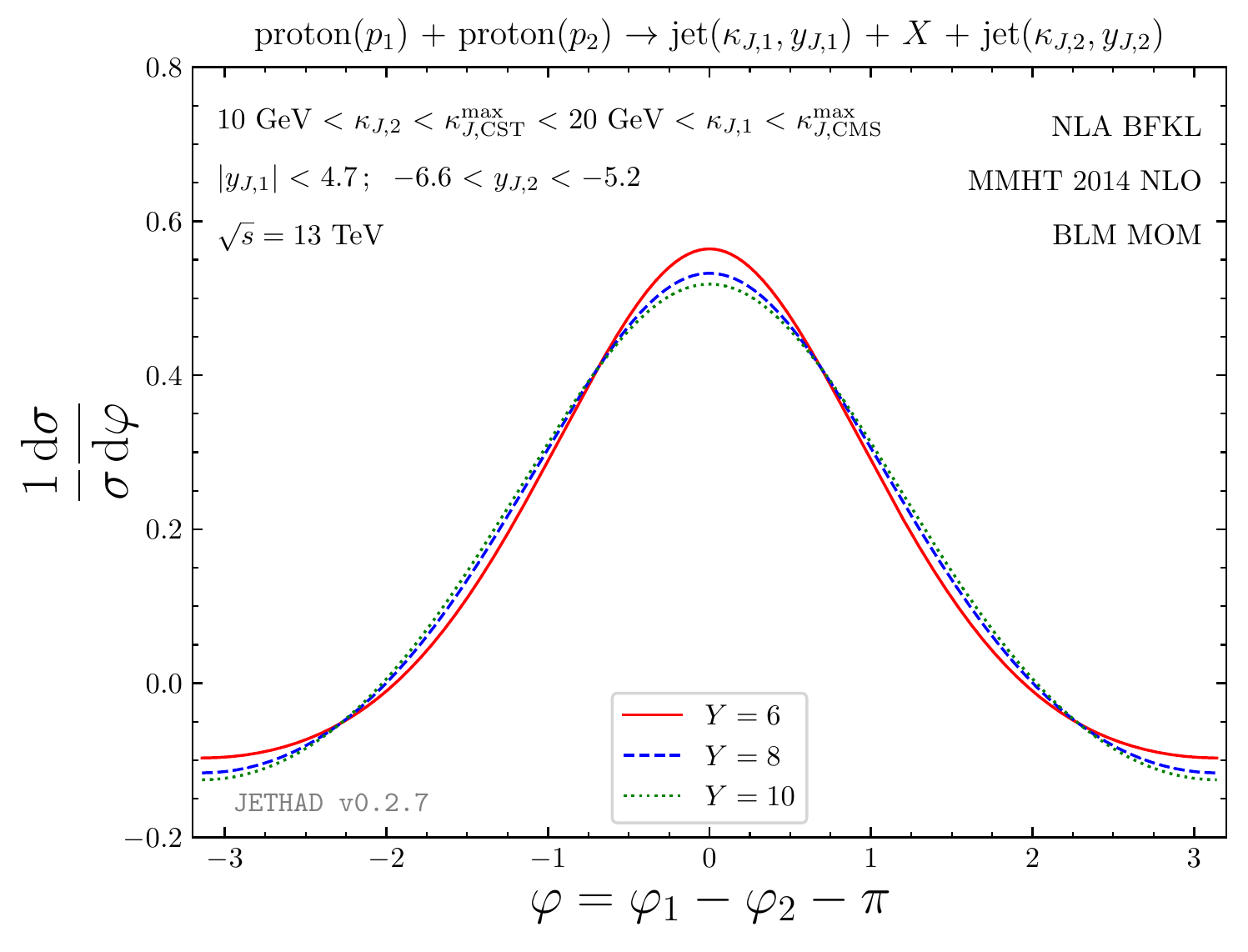}
   \hspace{-0.25cm}
   \includegraphics[scale=0.56,clip]{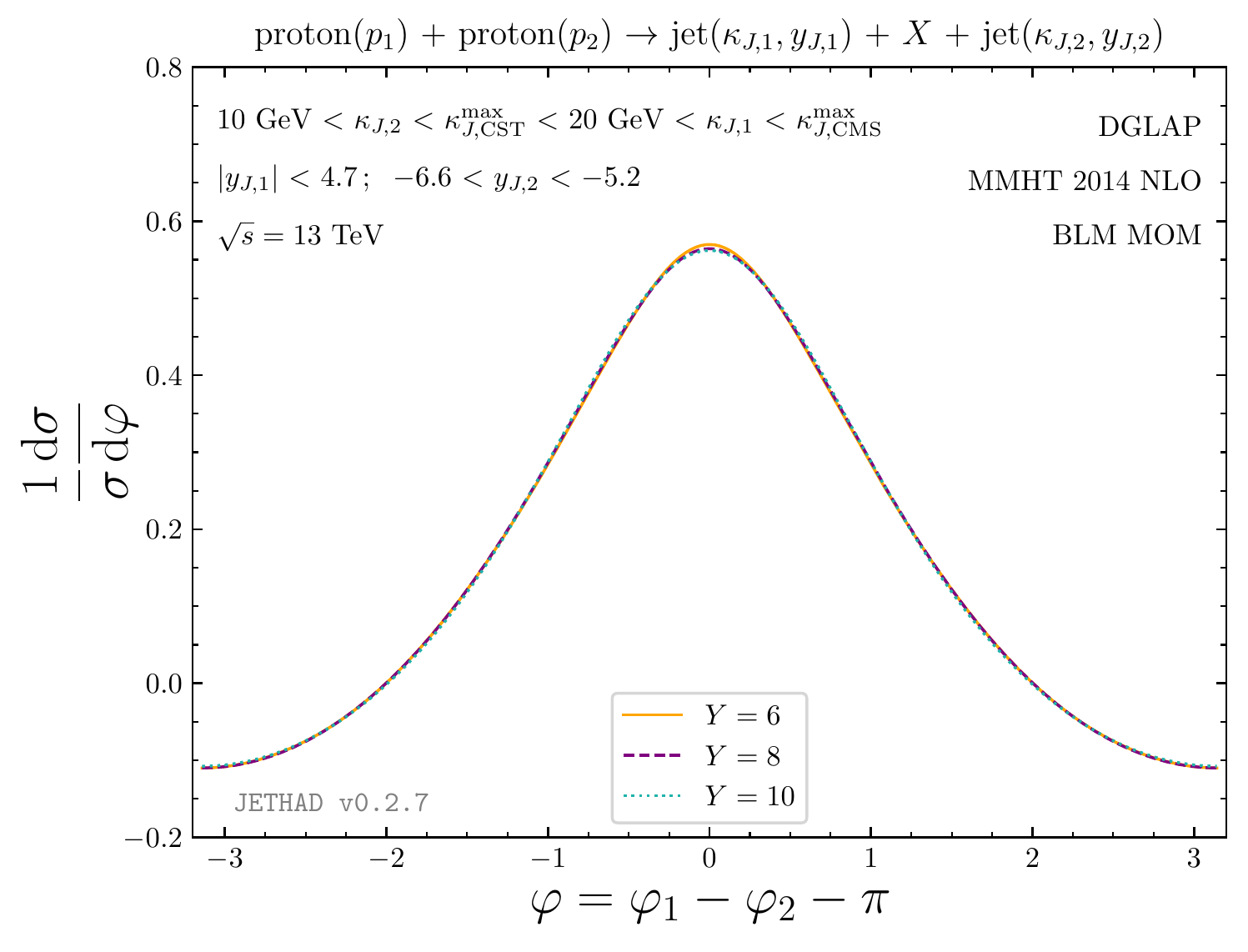}

\caption{NLA BFKL (left) and high-energy DGLAP (right) predictions for the azimuthal distribution of the Mueller--Navelet jet production (left panel of Fig.~\ref{fig:processes}), for three distinct values of the final-state rapidity interval, $Y$, and $\sqrt{s} = 13$ TeV. Results in the \textit{asymmetric CMS} (\textit{CASTOR-jet}) configuration are given in upper (lower) panels.}
\label{fig:MN-PHI-CMS-CST}
\end{figure}
\begin{figure}[t]
\centering

   \includegraphics[scale=0.56,clip]{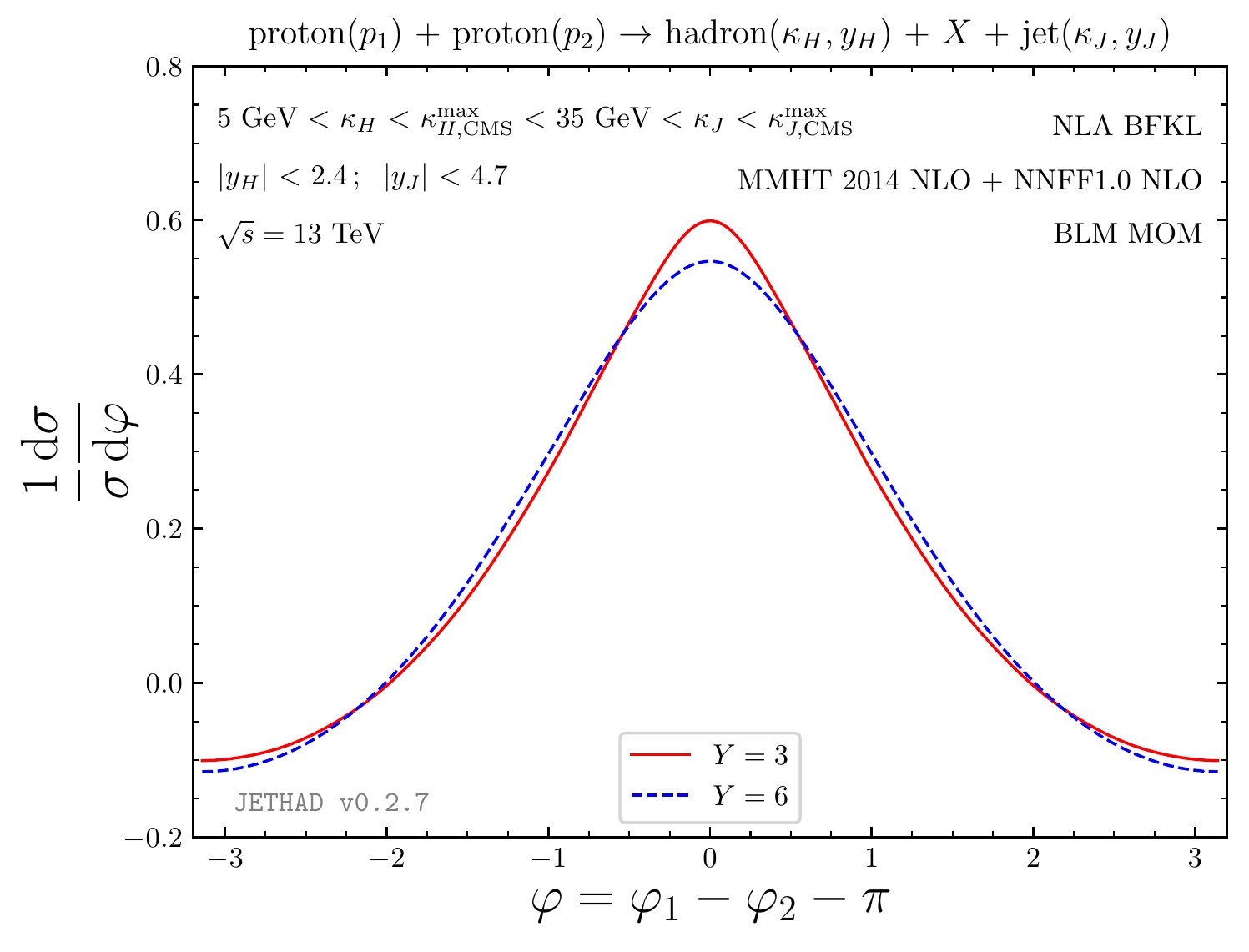}
   \hspace{-0.25cm}
   \includegraphics[scale=0.56,clip]{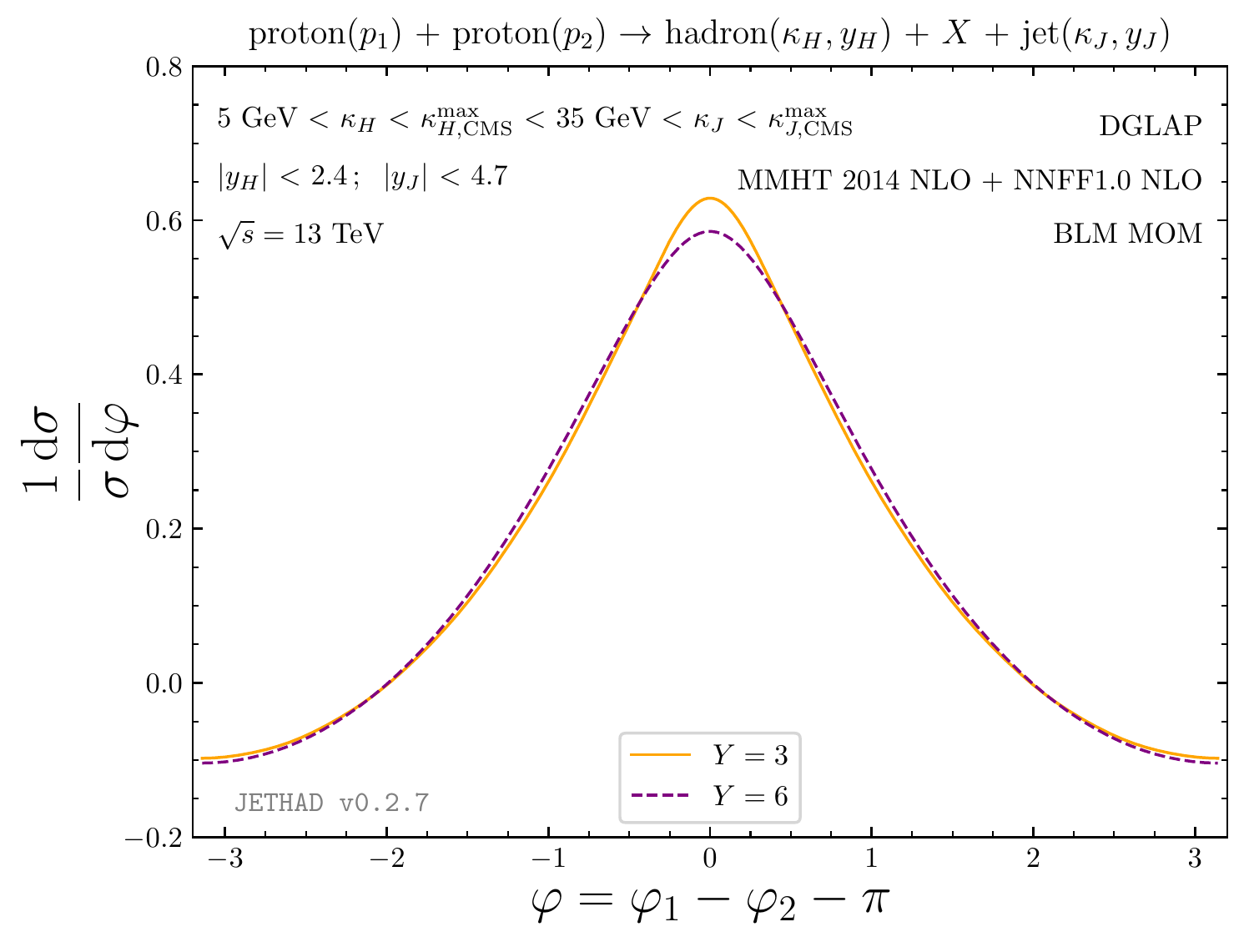}

   \includegraphics[scale=0.56,clip]{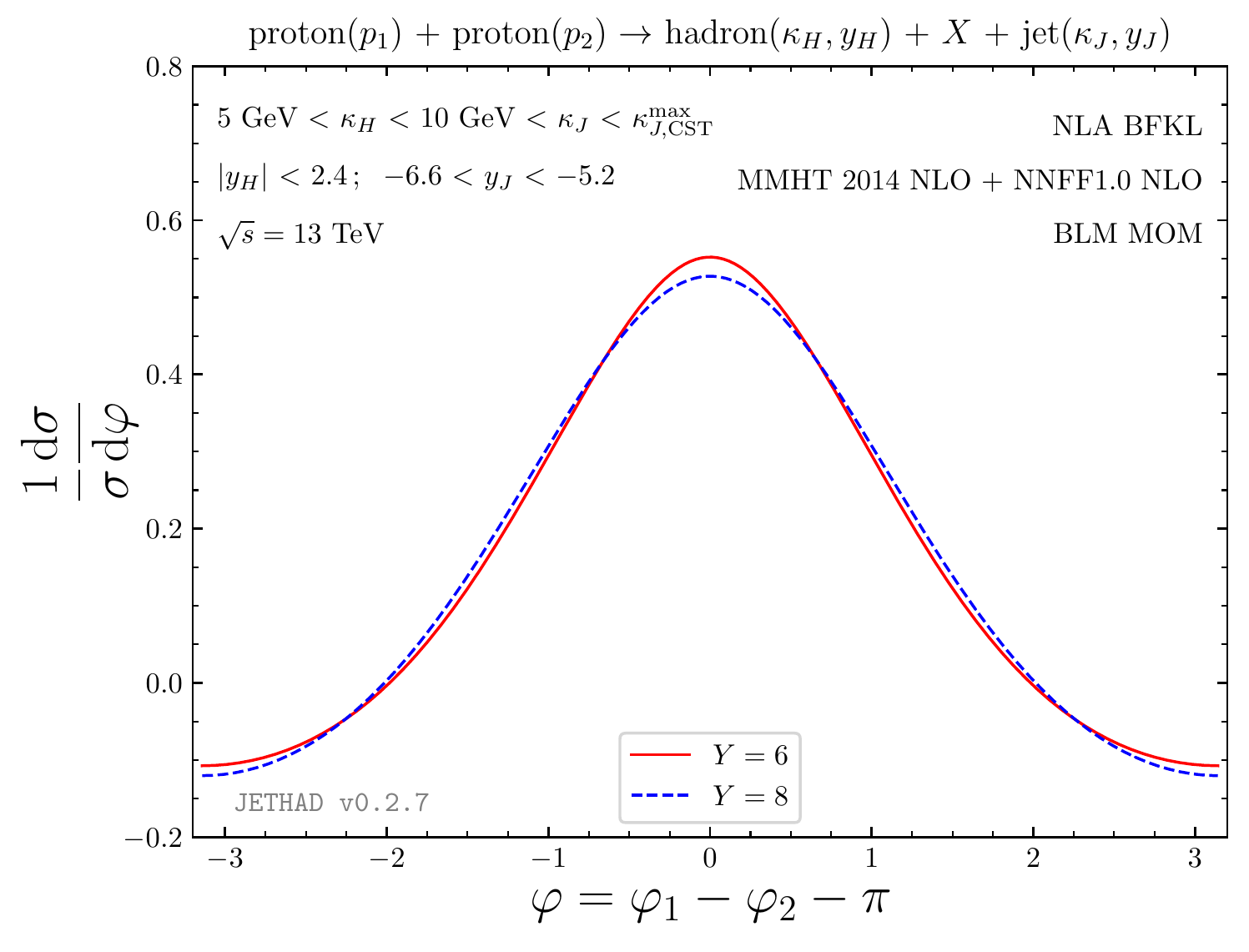}
   \hspace{-0.25cm}
   \includegraphics[scale=0.56,clip]{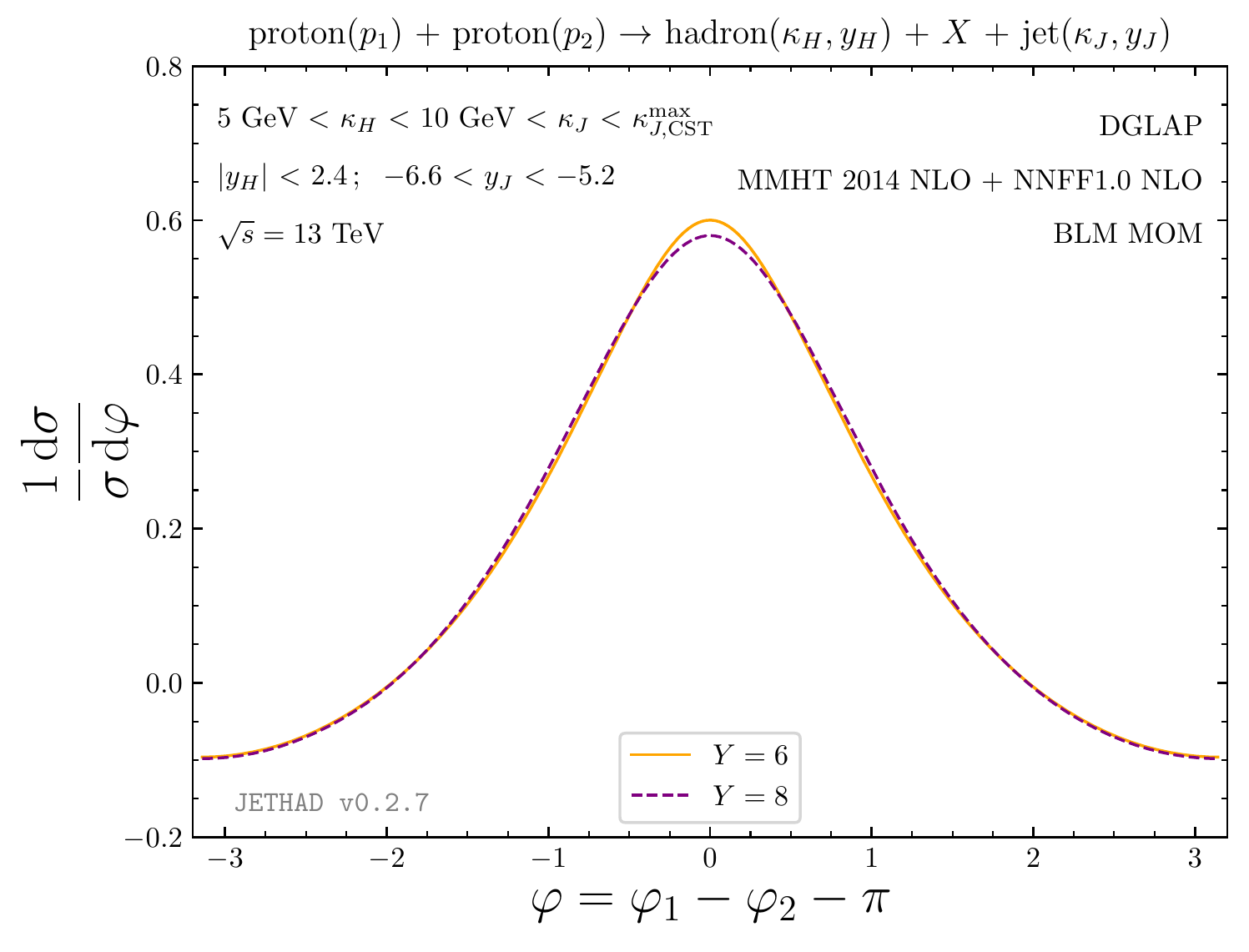}

\caption{NLA BFKL (left) and high-energy DGLAP (right) predictions for the azimuthal distribution of the inclusive hadron-jet production (central panel of Fig.~\ref{fig:processes}), for two distinct values of the final-state rapidity interval, $Y$, and $\sqrt{s} = 13$. Results in the \textit{asymmetric CMS} (\textit{CASTOR-jet}) configuration are given in upper (lower) panels.}
\label{fig:HJ-PHI-CMS-CST}
\end{figure}
\begin{figure}[t]
\centering

   \makebox[\linewidth]{
   \includegraphics[scale=0.56,clip]{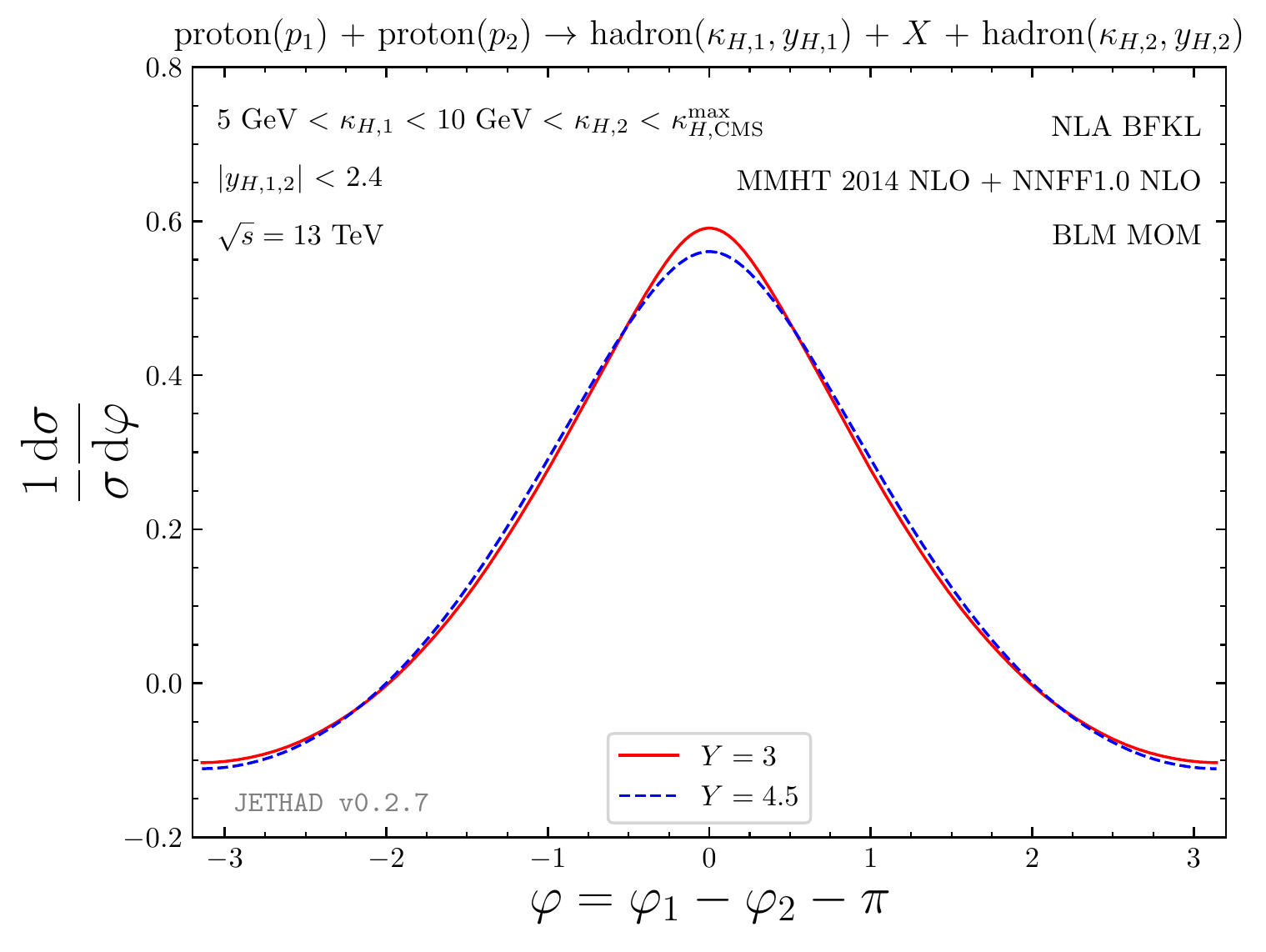}
   \hspace{-0.25cm}
   \includegraphics[scale=0.56,clip]{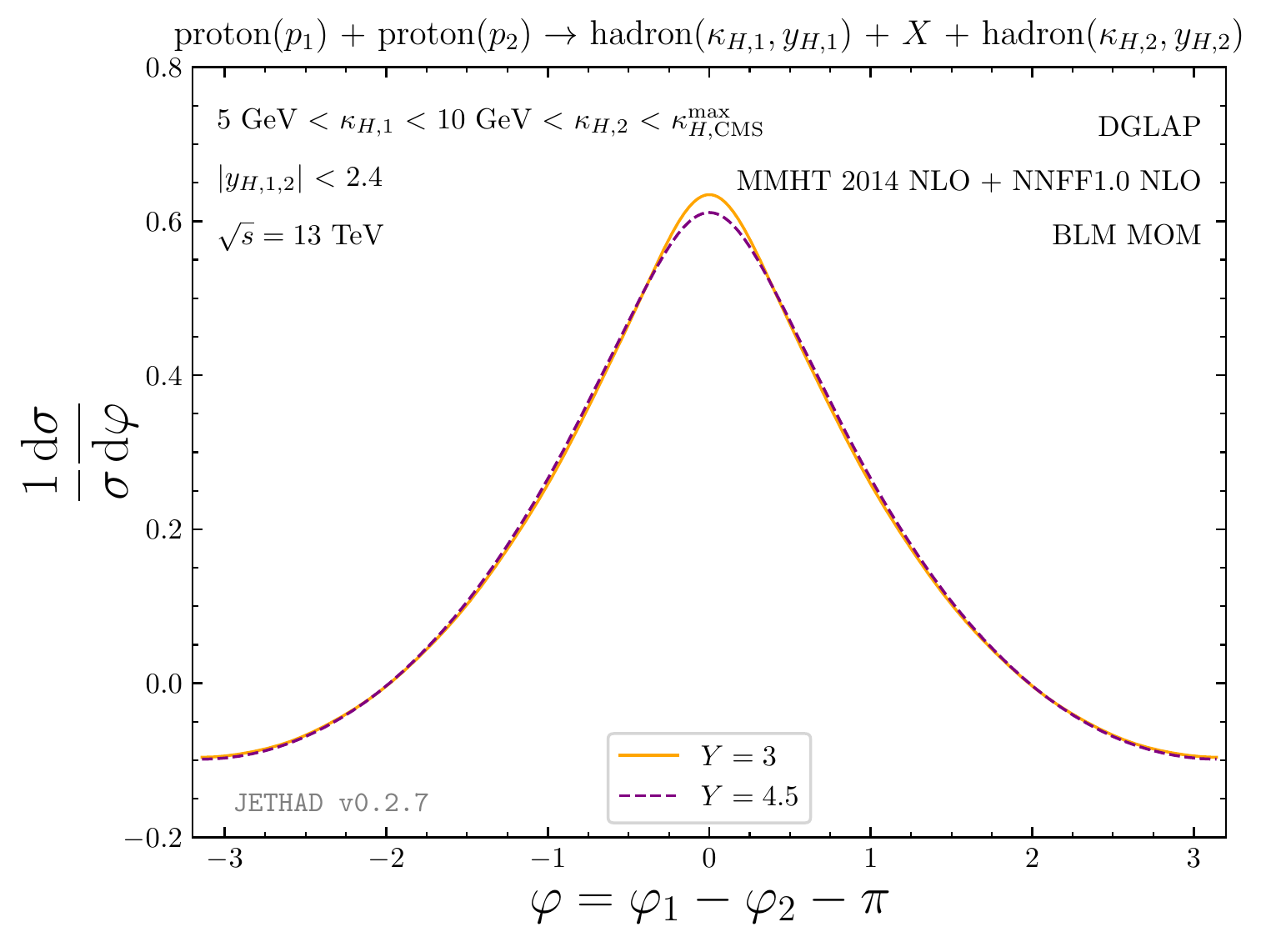}
   }

\caption{NLA BFKL (left) and high-energy DGLAP (right) predictions for the azimuthal distribution of the inclusive di-hadron production (right panel of Fig.~\ref{fig:processes}), for two distinct values of the final-state rapidity interval, $Y$, and $\sqrt{s} = 13$. Results are given in the \textit{asymmetric CMS} configuration.}
\label{fig:HH-PHI-CMS}
\end{figure}

We show predictions for the $R_{nm}$ ratios in the hadron-jet channel (central panel of Fig.~\ref{fig:processes}) in Fig.~\ref{fig:HJ-BvD-CMS} (\textit{asymmetric CMS}) and in Fig.~\ref{fig:HJ-BvD-CST} (\textit{CASTOR-jet}), whereas analogous results for the di-hadron production (right panel of Fig.~\ref{fig:processes}) are given in Fig.~\ref{fig:HH-BvD-CMS} (\textit{asymmetric CMS}). Here, presence of identified hadrons in the final state has a dual impact. On the one part, the effect of further splittings (of parton momenta) entering the definition of hadron FFs prevails over the azimuthal recorrelation of the emitted objects at large $Y$, thus shadowing the turn-up tail observed in the Mueller--Navelet case. On the other part, the convolution between PDFs and FFs in the LO/NLO hadron impact factor(s) (Eqs.~(\ref{hadron_IF_LO})~and~(\ref{hadron_IF_NLO})) stabilizes the oscillations of the $\nu$-integrand, thus reducing the numeric uncertainty in the calculation of the high-energy DGLAP series, whose error bands are now of the same magnitude of the NLA BFKL ones.

The unphysical effect that brings $R_{10}$ above one for small values of the rapidity interval, $Y$, is well known in the context of semi-hard reactions and has a straightforward explanation. In this kinematic limit, the hard-subprocess energy, $\sqrt{\hat s}$, is small and the employment of BFKL resummation, which systematically neglects terms damped by powers of $\hat s$, becomes inadequate. The fact that, in the DGLAP case, values of $R_{10}$ larger than one are still present for larger $Y$-values originates from the nature of our approach. Being it, \emph{de facto}, an high-energy calculation where the NLA resummation is truncated to a given order in the running coupling~(see Section~\ref{DGLAP}), \emph{collinear contaminations}, which are largely present at low values of the conformal spin and at small-$Y$, may survive and manifest themselves also in the remaining part of the $Y$-range.

The overall outcome of the analysis conducted in this Section is a clear separation between BFKL and DGLAP predictions on azimuthal correlations of the two detected objects. This effects holds for all the considered reactions and becomes more and more evident as the rapidity interval, $Y$, raises. The found pattern matches \emph{in toto} the farsighted idea of Mueller and Navelet~\cite{Mueller:1986ey}, namely that the wealth of undetected gluons radiated in the final state, theoretically described at the hand of the high-energy resummation, markedly heighten the decorrelation in the azimuthal plane between the emitted particles.
This makes a substantial difference with respect to the DGLAP case, where only a limited number of gluon emissions, fixed by the truncation order of the perturbatuve series, is allowed. The adoption of asymmetric cuts for the transverse momenta in the final state fades the Born contribution, thus spotlighting the discrepancy between the two approaches.

\subsubsection{Azimuthal distribution}
\label{azimuthal_distribution}

Here we give predictions for azimuthal distributions, which, as anticipated, are directly accessible observables in the experimental studies. The first examination of these quantities was performed few years ago~\cite{Ducloue:2013wmi} in the context of the Mueller--Navelet jet production, for symmetric ranges of transverse momenta, and with partial and full NLA BFKL accuracy.

Results for the azimuthal distribution in the di-jet channel are shown with NLA BFKL (high-energy DGLAP)
accuracy in the left (right) panels of Fig.~\ref{fig:MN-PHI-CMS-CST}, for three distinct values of the rapidity interval, $Y$. Predictions in the \textit{asymmetric CMS} configuration are given in upper panels, while the lower ones refer to the \textit{CASTOR-jet} selection. Fig.~\ref{fig:HJ-PHI-CMS-CST} shows, in the same way, the azimuthal distribution in the hadron-jet emission for two distinct values of $Y$, whareas plots of Fig.~\ref{fig:HH-PHI-CMS} correspond to the di-hadron distribution with \textit{asymmetric CMS} cuts, for two distinct values of $Y$.

The peculiar shape of our series represent a further manifestation of the BFKL-versus-DGLAP ``dicotomy''. All distributions feature a clear peak in correspondence of the value of the azimuthal-angle difference, $\varphi$~$\equiv$~$\varphi_1$~$-$~$\varphi_2$~$-$~$\pi$, for which the two final-state objects are emitted back-to-back, \emph{i.e.} $\varphi = 0$. Let us start by considering, for each panel, the first entry, which corresponds to the lower value of $Y$ among the selected ones. Both in the BFKL and in the DGLAP cases, this peak dominates over the larger-$Y$ ones, the DGLAP being narrower in comparison to the BFKL one. Then, when one moves towards the larger values of $Y$, peaks visibly shrink and widths moderately widen in the BFKL case, while that switch is much less evident for the DGLAP series. This means that, when the rapidity interval grows, the number of back-to-back events predicted by BFKL decreases, while it remains relatively unchanged according to DGLAP. Our outcome is in perfect agreement with the main statement raised in Section~\ref{core}. 

As a side consideration, we notice that trends obtained in the \textit{CASTOR-jet} ranges always show a slightly larger decorrelation with respect to \textit{asymmetric CMS} patterns (lower panels versus upper ones in Figs.~\ref{fig:MN-PHI-CMS-CST} and~\ref{fig:HJ-PHI-CMS-CST}). The added value of the more exclusive kinematic configurations offered by tagging a jet in the CASTOR ultra-backward detector translates in an intrinsic decorrelation gained by the final state.

All these features brace the message that kinematic configurations suitable to heighten high-energy effects in the context of the LHC phenomenology exist and have been effectively detected.

\subsection{Numerical strategy}
\label{numerics}

\subsubsection{{\Jethad}: an object-based, process-independent interface}
\label{jethad}

All numerical calculations were performed using {\Jethad}, a hybrid \textsc{Fortran2008/Python3} modular code we recently developed, suited for the computation of cross sections and related observables for inclusive semi-hard reactions. Natively equipped with performance acceleration, by making extensive use of parallel computing techniques, and interfaced with the most advanced (multi)dimensional integration routines, {\Jethad} allowed us to dynamically select the best integration algorithm, depending on the behavior of the considered integrand. \emph{Monte-Carlo} inspired integrators with variance reduction via \emph{importance sampling}, like the {\tt Vegas} routine~\cite{Lepage:1977sw} as given in its concurrent version via the {\tt Cuba}  package 4.2~\cite{Hahn:2004fe,Hahn:2014fua}, were primarily selected to perform multidimensional integrations needed to calculate the $C_n^{\rm DGLAP}$ coefficients (Eq.~(\ref{DGLAP_Cn})) in the Mueller--Navelet channel (left panel of Fig.~\ref{fig:processes}). Conversely, \emph{adaptive-quadrature} based functions, like {\tt DADMUL} and {\tt WGauss} as implemented in the last version of \textsc{Cern} program libraries~\cite{cernlib}, were mainly preferred for the computation of all observables related to (di-)hadron emission (central and right panels of Fig.~\ref{fig:processes}) and for the one-dimensional integration over the longitudinal momentum fraction, $\zeta$, entering the expression for the NLO hadron/jet impact factors (Eqs.~(\ref{hadron_IF_NLO}) and~(\ref{jet_IF_NLO})), respectively. 
All PDF parametrizations, as well as the NNFF1.0 FF set, were calculated through the Les Houches Accord PDF interpolator (LHAPDF) 6.2.1~\cite{Buckley:2014ana}, while native routines for the remaining FFs were direclty linked to the corresponding module in our code. In order to dynamically select the considered reaction through a common interface, a \emph{structure}-based smart-managment system, where physical final-state particles are described in terms of \emph{object} prototypes (\emph{i.e.}, \textsc{Fortran} structures), was incorporated inside {\Jethad}. Particle objects carry all information about basic and kinematic properties of their physical counterparts, from mass and charge, to kinematic ranges and rapidity tag. They are first loaded by the {\Jethad} user routine from the master database through a specific \emph{particle generator} routine (custom-particle generation is allowed too). Then, they are \emph{cloned} to the final-state object array and thus \emph{injected} from the integrand routine, differential on the final-state variables, to the respective impact-factor module through the \emph{impact-factor controller}. To the strong flexibility in the final-state generation, a wide choice for the initial-state selection corresponds. Thanks to a peculiar \emph{particle-ascendancy} structure attribute, {\Jethad} is able indeed to recognize if an object is hadroproduced or emitted as subproduct of a leptonic interaction, and automatically determines which modules need to be initialized (PDFs, FFs, unintegrated densities, etc...), breaking down computing-time lags. Hence, {\Jethad} comes as an object-based interface, completely independent on the reaction being investigated. While inspired by the BFKL phenomenology, it is possible to perform calculations of the same observables in different approaches, by creating new, customized routines, which can be linked to the core structure of the code through a native {\it point-to-routine} system, thus making {\Jethad} a general, HEP-purposed tool. We pursue, as a medium-term goal, to release a first public version soon, providing the scientific community with a standard software for the analysis of inclusive semi-hard processes. Another code, {\Lexa}, based on the same framework and suited to the study of exclusive semi-hard reactions, is currently at an early-development stage.

A robust improvement of our technology would consist in successfully interfacing these codes with already existing software, suited to high-energy/small-$x$ studies. An incomplete list includes: novel software for NLA studies~\cite{Chevallier:2009cu,Hentschinski:2014lma,Hentschinski:2014esa,Royon:2020vfl} of Mueller--Tang (\emph{alias} jet-gap-jet) configurations~\cite{Mueller:1992pe}, the {\tt BFKLex} Monte Carlo~\cite{Chachamis:2011nz,Chachamis:2012qw,Chachamis:2016ejm}, designed to investigate the high-energy jet production in the \emph{multi-Regge} limit, the {\tt KaTie} generator~\cite{vanHameren:2016kkz} of off-shell matrix elements, and the {\tt TMDlib} library~\cite{Hautmann:2014kza}, where different models of the small-$x$ UGD are collected.

\subsubsection{Uncertainty estimate}
\label{uncertainty}

The most relevant uncertainty comes from the numerical 4-dimensional integration over the transverse momenta, $\kappa_{1,2}$, of the two final-state objects, the rapidity, $y_{1/2}$, of one of them (the other one is fixed by the condition $Y = y_1 - y_2$ enforced in  the definition of the integrated azimuthal coefficients~(\ref{int_Cn})), and over $\nu$. Its effect was directly estimated and given as output by master integrator routines. Further, secondary sources of uncertainty, are respectively: the one-dimensional integration over the parton longitudinal fraction, $x$, needed to perform the convolution between PDFs and, eventually, FFs in the LO/NLO hadron impact factors (Eq.~(\ref{hadron_IF_LO})) for LO, Eq.~(\ref{hadron_IF_NLO})), the one-dimensional integration over the longitudinal momentum fraction, $\zeta$, in the NLO impact factor corrections (Eq.~(\ref{hadron_IF_NLO}) for hadrons, Eq.~(\ref{jet_IF_NLO}) for jets), and the upper cutoff in the numerical integral over $\nu$. While the first two ones turn out to be negligible with respect to the multidimensional integration, the last one deserves particular attention. As pointed out in Section~3.3 of Ref.~{\cite{Celiberto:2015yba}}, the $C_n^{\rm DGLAP}$ coefficients are expected to exhibit a stronger sensitivity to the upper cutoff of the $\nu$-integration, $\nu^{\rm max}$, due to the fact that oscillations rising in the $\nu$-integrand in Eq.~(\ref{DGLAP_Cn}) are not quenched by the exponential factor as in the NLA and LLA BFKL expressions (Eqs.~(\ref{BFKL_Cn}) and~(\ref{LLA_Cn}), respectively). This turns to be particularly true in the Mueller--Navalet channel (left panel of Fig.~\ref{fig:processes}), while the peculiar structure of the $x$-integrand in the LO hadron impact factor (Eq.~(\ref{hadron_IF_LO})) generates, as the net result, a significant damping of the $\nu$-associated oscillations both in the DGLAP and in the NLA/LLA BFKL cases, allowing us to use smaller values of $\nu^{\rm max}$ in hadron-jet and di-hadron production (central and right panels of Fig.~\ref{fig:processes}) with respect to the di-jet case. Values for $\nu^{\rm max}$, given below (Tab.~\ref{tab:numax}), have been taken after checking that the uncertainty on the $R_{nm}$ ratios, coming from raising them by a value of 10, is negligible if compared to the valued error of the multidimensional integration.

Error bands of all predictions presented in this work are given in terms of the uncertainty on the final-state integration, combined, in case of hadron emission (central and right panels of Fig.~\ref{fig:processes}), with the one coming from averaging over different FF sets (see Section~\ref{PDF_FF} for further details). An exception is representend by the \emph{theory-versus-experiment} analysis in the Mueller--Navelet channel (Section~\ref{th_vs_exp}), where error bands in the two panels of  Fig.~\ref{fig:MN-tve-CMS} show the standard deviation of predictions for the $R_{nm}$ ratios calculated via the replica method.

%%%%%%%%%% nu max %%%%%%%%%%%%%%%
\begin{table}[h!]
\setlength{\tabcolsep}{0.40cm}
\centering
\caption{Values of the upper cutoff, $\nu^{\rm max}$, on the numerical $\nu$-integration of the $C_n$ coefficients in the NLA/LLA BFKL accuracy ((Eqs.~(\ref{BFKL_Cn}) and~(\ref{LLA_Cn})) and in the high-energy DGLAP limit (Eq.~(\ref{DGLAP_Cn})), for all considered reactions (Fig.~\ref{fig:processes}).}
\label{tab:numax}
\begin{tabular}{c|c|c|c}
\hline\noalign{\smallskip}
%\toprule
$\nu^{\rm max}$ & Mueller--Navelet & \phantom{....}hadron-jet\phantom{....} & \phantom{....}di-hadron\phantom{....} \\
\noalign{\smallskip}\hline\noalign{\smallskip}
%\midrule
NLA/LLA & 30 & 10 & 10 \\
\noalign{\smallskip}\hline\noalign{\smallskip}
%\midrule
DGLAP   & 50 & 10 & 10 \\
\noalign{\smallskip}\hline
%\bottomrule
\end{tabular}
\end{table}

\section{Closing statements}
\label{conclusions}

We brought evidence that distinctive signals of the high-energy resummation emerge in the NLA description of different semi-hard reactions. Taking advantage of the record energies and of the exclusive kinematic configurations provided by the LHC, their effects can be effectively disengaged from the ones arising from (pure) fixed-order, DGLAP-inspired calculations. In this direction, the use of (completely) asymmetric intervals for the transverse momenta of the detected objects plays a crucial role and definitely needs to be taken into account in the forthcoming experimental analyses on forward/backward final-state emissions. Among them, stringent measurements of the azimuthal-angle averaged cross section, $C_0$, turn out to be essential not only in the discrimination of BFKL from other theoretical approaches, but also in the assessment of the intrinsic ambiguities of the given approach. 

With the goal of providing, in the medium-term future, comparisons with genuine fixed-order calculations, the combined effect of a selection of potential uncertainties was gauged. 
Our preference fell on those uncertainties which are typically included in collinear-physics phenomenology, but they turn out to be novel in the semi-hard one. More in particular, we studied the sensitivity of azimuthal-correlation moments on PDF uncertainty via the so-called replica method~\cite{Forte:2002fg} (this was done for the Mueller--Navelet channel at~$\sqrt{s}~=~7$~TeV (Section~\ref{th_vs_exp})) and on FF uncertainty, by averaging among four distinct FF sets.
A detailed analysis of all the potential sources of uncertainty is postponed to future studies, when data for our disjoint $\kappa$-windows will be available.
Then, two complementary paths should be traced and concurrently treated in the context of our studies. 

Pursuing the goal of dealing with more exclusive final states, the investigation of heavy-flavored emissions has been taken into account in our program~\cite{Bolognino:2019yls,Celiberto:2017nyx,Bolognino:2019ouc}, with the medium-term target of including the case of quarkonia (for a recent work on the inclusive $J/\psi$-plus-jet hadroproduction, see Ref.~\cite{Boussarie:2017oae}). Here, the ultimate task relies upon a profound examination of the theoretical production mechanism~\cite{Brambilla:2010cs,Bodwin:2013nua,Andronic:2015wma} for these quark-bound states, which has not yet been completely outfound, and on the grading of the most popular models proposed so far~\cite{Fritzsch:1977ay,Halzen:1977rs,Bodwin:1994jh}. Various benefits can be gained from considering single forward emissions. First, the experimental statistics appreciably increases. Second, it allows us to probe different frameworks for the unintegrated gluon distribution (UGD), including the ones~\cite{Bacchetta:2020vty,Celiberto:2021zww} whose definition suitably embodies non-perturbative inputs driven by the \emph{transverse-momentum-dependent} (TMD) factorization (see, \emph{e.g.}, Refs.~\cite{Rogers:2015sqa,Diehl:2015uka,Angeles-Martinez:2015sea} and references therein for an overview on the general framework, Refs.~\cite{Boffi:2009sh,Pasquini:2011tk,Pasquini:2014ppa,Dai:2014ala,Echevarria:2015uaa,Boer:2016xqr,Lansberg:2017tlc,Bacchetta:2017gcc,Radici:2018iag,Boglione:2018dqd,Anselmino:2018psi,Buffing:2018ggv,Boglione:2019nwk,Bertone:2019nxa,Gutierrez-Reyes:2019vbx,Gutierrez-Reyes:2019rug,Echevarria:2019ynx,Pasquini:2019evu,DAlesio:2019qpk,Scarpa:2019fol, DAlesio:2019gnu,Fleming:2019pzj,Scimemi:2019cmh,Bacchetta:2019sam,Luo:2020hki,Boer:2020bbd,Bacchetta:2020gko,Bastami:2020asv} for quite recent applications) together with effective small-$x$ effects.

The analysis of more differential distributions covering broader kinematic ranges requires an ineludible effort into the enhancement of our formalism  to accommodate other resummation mechanisms. A major outcome of a recent work on the inclusive hadroproduction of a Higgs boson and a jet well separated in rapidity~\cite{Celiberto:2020tmb} is that an exhaustive study of the Higgs transverse-momentum distribution lies on the exploration of neighboring kinematic regions, each of them representing a preferred channel to probe a specific resummation. That calls for a net overhaul of our framework (first from the analytic and then from the numeric point of view) which would not anymore rely \emph{only} on the pure BFKL approach, but should rather evolve into an underlying \emph{staging} where distinct resummations are \emph{plugged-in} and play their part. 

One step forward is represented by the Altarelli--Ball--Forte (ABF) formalism~\cite{Ball:1995vc,Ball:1997vf,Altarelli:2001ji,Altarelli:2003hk,Altarelli:2005ni,Altarelli:2008aj,White:2006yh}, where DGLAP and BFKL inputs are combined to improve the perturbative accuracy of the resummed series, by imposing consistency conditions (duality aspects), symmetrizing the BFKL kernel in the (anti-)collinear phase-space regions, and incorporating those contributions to running coupling which affect the small-$x$ divergences. Thence, the interplay between collinear and high-energy factorization allows us to perform the resummation of coefficient functions, whereas the resummation of splitting functions is endowed by enforcing the consistency between the two evolution equations. The first determination of proton PDFs where NLO and next-to-NLO fixed-order calculations are supplemented by the NLA small-$x$ resummation has been recently realized~\cite{Ball:2017otu,Abdolmaleki:2018jln,Bonvini:2019wxf}, by making use of the {\Hell}~\cite{Bonvini:2016wki,Bonvini:2017ogt} numerical code interfaced to the {\Apfel}~\cite{Bertone:2013vaa,Carrazza:2014gfa,Bertone:2017gds} PDF-evolution library.

Another intriguing possibility is represented by the Catani--Ciafaloni--Fiorani--Marchesini (CCFM) \emph{branching} scheme~\cite{Ciafaloni:1987ur,Catani:1989sg,Catani:1989yc,Marchesini:1994wr}. Enforcing angular ordering (\emph{coherence}) of soft-parton emission from the large-$x$ to the small-$x$ kinematic ranges, the CCFM framework provides with a unified evolution pattern for unintegrated gluon densities. In the totally inclusive configuration, this evolution interpolates between the DGLAP equation at moderate-$x$ and the BFKL one at small-$x$.
For large values of $x$ and for high virtualities, CCFM dynamics matches DGLAP evolution, whereas in the asymptotic-energy limit it \emph{almost} corresponds to BFKL (see Refs.~\cite{Forshaw:1998uq,Webber:1998we,Salam:1999ft}).
It essentialy builds on the sum of ladder diagrams with angular ordering along the chain. 
Parton transverse momentum is generated via the genuine recoil effect due to gluon radiation controlling the UGD evolution.
The direct employment of CCFM to evolution equations for parton distributions was proposed by Jan Kwieci{\'n}ski~\cite{Kwiecinski:2002bx} through the so-called $\mbox{CCFM--K}$ equations.
The CCFM formalism was then generalized~\cite{Kutak:2011fu} in order to account for \emph{non-linear} effects in the gluon evolution, thus giving us a chance to gauge the impact of parton saturation on exclusive observables. On the phenomenological side, a description of the Drell--Yan production at the hand of $\mbox{CCFM--K}$-evolved distributions was recently proposed~\cite{Golec-Biernat:2019scr}.

We strongly believe the scientific community would largely benefit from  a \emph{multi-lateral} formalism in which several approaches (as the high-energy resummation, the threshold resummation at fixed transverse momentum~\cite{Bonciani:2003nt,deFlorian:2005fzc,Muselli:2017bad}, the transverse-momentum resummation~\cite{Dokshitzer:1978yd,Dokshitzer:1978hw,Parisi:1979se,Curci:1979bg,Collins:1981uk,Collins:1981va,Collins:1984kg,Kodaira:1981nh,Kodaira:1982cr,Kodaira:1982az,Catani:2000vq,Bozzi:2005wk,Bozzi:2008bb,Catani:2010pd,Catani:2011kr,Catani:2013tia,Catani:2015vma} and the resummation of Sudakov-type logarithms emerging when almost back-to-back final-state configurations occur in the small-$x$ limit~\cite{Mueller:2012uf,Mueller:2013wwa,Balitsky:2015qba,Marzani:2015oyb,Mueller:2015ael,Xiao:2018esv}) coexist and can be concurrently employed in the description of an increasing number of hadronic and lepto-hadronic reactions at the LHC as well as at new-generation colliding machines, as NICA-SPD~\cite{Arbuzov:2020cqg}, HL-LHC~\cite{Chapon:2020heu} and the EIC~\cite{AbdulKhalek:2021gbh}.

\vspace{0.75cm} \hrule %\vspace{0.75cm}

% \section{Acknowledgments}
\acknowledgments

This work was supported by the Italian Foundation ``Angelo Della Riccia'' and by the Italian Ministry of Education, Universities and Research under the FARE grant ``3DGLUE'' (n. R16XKPHL3N).

The author would like to express his gratitude to Alessandro~Papa, Dmitry~Yu.~Ivanov and Valerio~Bertone for a critical reading of the manuscript, for useful suggestions and for encouragement.

The author thanks Alessandro~Bacchetta, Joachim~Bartels, Chiara~Bissolotti, Andr{\`e}e~Dafne~Bolognino, Giuseppe Bozzi, Stanley~Brodsky, Francesco Caporale, Grigorios~Chachamis, Victor~S.~Fadin, Michael~Fucilla, Krzysztof~J.~Golec-Biernat, David~Gordo~G{\'o}mez, Krzysztof~Kutak, Leszek~Motyka, Mohammed~M.A.~Mohammed, Beatrice Murdaca, Barbara~Pasquini, Fulvio~Piacenza, Marco~Radici, Christophe~Royon, Douglas~A.~Ross, Deniz~Sunar~Cerci, Pieter~Taels, Agust{\'i}n~Sabio~Vera, Antoni~Szczurek and Lech~Szymanowski for inspiring discussions and for stimulating conversations.

Feynman diagrams in this work were realized via the {\tt JaxoDraw 2.0} interface~\cite{Binosi:2008ig}.

\vspace{0.75cm} \hrule

\appendix

\setcounter{appcnt}{0}
\hypertarget{app:hadron_IF_NLO_link}{
\section{Forward-hadron NLO impact factor}
                                    }
\label{app:hadron_IF_NLO}

We give below the analytic formula for the NLO impact-factor correction for the forward-hadron production:

\begin{equation}
  \label{hadron_IF_NLO}
  \hat c^{(H)}(n,\nu,\kappa_H,x_H)=
  2\sqrt{\frac{C_F}{C_A}}
  \left(\kappa_H^2\right)^{i\nu-\frac{1}{2}}\frac{1}{2\pi}
  \int_{x_H}^1\frac{\drv x}{x}
  \int_{\frac{x_H}{x}}^1\frac{\drv \zeta}{\zeta}
  \left(\frac{x\zeta}{x_H}\right)^{2i\nu-1}
\end{equation}
  \[ \times \,
  \left[
  \frac{C_A}{C_F}f_g(x)D_g^h\left(\frac{x_H}{x\zeta}\right)C_{gg}
  \left(x,\zeta\right)+\sum_{\alpha=q\bar q}f_\alpha(x)D_\alpha^h
  \left(\frac{x_H}{x\zeta}
  \right)C_{qq}\left(x,\zeta\right)
  \right.
  \]
  \[ \times \,
  \left.D_g^h\left(\frac{x_H}{x\zeta}\right)
  \sum_{\alpha=q\bar q}f_\alpha(x)C_{qg}
  \left(x,\zeta\right)+\frac{C_A}{C_F}f_g(x)\sum_{\alpha=q\bar q}D_\alpha^h
  \left(\frac{x_H}{x\zeta}\right)C_{gq}\left(x,\zeta\right)
  \right]\, ,
  \]

\begin{equation}
\stepcounter{appcnt}
\label{Cgg_hadron}
 C_{gg}\left(x,\zeta\right) =  P_{gg}(\zeta)\left(1+\zeta^{-2\gamma}\right)
 \ln \left( \frac {\kappa_H^2 x^2 \zeta^2 }{\mu_F^2 x^2}\right)
 -\frac{\beta_0}{2}\ln \left( \frac {\kappa_H^2 x^2 \zeta^2 }
 {\mu^2_R x^2}\right)
\end{equation}
\[
 + \, \delta(1-\zeta)\left[C_A \ln\left(\frac{s_0 \, x^2}{\kappa_H^2 \,
 x^2 }\right) \chi(n,\gamma)
 - C_A\left(\frac{67}{18}-\frac{\pi^2}{2}\right)+\frac{5}{9}n_f
 \right.
\]
\[
 \left.
 +\frac{C_A}{2}\left(\psi^\prime\left(1+\gamma+\frac{n}{2}\right)
 -\psi^\prime\left(\frac{n}{2}-\gamma\right)
 -\chi^2(n,\gamma)\right) \right]
 + \, C_A \left(\frac{1}{\zeta}+\frac{1}{(1-\zeta)_+}-2+\zeta\bar\zeta\right)
\]
\[
 \times \, \left(\chi(n,\gamma)(1+\zeta^{-2\gamma})-2(1+2\zeta^{-2\gamma})\ln\zeta
 +\frac{\bar \zeta^2}{\zeta^2}{\cal I}_2\right)
\]
\[
 + \, 2 \, C_A (1+\zeta^{-2\gamma})
 \left(\left(\frac{1}{\zeta}-2+\zeta\bar\zeta\right) \ln\bar\zeta
 +\left(\frac{\ln(1-\zeta)}{1-\zeta}\right)_+\right) \ ,
\]

\begin{equation}
\stepcounter{appcnt}
\label{Cgq_hadron}
 C_{gq}\left(x,\zeta\right)=P_{qg}(\zeta)\left(\frac{C_F}{C_A}+\zeta^{-2\gamma}\right)\ln \left( \frac {\kappa_H^2 x^2 \zeta^2 }{\mu_F^2 x^2}\right)
\end{equation}
\[
 + \, 2 \, \zeta \bar\zeta \, T_R \, \left(\frac{C_F}{C_A}+\zeta^{-2\gamma}\right)+\, P_{qg}(\zeta)\, \left(\frac{C_F}{C_A}\, \chi(n,\gamma)+2 \zeta^{-2\gamma}\,\ln\frac{\bar\zeta}{\zeta} + \frac{\bar \zeta}{\zeta}{\cal I}_3\right) \ ,
\]

\begin{equation}
\stepcounter{appcnt}
\label{qg}
 C_{qg}\left(x,\zeta\right) =  P_{gq}(\zeta)\left(\frac{C_A}{C_F}+\zeta^{-2\gamma}\right)\ln \left( \frac {\kappa_H^2 x^2 \zeta^2 }{\mu_F^2x^2}\right)
\end{equation}
\[
 + \zeta\left(C_F\zeta^{-2\gamma}+C_A\right) + \, \frac{1+\bar \zeta^2}{\zeta}\left[C_F\zeta^{-2\gamma}(\chi(n,\gamma)-2\ln\zeta)+2C_A\ln\frac{\bar \zeta}{\zeta} + \frac{\bar \zeta}{\zeta}{\cal I}_1\right] \ ,
\]

\begin{equation}
\stepcounter{appcnt}
\label{Cqq_hadron}
 C_{qq}\left(x,\zeta\right)=P_{qq}(\zeta)\left(1+\zeta^{-2\gamma}\right)\ln \left( \frac {\kappa_H^2 x^2 \zeta^2 }{\mu_F^2 x_H^2}\right)-\frac{\beta_0}{2}\ln \left( \frac {\kappa_H^2 x^2 \zeta^2 }{\mu^2_Rx_H^2}\right)
\end{equation}
\[
 + \, \delta(1-\zeta)\left[C_A \ln\left(\frac{s_0 \, x_H^2}{\kappa_H^2 \, x^2 }\right) \chi(n,\gamma)+ C_A\left(\frac{85}{18}+\frac{\pi^2}{2}\right)-\frac{5}{9}n_f - 8\, C_F \right.
\]
\[
 \left. +\frac{C_A}{2}\left(\psi^\prime\left(1+\gamma+\frac{n}{2}\right)-\psi^\prime\left(\frac{n}{2}-\gamma\right)-\chi^2(n,\gamma)\right) \right] + \, C_F \,\bar \zeta\,(1+\zeta^{-2\gamma})
\]
\[
 +\left(1+\zeta^2\right)\left[C_A (1+\zeta^{-2\gamma})\frac{\chi(n,\gamma)}{2(1-\zeta )_+}+\left(C_A-2\, C_F(1+\zeta^{-2\gamma})\right)\frac{\ln \zeta}{1-\zeta}\right]
\]
\[
 +\, \left(C_F-\frac{C_A}{2}\right)\left(1+\zeta^2\right)\left[2(1+\zeta^{-2\gamma})\left(\frac{\ln (1-\zeta)}{1-\zeta}\right)_+ + \frac{\bar \zeta}{\zeta^2}{\cal I}_2\right] \; ,
\]
where $s_0$ is an artificial normalization scale to be suitably fixed.
We define $\bar \zeta=1-\zeta$ and $\gamma=i\nu-1/2$, while $P_{i j}(\zeta)$ are LO DGLAP kernels:
\begin{eqnarray}
\stepcounter{appcnt}
\label{DGLAP_kernels}
 %\nonumber
 P_{gq}(z)&=&C_F\frac{1+(1-z)^2}{z} \; , \\ \nonumber
 P_{qg}(z)&=&T_R\left[z^2+(1-z)^2\right]\; , \\ \nonumber
 P_{qq}(z)&=&C_F\left( \frac{1+z^2}{1-z} \right)_+= C_F\left[ \frac{1+z^2}{(1-z)_+} +{3\over 2}\delta(1-z)\right]\; , \\ \nonumber
 P_{gg}(z)&=&2C_A\left[\frac{1}{(1-z)_+} +\frac{1}{z} -2+z(1-z)\right]+\left({11\over 6}C_A-\frac{n_f}{3}\right)\delta(1-z) \; .
\end{eqnarray}
As for the ${\cal I}_{1,2,3}$ functions, one can write:
\begin{equation}
\stepcounter{appcnt}
\label{I2}
{\cal I}_2=
\frac{\zeta^2}{\bar \zeta^2}\left[
\zeta\left(\frac{{}_2F_1(1,1+\gamma-\frac{n}{2},2+\gamma-\frac{n}{2},\zeta)}
{\frac{n}{2}-\gamma-1}-
\frac{{}_2F_1(1,1+\gamma+\frac{n}{2},2+\gamma+\frac{n}{2},\zeta)}{\frac{n}{2}+
\gamma+1}\right)\right.
\end{equation}
\[
 \stepcounter{appcnt}
 \left.
 +\zeta^{-2\gamma}\left(\frac{{}_2F_1(1,-\gamma-\frac{n}{2},1-\gamma-\frac{n}{2},\zeta)}{\frac{n}{2}+\gamma}-\frac{{}_2F_1(1,-\gamma+\frac{n}{2},1-\gamma+\frac{n}{2},\zeta)}{\frac{n}{2} -\gamma}\right)
\right.
\]
\[
 \left.
 +\left(1+\zeta^{-2\gamma}\right)\left(\chi(n,\gamma)-2\ln \bar \zeta \right)+2\ln{\zeta}\right] \; ,
\]
\begin{equation}
\stepcounter{appcnt}
\label{I1}
 {\cal I}_1=\frac{\bar \zeta}{2\zeta}{\cal I}_2+\frac{\zeta}{\bar \zeta}\left[\ln \zeta+\frac{1-\zeta^{-2\gamma}}{2}\left(\chi(n,\gamma)-2\ln \bar \zeta\right)\right] \; ,
\end{equation}
\begin{equation}
\stepcounter{appcnt}
\label{I3}
 {\cal I}_3=\frac{\bar \zeta}{2\zeta}{\cal I}_2-\frac{\zeta}{\bar \zeta}\left[\ln \zeta+\frac{1-\zeta^{-2\gamma}}{2}\left(\chi(n,\gamma)-2\ln \bar \zeta\right)\right] \; .
\end{equation}
In Eqs.~(\ref{Cgg_hadron}) and~(\ref{Cqq_hadron}) the \emph{plus-prescription} is introduced:
\beq
\label{plus-prescription}
\stepcounter{appcnt}
\int\limits^1_a \drv \zeta \frac{F(\zeta)}{(1-\zeta)_+}
=\int\limits^1_a \drv \zeta \frac{F(\zeta)-F(1)}{(1-\zeta)}
-\int\limits^a_0 \drv \zeta \frac{F(1)}{(1-\zeta)}\; ,
\eeq
with $F(\zeta)$ a generic function, regular at $\zeta=1$.

\setcounter{appcnt}{0}
\hypertarget{app:jet_IF_NLO_link}{
\section{Forward-jet NLO impact factor}
                                 }
\label{app:jet_IF_NLO}

In the $(n,\nu)$-representation, the expression for the NLO correction to the forward-jet impact factor in the small-cone limit reads (see Ref.~\cite{Caporale:2012ih} for further details):
\hypertarget{jet_IF_NLO}{}
\begin{equation}
\stepcounter{appcnt}
\label{jet_IF_NLO}
 \hat c^{(J)}(n,\nu,\kappa_J,x_J)=
 \frac{1}{\pi}\sqrt{\frac{C_F}{C_A}}
 \left(\kappa_J^2 \right)^{i\nu-1/2}
 \int\limits^1_{x_J}\frac{\drv \zeta}{\zeta}
 \zeta^{-\bar\alpha_s(\mu_R)\chi(n,\nu)}
\end{equation}
\[
\left\{\sum_{\alpha=q,\bar q} f_\alpha \left(\frac{x_J}{ \zeta}\right)\left[\left(P_{qq}(\zeta)+\frac{C_A}{C_F}P_{gq}(\zeta)\right)
\ln\frac{\kappa_J^2}{\mu_F^2}\right.\right.
-2\zeta^{-2\gamma}\ln R\,
\left\{P_{qq}(\zeta)+P_{gq}(\zeta)\right\}-\frac{\beta_0}{2}
\ln\frac{\kappa_J^2}{\mu_R^2}\delta(1-\zeta)
\]
\[
+C_A\delta(1-\zeta)\left(\chi(n,\gamma)\ln\frac{s_0}{\kappa_J^2}
+\frac{85}{18}+\frac{\pi^2}{2}+\frac{1}{2}\left(\psi^\prime
\left(1+\gamma+\frac{n}{2}\right)
-\psi^\prime\left(\frac{n}{2}-\gamma\right)-\chi^2(n,\gamma)\right)
\right)
\]
\[
+(1+\zeta^2)\left\{C_A\left(\frac{(1+\zeta^{-2\gamma})\,\chi(n,\gamma)}
{2(1-\zeta)_+}-\zeta^{-2\gamma}\left(\frac{\ln(1-\zeta)}{1-\zeta}\right)_+
\right)+\left(C_F-\frac{C_A}{2}\right)\left[ \frac{\bar \zeta}{\zeta^2}{\cal I}_2
-\frac{2\ln\zeta}{1-\zeta}
+2\left(\frac{\ln(1-\zeta)}{1-\zeta}\right)_+ \right]\right\}
\]
\[
+\delta(1-\zeta)\left(C_F\left(3\ln 2-\frac{\pi^2}{3}-\frac{9}{2}\right)
-\frac{5n_f}{9}\right)
+C_A\zeta+C_F\bar \zeta
\]
\[
\left. +\frac{1+\bar \zeta^2}{\zeta}
\left(C_A\frac{\bar \zeta}{\zeta}{\cal I}_1+2C_A\ln\frac{\bar\zeta}{\zeta}
+C_F\zeta^{-2\gamma}(\chi(n,\gamma)-2\ln \bar \zeta)\right)\right]
+f_{g}\left(\frac{x_J}{ \zeta}\right)\frac{C_A}{C_F}
\left[
\left(P_{gg}(\zeta)+2 \,n_f \frac{C_F}{C_A}P_{qg}(\zeta)\right)
\ln\frac{\kappa_J^2}{\mu_F^2}
\right.
\]
\[
\left.
-2\zeta^{-2\gamma}\ln R \left(P_{gg}(\zeta)+2 \,n_f P_{qg}(\zeta)\right)
-\frac{\beta_0}{2}\ln\frac{\kappa_J^2}{4\mu_R^2}\delta(1-\zeta)
+\, C_A\delta(1-\zeta)
\left(
\chi(n,\gamma)\ln\frac{s_0}{\kappa_J^2}+\frac{1}{12}+\frac{\pi^2}{6}
\right.\right.
\]
\[
\left.
+\frac{1}{2}\left(\psi^\prime\left(1+\gamma+\frac{n}{2}\right)
-\psi^\prime\left(\frac{n}{2}-\gamma\right)-\chi^2(n,\gamma)\right)
\right)
%+\,
%\delta(1-\zeta)\, \ln 2 \left(\frac{11 C_A}{3}-\frac{2n_f}{3}\right)
+\, 2 C_A (1-\zeta^{-2\gamma})\left(\left(\frac{1}{\zeta}-2
+\zeta\bar\zeta\right)\ln \bar \zeta + \frac{\ln (1-\zeta)}{1-\zeta}\right)
\]
\[
+ \, C_A\, \left[\frac{1}{\zeta}+\frac{1}{(1- \zeta)_+}-2+\zeta\bar\zeta\right]
\left((1+\zeta^{-2\gamma})\chi(n,\gamma)-2\ln\zeta+\frac{\bar \zeta^2}
{\zeta^2}{\cal I}_2\right)
\]
\[
\left.\left.
+\, n_f\left[\, 2\zeta\bar \zeta \, \frac{C_F}{C_A} +(\zeta^2+\bar \zeta^2)
\left(\frac{C_F}{C_A}\chi(n,\gamma)+\frac{\bar \zeta}{\zeta}{\cal I}_3\right)
-\frac{1}{12}\delta(1-\zeta)\right]\right]\right\} \; ,
\]
with $R$ the jet-cone radius, $s_0$ an artificial normalization scale whose value needs to be appropriately chosen, $\bar \zeta=1-\zeta$ and $\gamma=i\nu-1/2$. $P_{i j}(\zeta)$ are LO DGLAP kernels defined in Eq.~(\ref{DGLAP_kernels}), the ${\cal I}_{1,2,3}$ functions are given in Eqs.~(\ref{I2})-(\ref{I3}) and the plus-prescription is written in Eq.~(\ref{plus-prescription}).

\vspace{0.75cm} \hrule

\end{document}